\documentclass[%
 reprint,
superscriptaddress,
showpacs,
 amsmath,amssymb,
 aps,
]{revtex4-1}

\usepackage{graphicx}% Include figure files
\usepackage{dcolumn}% Align table columns on decimal point
\usepackage{bm}% bold math
\usepackage{color}

\begin{document}

\preprint{APS/123-QED}

\title{Exclusive diffractive production of real photons and vector mesons in a factorized Regge-pole model with non-linear Pomeron trajectory
}% Force line breaks with \\

\author{S.~Fazio}
\email{sfazio@bnl.gov}
\affiliation{Brookhaven National Laboratory, Physics Department, 11973 Upton, NY - USA.}
 %\altaffiliation[Also at] {Brookhaven National Laboratory, Physics Department, 11973 Upton, NY - USA.}%Lines break automatically or can be forced with \\
\author{R.~Fiore}%
\email{roberto.fiore@cs.infn.it}
\affiliation{Dipartimento di Fisica, Universit\`a della Calabria\\ 
and\\
Istituto Nazionale di Fisica Nucleare, Gruppo collegato di Cosenza,\\
I-87036 Arcavacata di Rende, Cosenza - Italy}
\author{L.~Jenkovszky}
\email{jenk@bitp.kiev.ua}
\affiliation{Bogolubov Institute for Theoretical Physics, National Academy of Sciences of Ukraine, \\
 UA-03680 Kiev - Ukraine}
\author{A.~Lavorini}
\email{aadelmo.lavorini@gmail.com}
\affiliation{Dipartimento di Fisica, Universit\`a della Calabria\\ 
and\\
Istituto Nazionale di Fisica Nucleare, Gruppo collegato di Cosenza,\\
I-87036 Arcavacata di Rende, Cosenza - Italy}

%\affiliation{
%\affiliation{
%\affiliation{%


\date{\today}% It is always \today, today,

\begin{abstract}
Exclusive diffractive production of real photons and vector mesons in ep collisions has been studied at HERA in a wide kinematic range. Here we present and discuss a Regge-type model of real photon production (Deeply Virtual Compton Scattering), as well as production of vector mesons (VMP) treated on the same footing by using an extension of a factorized Regge-pole model proposed earlier. The model has been fitted to the HERA data. Despite the very small number of the free parameters, the model gives a satisfactory description of the experimental data, both for the total     cross section as a function of the photon virtuality $Q^2$ or the energy $W$ in the center of mass of the $\gamma^* p$ system, and the differential cross sections as a function of the squared four-momentum transfer $t$ with fixed $Q^2$ and $W$.
%\begin{description}
%\pacs{2.38.-t,12.39.St,13.60.Fz,13.60.Le} 
%May be entered using the \pacs{12.38.-t}
%\verb+\pacs{12.39.St}+\verb+\pacs{13.60.Fz}+\verb+\pacs{13.60.Le}+ command.
%\end{description}
\end{abstract}

\maketitle

%\tableofcontents

\section{\label{s1}Introduction}

Measurements of exclusive deep inelastic processes, such as the production of a real photon or a vector meson, processes known as Deeply Virtual Compton Scattering (DVCS) and Vector Meson Production (VMP), respectively, opened a new window in the study of the nucleon structure in three dimensions, namely in the virtuality $Q^2$, the energy $W$ in the center of mass of the $\gamma^* p$ system and the squared momentum transfer $t$. The construction of scattering amplitudes depending simultaneously on these variables is a challenge for the theory and its knowledge is necessary for the deconvolution of the relevant Generalized Parton Distributions (GPDs). These ones provide information on parton distributions in the coordinate space and, in a more restricted sense, the parton density in the 2D impact parameter space. They are complementary to the linear momentum distribution in the variable $x$ to give a tomographic picture of the nucleon.  

The aim of the present paper is to construct an explicit model for the DVCS and VMP amplitudes depending on the three independent variables $Q^2$, $W$ and $t$. The amplitude should satisfy Regge behaviour, scaling behaviour, be compatible with the quark counting rules and fit the experimental data on DVCS and VMP. In this paper we extend a model on DVCS~\cite{DVCS}, published earlier by some of the authors, to include VMP.
Besides the similarities between these two processes, there are also differences.
In a number of papers, Regge-pole models were successfully applied to VMP (for a review on VMP at HERA see Ref.~\cite{Nikolaev}). The main problem is how the photon virtuality $Q^2$ enters the scattering amplitude. In Ref.~\cite{Schild}, the $Q^2$ dependence is described via a generalization of the vector dominance model. According to Donnachie and Landshoff~\cite{L, DL}, the $Q^2$ evolution can be effectively mimicked by a properly chosen factor in front of the Regge-pole terms. Moreover the same authors argue that HERA data on DVCS and VMP indicate the existence of a soft and a hard Pomeron, whose relative
contribution changes with the hardness of the reaction, i.e. with the photon virtuality and the mass of produced vector mesons.

The paper is organized as follows. In Section~\ref{model} we remind the main features of the model. In Section~\ref{results} we illustrate our fitting strategy and present the results of fits of our model to experimental data on DVCS and VMP processes. In Section~\ref{co} there are our conclusions. Details of the calculation of the integrated cross section with a non-linear Pomeron trajectory are given in the Appendix.

\section{\label{model}The Model}

\subsection{Kinematics}

\begin{figure}[b]
\includegraphics[clip,scale=0.45]{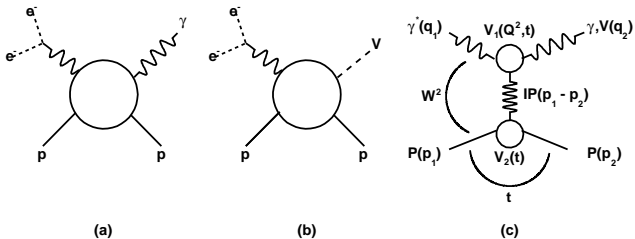}
\caption{ \label{fig:diagram1} Diagrams of DVCS (a) and VMP (b); (c) DVCS (VMP) amplitude in a Regge-factorized form.}
\end{figure}

The diagrams of the reactions in question, DVCS and VMP processes, with a single-photon exchange, are shown in Fig.~\ref{fig:diagram1}. Since we are interested in the nucleon structure, the precisely calculable electroweak vertex $e^{-}\gamma e^{-}$ of Fig.~\ref{fig:diagram1}(a,b) can be factorized out. In the remaining sub-process $\gamma^*p \rightarrow \gamma (V) p$, where $\gamma^*$ is the incoming virtual photon and the outgoing vector particle is a real photon $\gamma$ (Fig.~\ref{fig:diagram1} (a)) or a vector meson V
(Fig.~\ref{fig:diagram1}(b)), at high energies, typical of the HERA experiments, the amplitude is dominated by Regge exchanges, as shown in Fig.~\ref{fig:diagram1}(c).
In the center of mass of the $\gamma^* p$ system the three independent variables of the reactions are, as said above, the virtuality $Q^2=-q_1^2$, whose physical values are positive, the energy $W=(p_1+q_1)$ and squared momentum transfer $t=(q_2-q_1)^2$.
In the study of VMP it is customary to combine the virtuality $Q^2$ and the squared mass of the produced vector particle $M^2_V$ as $\tilde Q^2=Q^2+ M_V^2$. Note that there is no proof for this relation, it is rather a plausible assumption. Moreover, a weight factor might enter the game, namely we could perform the substitution $\tilde Q^2 \Rightarrow c \cdot \tilde Q^2$.

In Ref.~\cite{DVCS} one further step was made, introducing a new variable $z$ through the combination
\begin{equation}\label{z}
    z=t- Q^2.
\end{equation} The argument in favor of this relation is that both $t$ and $-Q^2$ have the meaning of the squared momentum transfer and are an indication of the {\it softness} - {\it hardness} of the dynamics.

\subsection{The Amplitude}

In Ref.~\cite{DVCS} a simple factorized Regge-pole model for DVCS was suggested and successfully fitted to the HERA data. Here we extend the analysis to VMP processes by using the main ideas of the model. The extension includes a more detailed analysis of the $Q^2$ and $M_V^2$ dependence on dynamics.
Note that at the HERA energies sub-leading (secondary Reggeon) contributions are negligible, so that a Pomeron exchange can account for the whole dynamics of the reaction.
The Pomeron pole contribution was defined in Ref.~\cite{DVCS} on the following grounds:

\begin{itemize}

\item it is a single and factorable Regge pole;

\item the dependence on the mass and virtuality of the external
particles enters via the relevant residue functions, which means
that the virtuality $Q^2$ and
the produced vector meson mass enter only via the upper
residue on Fig.~\ref{fig:diagram1}(c), $V_1$, while the Pomeron trajectory $\alpha(t)$ is universal and $Q^2$-independent;

\item following dual models (see, for instance,  Ref.~\cite{BGJPP}), we introduce a $t$ dependence in the residues that enters solely in terms of the trajectory;

\item the Pomeron trajectory has the form
\begin{equation}\label{trajectory}
\alpha(t)=\alpha_0-\alpha_1\ln(1-\alpha_2t),
\end{equation}where $\alpha_i,\
\ i=0-2$, are the $\alpha(t)$-trajectory parameters. This choice is unique for the trajectory, giving a nearly linear behaviour at small $|t|$, where $\alpha'=\alpha_1\alpha_2$ is the forward slope, whereas at large $|t|$ the amplitude and the cross section obey scaling behaviour governed by the quark counting rule. In fact, the logarithmic asymptotics of the trajectory is required by the scaling of the fixed angle scattering amplitude (see Refs.~\cite{FJMP, DVCS}, moreover it follows from perturbative Quantum Chromodynamic (pQCD) calculations (consider, for instance, the BFKL theory~\cite{BFKL}).
\end{itemize}

Fig.~{\ref{Log_Lin} shows the comparison of our logarithmic trajectory with a linear one, $\alpha_0+\alpha' t$, where $\alpha_0=1.09$ and $\alpha'=0.25$ GeV$^{-2}$ for the intercept and the slope respectively, typical of the soft processes \cite{L}, have been used.
The logarithmic asymptotics are important for physical reasons: at large $|t|$ the amplitude and the cross sections obey scaling behaviour governed by the quark counting rules, as seen in hadronic reactions \cite{FJMP}, where sufficiently large values of $|t|$ have been reached in $pp$ and $\bar pp$ scattering, confirming the quark counting rules. More arguments in favour of the logarithmic behaviour in $Q^2$ can be found in Ref.~\cite{a}. This is expected in future measurements ~\cite{Boeretal}, and should be implied, in DVCS and VMP as well. 
The high $Q^2$ region is governed by QCD evolution, and it is beyond the scope of the Regge-pole models. In any case, according to the DGLAP evolution equation~\cite{DGLAP}, the high $Q^2$ behaviour must be tempered with respect to that given by the linear trajectory, and be closer to the logarithmic one, or maybe even slower (double logarithmic?). This behaviour was studied in Ref.~\cite{CJKLMP}.

\begin{figure}[htbp]
\includegraphics[clip,scale=0.4]{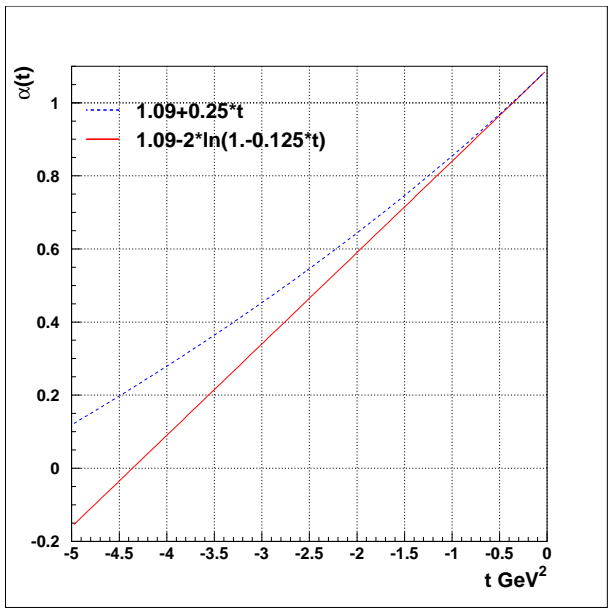}
\caption{\label{Log_Lin}Logarithmic vs linear trajectory as function of t.}
\end{figure}

Neglecting spin, the invariant scattering amplitude with a simple Regge pole exchange, as shown in Fig.~\ref{fig:diagram1}(c), can be written as

\begin{equation}\label{amp}
A(s,t,\tilde Q^2)_{\gamma^* p \rightarrow\gamma (V) p}=-A_0V_1(t,\tilde Q^2)V_2(t)(-is/s_0)^{\alpha(t)}.
\end{equation}
Here $A_0$ is a normalization factor, $V_1(t,\tilde Q^2)=$ exp$[b_2\beta(z)]$ is the $\gamma^*I\!\!P\gamma$ vertex, $V_2(t)=$ exp$[b_1\alpha(t)]$ is the $pI\!\!Pp$ vertex, being $\beta(z)$ and $\alpha(t)$ the exchanged Pomeron trajectory in the photon vertex and in the proton vertex, respectively.

Similarly to Ref.~\cite{Mullerfit}, in Ref.~\cite{DVCS} for DVCS only the helicity conserving amplitude was considered. For not too large $Q^2$ the
contribution from longitudinal photons is small (it vanishes for $Q^2=0$). Moreover, at high energies, typical of the HERA collider, the amplitude is dominated by the helicity conserving Pomeron exchange and, since the final photon is real and transverse, the initial one is also transverse. Electroproduction of vector mesons, discussed in the present paper, requires to take into account both the longitudinal and transverse cross sections.
For convenience, and following the arguments based on duality (see Ref.~\cite{DVCS} and references therein), the $t$ dependence of the $pI\!\!Pp$ vertex $V_2$ is introduced via the trajectory and a generalization of this concept is applied also to the $\gamma^*I\!\!P\gamma$ vertex $V_1$, which however, apart from $t$, depends also on $Q^2$ through the trajectory
\begin{equation}\label{beta}
\beta(z)=\beta_0-\beta_1\ln(1-\beta_2 z),
\end{equation}where $\beta_i,\
\ i=0-2$, are the $\beta(z)$-trajectory parameters, and $\beta_1\beta_2=\beta'$ is the forward slope of this trajectory.

Hence the scattering amplitude in Eq.~(\ref{amp}) can be written in the form
\begin{equation}
A(s,t,\tilde Q^2)_{\gamma^* p\rightarrow\gamma (V) p}=
-A_0e^{b_2\beta(z)}e^{b_1 \alpha(t)}(-is/s_0)^{\alpha(t)}.
\label{amplitude}
\end{equation}
Although the model has many parameters, most of them are constrained by plausible assumptions. First, we fix the intercepts of both $\alpha(t)$ and $\beta(z)$ to the value of $1.09$. The {\it Hardening} of the dynamics with increasing $\tilde Q^2$ may be accounting for either by letting the intercept to be $\tilde Q^2$-dependent, unacceptable by Regge-factorization, or by introducing one more, {\it hard} component in the Pomeron (still unique!) with a $\tilde Q^2$-dependent residue, as suggested e.g. in Refs.~\cite{DL} and \cite{FFJL}. In any case, the trajectories and their parameters are the same for DVCS and for VMP. The other two parameters of the trajectories, $\alpha_1$ and $\alpha_2$ ($\beta_1$ and $\beta_2$) are fixed in the following way: their product $\alpha'=\alpha_1\alpha_2$ {\bf ($\beta'=\beta_1\beta_2$)} is their forward slope, that we set equal to the value $\alpha'=0.25$ GeV$^{-2}$. Furthermore, since $\alpha_1\approx 2$ from the quark counting rules (see Ref.~\cite{DVCS}), we get {\bf $\alpha_2=\alpha'/\alpha_1=0.125$} GeV$^{-2}$.
The same values are used also for the correspondent $\beta_1$ $\beta_2$ parameters of the $\beta(z)$-trajectory.

The parameter $s_0$ is not fixed by the Regge-pole theory. The nice and plausible relation
$s_0=1/\alpha'\approx (1/4)m^2_p$ follows from the hadronic string model~\cite{BaPre}, other values, however, cannot be excluded. We set, for sake of simplicity, $s_0=m^2_p\approx1$ GeV$^2$.

Finally, we set the parameter $b_{1}$ entering the proton vertex (lower vertex of Fig.~\ref{fig:diagram1}(c)) to $b_1=2.0$. In fact, this (p$I\!\!P$p) vertex is know from the analysis of the $pp$ and $\bar{p}p$ scattering to be of the form exp$(bt)$, and an estimate of $b$ is $b\approx2$
GeV$^{-2}$ (see for this Ref.~\cite{b} and references therein).

In Table~\ref{param} we summarize the values of all parameters fixed following the arguments above. Thus, the only free parameters we remain with are the parameter $b_2$, entering the photon vertex $\gamma^*I\!\!P\gamma (V)$, and the squared modulus of the normalization factor, $|A_0|^2$.
\begin{table*}
\caption{\label{param}Values of fixed parameters of our model.}
\begin{ruledtabular}
\begin{tabular}{ccccccc}
$\alpha_0=\beta_0$ & $\alpha_1=\beta_1$  & $\alpha_2=\beta_2~[GeV^{-2}]$  & $s_0~[GeV^2]$ & $b_1$ & $\alpha'=\beta'~[GeV^{-2}]$ & $M^2_V~[GeV^2]$ \\
\colrule
1.09 & 2.00 & 0.125 & 1.00 & 2.00 & 0.25 & $M^2_{\gamma}=0$;~$M^2_{\rho,\phi,...}$; \\
\end{tabular}
\end{ruledtabular}
\end{table*}

\subsection{Cross Sections}

Having fixed most of the parameters of the model as explained above, from the amplitude (\ref{amplitude}) we can now construct the physical quantities to be fitted to the experimental data.

Notice that in this amplitude there is no room for  sub-leading trajectories (e.g. the $f$-trajectory) contribution, because they are suppressed at the Hera energies \cite{DL}.
This assumption has been confirmed by an exploratory fit with an amplitude containing also an $f$-Reggeon contribution ($\alpha_0=0.5$  and $\alpha'= 1~$GeV$^{-2}$).
We found that the $f$-Reggeon contribution is totally negligible.

The differential cross-section $d\sigma (\gamma^*~p\rightarrow \gamma(V) ~p)/dt$ is defined as
\begin{equation}\label{dsigma}
{d\sigma\over{dt}}(\tilde Q^2,W,t)={\pi\over{s^2}}|A(\tilde Q^2,W,t)|^{2}.
\end{equation}

The slope of the differential cross section as a function of t is given by
\begin{equation}\label{slopedcs}
B(\tilde Q^2,W,t)={d\over{dt}}\ln|A(\tilde Q^2,W,t)|^2.
\end{equation}

The total cross section can be approximated\footnote{For the differential cross section the expression $d\sigma(\tilde Q^2,W,t)/dt\sim e^{B(\tilde Q^2,W)\cdot|t|}$ with t$~<~$0 can be used, where $B(\tilde Q^2,W)=B(\tilde Q^2,W,t=0)$ (see. Eq.~\ref{slopedcs}). This approximated expression becomes exact for linear Regge-trajectories} as follows:
\begin{eqnarray}\label{sigmael}
\sigma(s,\tilde Q^2) &=& \int_{t_{min}\approx
-s/2}^{t_{threshold}\approx -4m^2_p} dt {d\sigma(s,t,\tilde Q^2)\over{dt}}\nonumber\\
&\approx&
\left[{1\over B(s,t,\tilde Q^2)}{d\sigma(s,t,\tilde Q^2)\over{dt}}\right]_{t=0} \nonumber \\
&=& \frac{|A_0|^2\pi e^{2\alpha_0(b_1+b_2)}}{[b_1+\ln(s/s_0)](1+\alpha_2\tilde Q^2)+b_2} \cdot \nonumber\\
&&\cdot \frac{(s/s_0)^{2\alpha_0}}
{2\alpha's^2(1+\alpha_2 \tilde Q^2)^{2b_2\alpha_1-1}}.  \label{totalxsec}
\end{eqnarray}
Expression~(\ref{sigmael}) was obtained in the limit $s \rightarrow \infty$.
Alternatively, the integral can be analytically calculated; this is done in Appendix and the result is (see Eq.~(\ref{A5}))
\begin{equation}
\sigma(s,\tilde Q^2)=\frac{K}{\mu-1}~{_2}F_1\left(2b_2\alpha_1,\mu-1;\mu;-\alpha_2 \tilde Q^2\right),
\label{sigma-exact}
\end{equation}
where
\begin{eqnarray}
K &=& \frac{\pi|A_0|^2}{\alpha_2~s^2}e^{2\alpha_0(b_1+b_2)}(s/s_0)^{2\alpha_0},\\
\mu &=& 2\alpha_1\left[b_1+b_2+\ln(s/s_0) \right].
\label{constants}
\end {eqnarray}
Performing an exploratory fit both expressions for the total cross sections lead to results consistent with each other. This means that the main contribution to the total cross section comes  from the region close to $t = 0$. We decided to use the exact analytical expression in the fit procedure.

\section{Fitting Strategy and Results}\label{results}

In the present paper the differential and the total cross sections have been fitted  to the HERA data on DVCS~\cite{d1,d2,d3,d4} and electroproduction of vector mesons $\rho^0$~\cite{r1,r2}, $\phi$~\cite{r1,phi_zeus}, $\omega$~\cite{omega_zeus} and $J/\Psi$~\cite{j1,j2}. The fit on $\sigma(\tilde Q^2)$ is the most sensitive to the parameter $b_2$ and gives a precise estimation of it. For these reason  we first performed preliminary fits of the total cross section as a function of $\tilde Q^2$ at fixed $W$, to HERA data on DVCS and VMP collected by the H1 and ZEUS Collaborations, in order to get weighted average values for $b_2$. Then,  keeping $b_{2}$ fixed to these average values, we performed one-parameter fits to the data for
$\sigma(Q^2)$, $\sigma(W)$ and $d\sigma/dt$, the only free parameter being the normalization $A_0$.

\subsection{DVCS}\label{fitdvcs}

Experimental data for the fits are taken from Refs.~\cite{d1,d2,d3,d4}. The exploratory fit to determine the weighted average value of $b_2$ gives globally a rather satisfactory result, which is shown in Table~\ref{tab:dvcsq2}. Here and in the following Tables $\tilde\chi^2$  means $\chi^2$/d.o.f.

\begin{table}[htb]
\caption{\label{tab:dvcsq2} Values of the two free parameters $|A_0|$, $b_2$ and $\tilde\chi^2$, from the two-parameter fit of the total cross section to data from Refs.~\cite{d1,d3,d4} of $\gamma^*p\rightarrow\gamma p$ as function of $Q^2$ and fixed $W$. The average value of $b_2$ was found to be: $<b_2> =  0.690 \pm 0.021$}
\begin{ruledtabular}
\begin{tabular}{cccccc}
&\multicolumn{4}{c}{$\sigma_{(\gamma^{*}~p~->~\gamma~p)}(Q^2)$}\\
\colrule
Coll. & Years & $W$ [GeV] &$|A_{0}|$ [nb]$^{1/2}$ & $b_2$ & $\tilde\chi^2$ \\
\colrule
H1 & 04-07 & 82 & 0.164 $\pm$ 0.012 & 0.641 $\pm$ 0.055 & 1.1 \\
H1 & 96-00 & 82 & 0.162 $\pm$ 0.011 & 0.656 $\pm$ 0.069 & 0.7\\
ZEUS$(e^-)$ & 96-00 & 89 & 0.177 $\pm$ 0.013 & 0.703 $\pm$ 0.091 & 0.6 \\
ZEUS$(e^+)$ & 96-00 & 89 & 0.170 $\pm$ 0.005 & 0.596 $\pm$ 0.026 & 0.4  \\
ZEUS & 99-00 & 104 & 0.209 $\pm$ 0.010 & 0.769 $\pm$ 0.077 & 3.3 \\
%\end{tabular}
%\begin{tabular}{c}
%$<b_2> =  0.690 \pm 0.021$ \\
\end{tabular}
\end{ruledtabular}
\end{table}

Having fixed the parameter $b_2$ to the weighted average value 0.690(21), all subsequent fits are performed with only $|A_0|$ as free parameter. In Figs.~\ref{fig:dvcsq2},~\ref{fig:dvcsw} and Tables~\ref{tab:dvcsq2fix},~\ref{tab:dvcsw} we present the result for the total cross section as a function of $Q^2$ at fixed $W$ and as a function of $W$ at fixed $Q^2$, respectively, while in Fig.~\ref{fig:dvcsdsdth1} and Table~\ref{tab:dvcsdsdt} we present the results for the differential cross section. The result appears fairly good, except for high value of $Q^2$, for which we can observe some discrepancy between experimental data and our description.

\begin{table}[htb]
\caption{\label{tab:dvcsq2fix}Values of the free parameter $|A_0|$ and $\tilde\chi^2$ from the fit to data from Refs.~\cite{d1,d3,d4} of $\gamma^*p\rightarrow\gamma p$ total cross section as function of $Q^2$ for fixed values of $W$ and of $<b_2> =  0.690 \pm 0.021$.}
\begin{ruledtabular}
\begin{tabular}{ccccc}
&\multicolumn{3}{c}{$\sigma_{(\gamma^{*}~p~->~\gamma~p)}(Q^2)$}\\
\colrule
Coll. & Years & $W$ [GeV] &$|A_{0}|$ [nb]$^{1/2}$ & $\tilde\chi^2$ \\
\colrule
H1 & 04-07 & 82 & 0.172 $\pm$ 0.006  & 1.0 \\
H1 & 96-00 & 82 & 0.165 $\pm$ 0.008  & 0.6 \\
ZEUS$(e^-)$ & 96-00 & 89 & 0.176 $\pm$ 0.006 & 0.4 \\
ZEUS$(e^+)$ & 96-00 & 89 & 0.183 $\pm$ 0.005  & 1.1 \\
ZEUS & 99-00 & 104 & 0.204 $\pm$ 0.009  & 3.3 \\
\end{tabular}
\end{ruledtabular}
\end{table}

\begin{table} [htb]
\caption{\label{tab:dvcsw}Values of the free parameter $|A_0|$ and $\tilde\chi^2$ from the fit to data from Refs.~\cite{d1,d2,d4} of $\gamma^*p\rightarrow\gamma p$ integrated cross section as function of $W$ for fixed values of $Q^2$.}
\begin{ruledtabular}
\begin{tabular}{ccccc}
&\multicolumn{3}{c}{$\sigma_{(\gamma^{*}~p~->~\gamma~p)}(W)$}\\
\colrule
Coll. & Years & $Q^2$ [GeV$^2$] &$|A_{0}|$ [nb]$^{1/2}$ & $\tilde\chi^2$ \\
\colrule
H1 & 05-06 & 8 & 0.163 $\pm$ 0.008 & 6.0 \\
ZEUS & 99-00 & 2.4 & 0.240 $\pm$ 0.008  & 1.7 \\
ZEUS & 99-00 & 3.2 & 0.222 $\pm$ 0.007 & 3.8 \\
ZEUS & 96-00 & 6.2 & 0.181 $\pm$ 0.007  & 1.0 \\
ZEUS & 96-00 & 9.6 & 0.173 $\pm$ 0.007  & 3.2 \\
ZEUS & 96-00 & 9.9 & 0.170 $\pm$ 0.010  & 2.9 \\
ZEUS & 96-00 & 18.0 & 0.182 $\pm$ 0.010  & 1.9 \\
\end{tabular}
\end{ruledtabular}
\end{table}

\begin{table}[htb]
\caption{\label{tab:dvcsdsdt}Values of the free parameter $|A_0|$ and $\tilde\chi^2$ from the fit to data from Ref.~\cite{d3} of $\gamma^*~p\rightarrow \gamma p$ differential cross section as function of t for fixed values of Q$^2$ and W.}
\begin{ruledtabular}
\begin{tabular}{cccccc}
&\multicolumn{4}{c}{$\sigma_{(\gamma^{*}~p~->~\gamma~p)}(t)$}\\
\colrule
Coll. & Years & $W$ [GeV] & $Q^2$ [GeV$^2$] &$|A_{0}|$ [nb]$^{1/2}$ & $\tilde\chi^2$ \\
\colrule
H1 & 04-07 & 40 & 10 & 0.122 $\pm$ 0.007 & 1.5 \\
H1 & 04-07 & 70 & 10 & 0.157 $\pm$ 0.003 & 0.3 \\
H1 & 04-07 & 82 & 8 & 0.168 $\pm$ 0.006 & 0.8 \\
H1 & 04-07 & 82 & 15.5 & 0.161 $\pm$ 0.004 & 0.3 \\
H1 & 04-07 & 82 & 25 & 0.163 $\pm$ 0.008 & 0.4 \\
H1 & 04-07 & 100 & 10 & 0.185 $\pm$ 0.003 & 0.2 \\
H1 & 05-06 & 40 & 8 & 0.118 $\pm$ 0.0128 & 2.2 \\
H1 & 05-06 & 40 & 20 & 0.109 $\pm$ 0.005 & 0.3 \\
H1 & 05-06 & 70 & 8 & 0.146 $\pm$ 0.012 & 1.8 \\
H1 & 05-06 & 70 & 20 & 0.150 $\pm$ 0.005 & 0.4 \\
H1 & 05-06 & 100 & 8 & 0.181 $\pm$ 0.008 & 0.5 \\
H1 & 05-06 & 100 & 20 & 0.171 $\pm$ 0.008 & 0.4 \\
ZEUS & 99-00 & 104 & 3.2 & 0.204 $\pm$ 0.018 & 0.9 \\
\end{tabular}
\end{ruledtabular}
\end{table}

\subsection{Exclusive VM electroproduction ($\gamma^*~p\rightarrow
V~p$)}\label{fitvmp}

To describe vector meson production (VMP) we used the same model as for DVCS. 
 We considered exclusive electroproduction of $\rho^0$, $\phi$, $\omega$ and $J/\Psi$ mesons.
In VMP, contrary to DVCS, apart from transversely polarized photon amplitude, the longitudinal component is also important.
We have performed a fit for each reaction separately, applying the same strategy as in DVCS, using the same set of fixed parameters. First,  leaving as free parameters the normalization $|A_0|$ and $b_2$, through a preliminary fit we have determined the weighted average value of the parameter $b_2$ for each process. Then we have fitted our model to the data having only $|A_0|$ as free parameter to be determined by the fit. From the results one can see how the weighted average value of the parameter $b_2$ of the Regge-pole in the $\gamma^*I\!\!P\gamma$ vertex obviously depends on the specific reaction.

%integrated cross section vs Q2

\subsubsection{$\gamma^*~p~\rightarrow~\rho^0~p$}

Experimental data for the fits are taken from Refs.~\cite{r1,r2}. The exploratory fit to determine the weighted average value of $b_2$ is shown in Table~\ref{tab:rhoq2}.

\begin{table}[htb]
\caption{\label{tab:rhoq2} Values of the two free parameters $|A_0|$, $b_2$ and $\tilde\chi^2$, from the two-parameter fit to data from Refs.~\cite{r1,r2} of total cross section of $\gamma^*p\rightarrow\rho^0 p$ as function of $Q^2$ and fixed $W$.  The average value of $b_2$ was found to be: $<b_2>=1.087 \pm 0.025$.}
\begin{ruledtabular}
\begin{tabular}{cccccc}
&\multicolumn{4}{c}{$\sigma_{(\gamma^{*}~p~->~\rho^0~p)}(Q^2)$}\\
\colrule
Coll. & Years & $W$ [GeV] &$|A_{0}|$ [nb]$^{1/2}$ & $b_2$ & $\tilde\chi^2$ \\
\colrule
H1 & 96-00 & 75 & 0.887 $\pm$ 0.017 & 1.091 $\pm$ 0.025 & 1.5 \\
ZEUS & 96-00 & 90 & 0.916 $\pm$ 0.030 & 1.084 $\pm$ 0.044 & 7.5 \\
\end{tabular}
%\begin{tabular}{|c|}
%$<b_2>$   \\
%1.087 $\pm$ 0.025 \\
%\end{tabular}
\end{ruledtabular}
\end{table}

All subsequent fits are performed with only $|A_0|$ as free parameter. The results for the total cross section as a function of $Q^2$ at fixed $W$ and as a function of $W$ at fixed $Q^2$, are respectively shown in Figs.~\ref{fig:rhoq2}, ~\ref{fig:rhow}, and Tables~\ref{tab:dvcsq2fix},~\ref{tab:rhow}. The results for the differential cross section as function of t are shown in Fig.~\ref{fig:rhodsdt}, and Table~\ref{tab:rhodsdt}.

\begin{table}[htb]
\caption{\label{tab:rhoq2fix}Values of the free parameter $|A_0|$ and $\tilde\chi^2$ from the fit to data from Refs.~\cite{r1,r2} of $\gamma^*p\rightarrow\rho^0 p$ total cross section as function of $Q^2$ for fixed values of $W$.}
\begin{ruledtabular}
\begin{tabular}{ccccc}
&\multicolumn{3}{c}{$\sigma_{(\gamma^{*}~p~->~\rho^0~p)}(Q^2)$}\\
\colrule
Coll. & Years & $W$ [GeV] &$|A_{0}|$ [nb]$^{1/2}$ & $\tilde\chi^2$ \\
\colrule
H1 & 96-00 & 75 & 0.885 $\pm$ 0.013 & 1.4 \\
ZEUS & 96-00 & 90 & 0.918 $\pm$ 0.021 & 6.8 \\
\end{tabular}
\end{ruledtabular}
\end{table}

\begin{table}[htb]
\caption{\label{tab:rhow}Values of the free parameter $|A_0|$ and $\tilde\chi^2$ from the fit to data from Ref.~\cite{r1} of $\gamma^*p\rightarrow\rho^0 p$ integrated cross section as function of $W$ for fixed values of $Q^2$.}
\begin{ruledtabular}
\begin{tabular}{ccccc}
&\multicolumn{4}{c}{$\sigma_{(\gamma^{*}~p~->~\rho^0~p)}(W)$}\\
\colrule
Coll. & Years & $Q^2$ [GeV$^2$] &$|A_{0}|$ [nb]$^{1/2}$ & $\tilde\chi^2$ \\
\colrule
H1 & 96-00 & 3.3 & 0.916 $\pm$ 0.036 & 3.1 \\
H1 & 96-00 & 6.6 & 0.837 $\pm$ 0.027 & 2.1 \\
H1 & 96-00 & 11.9 & 0.883 $\pm$ 0.021 & 0.7 \\
H1 & 96-00 & 19.5 & 0.937 $\pm$ 0.054 & 4.2 \\
H1 & 96-00 & 35.6 & 1.082 $\pm$ 0.112 & 3.7 \\
ZEUS & 96-00 & 2.4 & 1.023 $\pm$ 0.018 & 1.2 \\
ZEUS & 96-00 & 3.7 & 0.946 $\pm$ 0.023 & 4.0 \\
ZEUS & 96-00 & 6.0 & 0.837 $\pm$ 0.017 & 2.2 \\
ZEUS & 96-00 & 8.3 & 0.854 $\pm$ 0.024 & 4.5 \\
ZEUS & 96-00 & 13.5 & 0.866 $\pm$ 0.026 & 5.8 \\
ZEUS & 96-00 & 32.0 & 1.109 $\pm$ 0.053 & 3.9 \\
\end{tabular}
\end{ruledtabular}
\end{table}

\begin{table}[htb]
\caption{\label{tab:rhodsdt}Values of the free parameter $|A_0|$ and $\tilde\chi^2$ from the fit to data from Ref.~\cite{r1} of $\gamma^*~p\rightarrow \rho^0 p$ differential cross section as function of t for  fixed values of Q$^2$ and W. }
\begin{ruledtabular}
\begin{tabular}{cccccc}
&\multicolumn{4}{c}{$\sigma_{(\gamma^{*}~p~->~\rho^0~p)}(t)$}\\
\colrule
Coll. & Years & $W$ [GeV] & $Q^2$ [GeV$^2$] &$|A_{0}|$ [nb]$^{1/2}$ & $\tilde\chi^2$ \\
\colrule
H1 & 96-00 & 75 & 3.3 & 0.885 $\pm$ 0.048 & 5.1 \\
H1 & 96-00 & 75 & 6.6 & 0.801 $\pm$ 0.039 & 5.0 \\
H1 & 96-00 & 75 & 11.5 & 0.872 $\pm$ 0.030 & 1.1 \\
H1 & 96-00 & 75 & 17.4 & 0.847 $\pm$ 0.022 & 0.7 \\
H1 & 96-00 & 75 & 33 & 0.901 $\pm$ 0.023 & 0.4 \\
ZEUS & 96-00 & 90 & 2.7 & 1.002 $\pm$ 0.038 & 4.3 \\
ZEUS & 96-00 & 90 & 5.0 & 0.867 $\pm$ 0.026 & 2.8 \\
ZEUS & 96-00 & 90 & 7.8 & 0.824 $\pm$ 0.025 & 2.4 \\
ZEUS & 96-00 & 90 & 11.9 & 0.834 $\pm$ 0.019 & 1.1 \\
ZEUS & 96-00 & 90 & 19.7 & 0.946 $\pm$ 0.016 & 0.4 \\
ZEUS & 96-00 & 90 & 41.0 & 1.166 $\pm$ 0.050 & 0.4 \\
\end{tabular}
\end{ruledtabular}
\end{table}

\subsubsection{$\gamma^*~p~\rightarrow~\phi~p$}

Experimental data for the fits are taken from Refs.~\cite{r1,phi_zeus}. The exploratory fit to determine the weighted average value of $b_2$ is shown in Table~\ref{tab:phiq2}.

\begin{table}[htb]
\caption{\label{tab:phiq2}Values of the two free parameters $|A_0|$, $b_2$ and $\tilde\chi^2$, from the two-parameter fit to data from Refs.~\cite{r1,phi_zeus} of total cross section of $\gamma^*p\rightarrow\phi p$ as function of $Q^2$ and fixed values of $W$. The average value of $b_2$ was found to be:$<b_2>=1.131 \pm 0.033 $.}
\begin{ruledtabular}
\begin{tabular}{cccccc}
&\multicolumn{4}{c}{$\sigma_{(\gamma^{*}~p~->~\phi~p)}(Q^2)$}\\
\colrule
Coll. & Years & $W$ [GeV] &$|A_{0}|$ [nb]$^{1/2}$ & $b_2$ & $\tilde\chi^2$ \\
\colrule
H1 & 96-00 & 75 & 0.390 $\pm$ 0.010 & 1.155 $\pm$ 0.044 & 0.9 \\
ZEUS & 98-00 & 75 & 0.433 $\pm$ 0.011 & 1.110 $\pm$ 0.050 & 2.6 \\
\end{tabular}
%\begin{tabular}{c}
%$<b_2>$   \\
%\hline \hline
%1.131 $\pm$ 0.033 \\
%\hline \hline
%\end{tabular}
\end{ruledtabular}
\end{table}

All subsequent fits are performed with only $|A_0|$ as free parameter. The results for the total cross section as a function of $Q^2$ at fixed $W$ and as a function of $W$ at fixed $Q^2$ are respectively shown in Figs.~\ref{fig:phiq2},~\ref{fig:phiw}, and Tables~\ref{tab:phiq2fix},~\ref{tab:phiw}. The results for the differential cross section as function of $t$ are shown in Figs.~\ref{fig:phidsdt_h1},~\ref{fig:phidsdt_zeus} and Tables~\ref{tab:phidsdt}.

\begin{table}[htb!]
\caption{\label{tab:phiq2fix}Values of the free parameter $|A_0|$ and $\tilde\chi^2$ from the fit to data from Refs.~\cite{r1, phi_zeus} of $\gamma^*p\rightarrow\phi p$ total cross section as function of $Q^2$ and fixed value of $W$.}
\begin{ruledtabular}
\begin{tabular}{ccccc}
&\multicolumn{3}{c}{$\sigma_{(\gamma^{*}~p~->~\phi~p)}(Q^2)$}\\ 
\colrule
Coll. & Years & $W$ [GeV] &$|A_{0}|$ [nb]$^{1/2}$ & $\tilde\chi^2$ \\
\colrule
H1 & 96-00 & 75 & 0.387 $\pm$ 0.008 & 0.8 \\
ZEUS & 96-00 & 75 & 0.435 $\pm$ 0.010 & 2.3 \\
\end{tabular}
\end{ruledtabular}
\end{table}

\begin{table}[htb!]
\caption{\label{tab:phiw}Values of the free parameter $|A_0|$ and $\tilde\chi^2$ from the fit to data from Ref.~\cite{r1} of $\gamma^*p\rightarrow\phi p$ integrated cross section as function of $W$ for fixed values of $Q^2$.}
\begin{ruledtabular}
\begin{tabular}{ccccc}
&\multicolumn{3}{c}{$\sigma_{(\gamma^{*}~p~->~\phi~p)}(W)$}\\
\colrule
Coll. & Years & $Q^2$ [GeV$^2$] &$|A_{0}|$ [nb]$^{1/2}$ & $\tilde\chi^2$ \\
\colrule
H1 & 96-00 & 3.3 & 0.397 $\pm$ 0.013 & 0.9 \\
H1 & 96-00 & 6.6 & 0.362 $\pm$ 0.017 & 1.7 \\
H1 & 96-00 & 15.8 & 0.423 $\pm$ 0.026 & 2.1 \\
ZEUS & 96-00 & 2.4 & 0.462 $\pm$ 0.010 & 1.1 \\
ZEUS & 96-00 & 3.8 & 0.431 $\pm$ 0.009 & 0.8 \\
ZEUS & 96-00 & 6.5 & 0.401 $\pm$ 0.007 & 0.4 \\
ZEUS & 96-00 & 13.0 & 0.470 $\pm$ 0.009 & 0.4 \\
\end{tabular}
\end{ruledtabular}
\end{table}

\begin{table}[htb]
\caption{\label{tab:phidsdt}Values of the free parameter $|A_0|$ and $\tilde\chi^2$ from the fit to data from Refs.~\cite{r1} of $\gamma^*~p\rightarrow \phi p$ differential cross section as function of t for fixed values of Q$^2$ and W.}
\begin{ruledtabular}
\begin{tabular}{cccccc}
&\multicolumn{4}{c}{$\sigma_{(\gamma^{*}~p~->~\phi~p)}(t)$}\\
\colrule
Coll. & Years & $W$ [GeV] & $Q^2$ [GeV$^2$] &$|A_{0}|$ [nb]$^{1/2}$ & $\tilde\chi^2$ \\
\colrule
H1 & 96-00 & 75 & 3.3 & 0.396 $\pm$ 0.026 & 3.2 \\
H1 & 96-00 & 75 & 6.6 & 0.358 $\pm$ 0.019 & 2.1 \\
H1 & 96-00 & 75 & 15.8 & 0.329 $\pm$ 0.024 & 2.1 \\
ZEUS & 98-00 & 75 & 0.0 & 0.438 $\pm$ 0.015 & 2.0 \\
\end{tabular}
\end{ruledtabular}
\end{table}

\subsubsection{$\gamma^*~p~\rightarrow~\omega~p$}

Experimental data for the fits are taken from Ref.~\cite{omega_zeus}. There only two set of data have been published by the ZEUS Collaboration, one at fixed $W$, the other at fixed $Q^2$. No data for the differential cross section as function of $t$ are available. The results for the total cross section as a function of $Q^2$ at fixed $W$ and as a function of $W$ at fixed $Q^2$ are respectively shown in Fig.~\ref{fig:omega} and Table~\ref{tab:omegaq2fix},~\ref{tab:phiw}.

\begin{table}[htb]
\caption{\label{tab:omegaq2fix}Values of the free parameter $|A_0|$ and $\tilde\chi^2$ from the fit to data from Ref.~\cite{omega_zeus} of $\gamma^*~p\rightarrow \omega p$ total cross section as function of Q$^2$ for fixed value of W.}
\begin{ruledtabular}
\begin{tabular}{ccccc}
&\multicolumn{3}{c}{$\sigma_{(\gamma^{*}~p~->~\omega~p)}(Q^2)$}\\
\colrule
Coll. & Years & $W$ [GeV] &$|A_{0}|$ [nb]$^{1/2}$ & $\tilde\chi^2$ \\
\colrule
ZEUS & 96-97 & 80 & 0.273 $\pm$ 0.012 & 0.3 \\
\end{tabular}
\end{ruledtabular}
\end{table}

\begin{table}[htb]
\caption{\label{tab:phiw}Values of the free parameter $|A_0|$ and $\tilde\chi^2$ from the fit to data from Ref.~\cite{omega_zeus} of $\gamma^*~p\rightarrow \omega p$ total cross section as function of $W$ for fixed value of Q$^2$.}
\begin{ruledtabular}
\begin{tabular}{ccccc}
&\multicolumn{3}{c}{$\sigma_{(\gamma^{*}~p~->~\omega~p)}(W)$}\\
\colrule
Coll. & Years & Q$^2$ [GeV$^2$] &$|A_{0}|$ [nb]$^{1/2}$ & $\tilde\chi^2$ \\
\colrule
ZEUS & 96-97 & 7.0 & 0.265 $\pm$ 0.022 & 0.5 \\
\end{tabular}
\end{ruledtabular}
\end{table}

\subsubsection{$\gamma^*~p~\rightarrow~J/\psi~p$}

Experimental data for the fits are taken from Refs.~\cite{j1, j2}. The exploratory fit to determine the weighted average value of $b_2$ gives  a result which is shown in Table~\ref{tab:jpsiq2}.

\begin{table}[htb]
\caption{\label{tab:jpsiq2} Values of the two free parameters $|A_0|$, $b_2$ and $\tilde\chi^2$, from the two-parameter fit to data from Refs.~\cite{j1,j2} of the total cross section of $\gamma^*p\rightarrow J/\psi p$ as function of $Q^2$ and fixed $W$. The average value of $b_2$ was found to be: $<b_2>=0.890 \pm 0.033$.}
\begin{ruledtabular}
\begin{tabular}{cccccc}
&\multicolumn{4}{c}{$\sigma_{(\gamma^{*}~p~->~J/\psi~p)}(Q^2)$}\\
\colrule
Coll. & Years & $W$ [GeV] &$|A_{0}|$ [nb]$^{1/2}$ & $b_2$ & $\tilde\chi^2$ \\
\colrule
H1 & 99-00 & 90 & 0.855 $\pm$ 0.031 & 0.898 $\pm$ 0.033 & 0.4 \\
ZEUS & 98-00 & 90 & 0.853 $\pm$ 0.040 & 0.879 $\pm$ 0.035 & 0.7 \\
\end{tabular}
%\begin{tabular}{|c|}
%$<b_2>$   \\
%0.890 $\pm$ 0.033 \\
%\end{tabular}
\end{ruledtabular}
\end{table}

All subsequent fits are performed with only $|A_0|$ as free parameter. The result for the total cross section as a function of $Q^2$ at fixed $W$ and as a function of $W$ at fixed $Q^2$ are respectively shown in Figs.~\ref{fig:jpsiq2},~\ref{fig:jpsiw}, and Tables~\ref{tab:jpsiq2},~\ref{tab:jpsiw}. The results for the differential cross section as function of $t$ are shown  in Fig.~\ref{fig:jpsidsdt} and Table~\ref{tab:jpsidsdt}.

\begin{table}[htb]
\caption{\label{tab:jpsiq2fixed}Values of the free parameter $|A_0|$ and $\tilde\chi^2$ from the fit to data from Refs.~\cite{j1,j2} of the $\gamma^*p\rightarrow J/\psi p$ total cross section as function of $Q^2$ for fixed value of $W$.}
\begin{ruledtabular}
\begin{tabular}{ccccc}
&\multicolumn{3}{c}{$\sigma_{(\gamma^{*}~p~->~J/\psi~p)}(Q^2)$}\\
\colrule
Coll. & Years & $W$ [GeV] &$|A_{0}|$ [nb]$^{1/2}$ & $\tilde\chi^2$ \\
\colrule
H1 & 99-00 & 90 & 0.857 $\pm$ 0.013 & 0.4 \\
ZEUS & 98-00 & 90 & 0.875 $\pm$ 0.014 & 0.6 \\
\end{tabular}
\end{ruledtabular}
\end{table}

\begin{table}[htb!]
\caption{\label{tab:jpsiw}Values of the free parameter $|A_0|$ and $\tilde\chi^2$ from the fit to data from Refs.~\cite{j2} of $\gamma^*~p\rightarrow J/\psi p$ total cross section as function of $W$ for fixed values of Q$^2$.}
\begin{ruledtabular}
\begin{tabular}{ccccc}
&\multicolumn{3}{c}{$\sigma_{(\gamma^{*}~p~->~J/\psi~p)}(W)$}\\
\colrule
Coll. & Years & Q$^2$ [GeV$^2$] &$|A_{0}|$ [nb]$^{1/2}$ & $\tilde\chi^2$ \\
\colrule
H1 & 99-00 & 3.2 & 0.790 $\pm$ 0.037 & 1.3 \\
H1 & 99-00 & 7.0 & 0.768 $\pm$ 0.050 & 1.4 \\
H1 & 99-00 & 22.4 & 0.916 $\pm$ 0.044 & 0.7 \\
ZEUS & 98-00 & 0.4 & 0.867 $\pm$ 0.111 & 4.0 \\
ZEUS & 98-00 & 3.1 & 0.840 $\pm$ 0.043 & 2.1 \\
ZEUS & 98-00 & 6.8 & 0.844 $\pm$ 0.033 & 1.4 \\
ZEUS & 98-00 & 16.0 & 0.815 $\pm$ 0.078 & 5.8 \\
\end{tabular}
\end{ruledtabular}
\end{table}

\begin{table}[htb!]
\caption{\label{tab:jpsidsdt}Values of the free parameter $|A_0|$ and $\tilde\chi^2$ from the fit to data from Ref.~\cite{j2} of $\gamma^*~p\rightarrow J/\psi p$ differential cross section as function of
$t$ for fixed values of Q$^2$ and W.}
\begin{ruledtabular}
\begin{tabular}{cccccc}
&\multicolumn{3}{c}{$\sigma_{(\gamma^{*}~p~->~J/\psi~p)}(t)$}\\
\colrule
Coll. & Years & $W$ [GeV] & $Q^2$ [GeV$^2$] &$|A_{0}|$ [nb]$^{1/2}$ & $\tilde\chi^2$ \\
\colrule
H1 & 99-00 & 90 & 0.05 & 0.860 $\pm$ 0.035 & 5.1 \\
H1 & 99-00 & 90 & 3.2 & 0.778 $\pm$ 0.066 & 4.1 \\
H1 & 99-00 & 90 & 7.0 & 0.710 $\pm$ 0.083 & 4.3 \\
H1 & 99-00 & 90 & 22.4 & 0.861 $\pm$ 0.090 & 3.0 \\
ZEUS & 98-00 & 90 & 3.1 & 0.868 $\pm$ 0.031 & 0.9 \\
%ZEUS & 98-00 & 90 & 6.8 in the bin 2 $<$ Q$^2$ $<$ 100 & 0.841 $\pm$ 0.033 & 2.9 \\
%ZEUS & 98-00 & 90 & 6.8 in the bin 5 $<$ Q$^2$ $<$ 10 & 0.824 $\pm$ 0.049 & 3.3 \\
ZEUS & 98-00 & 90 & 6.8 & 0.841 $\pm$ 0.033 & 2.9 \\
ZEUS & 98-00 & 90 & 6.8 & 0.824 $\pm$ 0.049 & 3.3 \\
ZEUS & 98-00 & 90 & 16.0 & 0.863 $\pm$ 0.023 & 0.5 \\
\end{tabular}
\end{ruledtabular}
\end{table}

\section{Conclusions and Outlooks}\label{co}
In this paper we have revised and extended the model of Ref. \cite{DVCS} to include, apart from DVCS, vector meson production as well. The basic features of the model here remain intact, but the fitting procedure has changed. On one hand, the parameters entering the Pomeron trajectory and the coefficient $b_1$ in the residue of the proton ($p I\!\!P p$) vertex are the same for DVCS and VMP processes. In particular, the parameter $b_1$  is  known, due to Regge factorization, from $pp$ and $\bar{p}$p scattering and set to the value $b_1=2.0$, related to the {\it proton radius}. On the other hand, the normalization parameter $A_{0}$ and the coefficient $b_2$ in the residue of the photon $\gamma^*I\!\!P\gamma$ vertex are different for
each reaction we considered in this paper (production of a real $\gamma$ or $\rho$, $\ \ \phi$, \ \ $\omega$ and $J/\Psi$ vector mesons). In particular, for each reaction the parameter $b_2$ first has been fitted to the existing sets of experimental data on the total cross section as a function of the virtuality $Q^2$, then it has been fixed to the average value among those obtained from the fits. Consequently, fits to experimental data on differential cross section and total cross section as a function of the energy $W$ have been performed, the normalization $A_{0}$ being the only free parameter. All experimental data used in the fitting procedure were selected by the H1 and ZEUS collaborations as diffractive ones, therefore there is no place for any secondary (nonleading) Regge contribution, the Pomeron trajectory being the only $t$ channel contribution, and hence the often used notion of an {\it effective} trajectory, in this paper means the genuine Pomeron.

It is always instructive to compare the Pomeron trajectory deduced from DVCS and VMP production data with
that extracted from hadronic scattering. Since the Pomeron trajectory is universal and the precision of the
high-energy $pp$ and $\bar pp$ data exceeds those of DVCS or VMP, it makes sense to use for it values of
parameters resulting from fits to the former data.

Here we come to the important question of {\it how many Pomerons exist} in nature.
Our answer is that there is only one Pomeron and it is universal, which does not mean that it is simple. Moreover, it may have more components, whose relative weight is governed by their residue.

%\textcolor{red}{
By assuming the universality of the Pomeron trajectory in lepton-hadron
and hadron-hadron reactions, one expects that the effect of its
non-linearity be visible also in $pp$ scattering, say in the ISR energy
region comparable to typical HERA energies. Indeed, as shown in a series
of papers \cite{Enrico}, the flattening of the differential cross section
of $pp$ scattering beyond the dip, fitted by Donnachie and Landshoff by a
power $t^{-n} $ can equally well be attributed by the logarithmic behavior
of the Pomeron trajectory, mimicking this {\it hard} power behavior.

In the present paper we adopted the simple case of a {\it soft} Pomeron. The presence of another {\it hard} component here was  not considered. The delicate interplay of {\it soft} and {\it hard} components (in a single Pomeron) is governed by the external masses and virtualities in the residues as shown in Ref \cite{DL}. We intend to come back to this point in a forthcoming study. As expected, our model leads to results on the whole satisfactory for moderate values of $\tilde Q^2$ and $|t|$.  Instead, without a {\it hard} component in the Pomeron, fits to the data for high $\tilde Q^2$ and $|t|$ definitely deteriorate.

Finally, let us come back to one of the main ingredients of the model we considered in the present paper, namely to the variable
$z=t-Q^2$. This is an unusual combination of the squared momentum transfer $t$ and virtuality $Q^2$ ; it does not follow from the theory, although it appears also e.g. in the expression for { the slope $B(Q^2,t)$ in Refs.~{\cite{R1, R2}}. These two variables not only have similar dimensions, but have also close physical meanings, so we could say that high values of the variable $z$ are correlated with a {\it hard} component in the Pomeron.

%\textcolor{red}{
In fact, as seen from Fig.~\ref{Log_Lin}, our non-linear Pomeron trajectory, in the region
$-1.0<t<0$ GeV$^2$ does not differ significantly from the linear one
  e.g. in Refs. \cite{L}. The use of
the non-linear trajectories is motivated: 1) conceptually, by the
expected large-$|t|$ scaling behavior (not reached yet experimentally) and 2) by the
large-$Q^2$ data, already experimentally accessible.

In the spirit of the Regge-pole theory, we have taken into account Regge-factorization of the lower vertex and the propagator by keeping them $t-$dependent only, while the upper $\gamma^*I\!\!P\gamma$ vertex depends also on $Q^2.$ At the same time, considering the Regge-exchange as an effective one,
one must not respect this factorization since the corresponding effective amplitude absorbs various Regge-exchanges anyway.

Since the H1 and ZEUS data, both on DVCS and VMP are well within the Regge kinematical region, we used the Regge-factorized form, according to which the scattering amplitude corresponding to the exchange of a single Regge-trajectory is the product of the two vertices and the Pomeron propagator.
Actually, the  sum of different Regge-pole exchanges often is comprised in a single effective pole, which however should not be confused with a true Regge-pole.

%\textcolor{red}{
Most of our figures presenting the $W$ dependence of the various
channels underestimate the high-energy tail of the data. In our opinion,
this is a strong evidence in favor of a Pomeron having two components, hard and soft, their
relative weights depending on $\tilde Q^2$, as advocated in Ref.
\cite{DL}. The presence of a hard component, with a Pomeron intercept as
high as $1.3-1.4$ will require unitarization.
In a forthcoming study, based on {\it Reggeometry}~\cite{FFJL}, we plan to extend the model to high values of $Q^2$ and $|t|$ by introducing a {\it hard} component in the single universal Pomeron.

Our final comment is that our model should be used as a guide in building explicit expressions for General Parton Distributions.

\begin{acknowledgments}
We thank M. Capua, A. Papa, E. Tassi, L. Primavera and L.~Favart for useful discussions, and G. Ciappetta for his collaboration at the earlier stage of this work. L.J. thanks the Department of Physics of the University of Calabria and the Istituto Nazionale di 
Fisica Nucleare - Sezione di Torino and Gruppo Collegato di Cosenza, where part of this work was done, for their hospitality and support. S.F thanks the organizers of the DIS2011 conference where the preliminary results of this work were presented, for their hospitality, K. Goulianos and D. Mueller for fruitful discussions and E. Aschenauer for her critical remarks.
\end{acknowledgments}

%%%%%%%%%%%%%%%%%%%%%%%%%%%%%%%%%%%%%%%%%%%%
\appendix*

\section{}

Here we present the calculation of the integrated cross section with the non linear trajectory given in Eqs.~(\ref{trajectory}) and (\ref{beta}) and entering the
amplitude~(\ref{amplitude}). The cross section is defined as

\begin{equation}
\sigma(s,\tilde Q^2)=\int^{-4m^2_p}_{-s/2}~dt~{\frac{d\sigma(s,t,\tilde Q^2)}{dt}}.
\label{A1}
\end {equation}
In the limit of very high energy ($s \rightarrow \infty$) and negligible proton mass it becomes
\begin{equation}
\sigma(s,\tilde Q^2)=\int^{\infty}_0~dt~\frac{d\sigma(s,t,\tilde Q^2)}{dt},
\label{A2}
\end{equation}
where the change $t \rightarrow -t$ has been applied.

Substituting the expression (\ref{dsigma}) for the differential cross section, using Eq.~(\ref{amplitude}) for the amplitude
%we get
%\begin{equation}
%\sigma(s,\tilde Q^2) =
% \int^{\infty}_0dt~\frac{\pi|A_0|^2(s/s_0)^{2\alpha_0}}{s^2}e^{2\alpha_0(b_1+b_2)}(1+\alpha_2)^{-\{ 2\alpha_1[b_1+\ln(s/s_0)]\}}(1+\alpha_2 \tilde Q^2+\alpha_2t)^{-2b_2\alpha_1}.
%\label{A3}
%\end{equation}
and with  the replacement $x = \alpha_2t$, this cross section assumes the form~\cite{GrRy}
\begin{eqnarray}
\sigma(s,\tilde Q^2) &=& \int^{\infty}_0dx~(1+x)^{-\mu+\nu}(\beta+x)^{-\nu} \nonumber\\
&=& K B(\mu-1,1)_{~2}F_1\left(\nu, \mu-1;\mu;1-\beta \right)~~~~~~~~
\label{A3}
\end {eqnarray}
with
\begin{eqnarray}
K &=& \frac{\pi|A_0|^2}{\alpha_2~s^2}e^{2\alpha_0(b_1+b_2)}(s/s_0)^{2\alpha_0},\nonumber\\
\beta &=& 1+\alpha_2 \tilde Q^2,\nonumber\\
\nu &=& 2b_2\alpha_1,\nonumber\\
\mu &=& 2\alpha_1\left[b_1+b_2+\ln(s/s_0) \right] > 1.\nonumber
\end{eqnarray}

Here $B(\mu-1,1) = 1/(\mu - 1)$ is our $Beta$ function and $_{~2}F_1\left(\nu,\mu-1;\mu;1-\beta \right)$ is the Gauss hypergeometric function. Then our final expression for the integrated cross section is
\begin{equation}
\sigma(s,\tilde Q^2)=\frac{K}{\mu-1}{~_2}F_1\left(\nu,\mu-1;\mu;1-\beta \right).
\label {A5}
\end{equation}

%%%%%%%%%%%%%%%%%%%%%%%%%%%%%%%%%%%%%%%

%%%%%\nocite{*}

%\bibliography{FFL_exclusive_regge}% Produces the bibliography via BibTeX.

%%%%%%%%%%%%%%%%%%%%%%%%%%%%%%%%%%%%%%%
%  FIGURES
%\section{figures}
% DVCS

\begin{figure*}[htbp!]
\includegraphics[clip,scale=0.45]{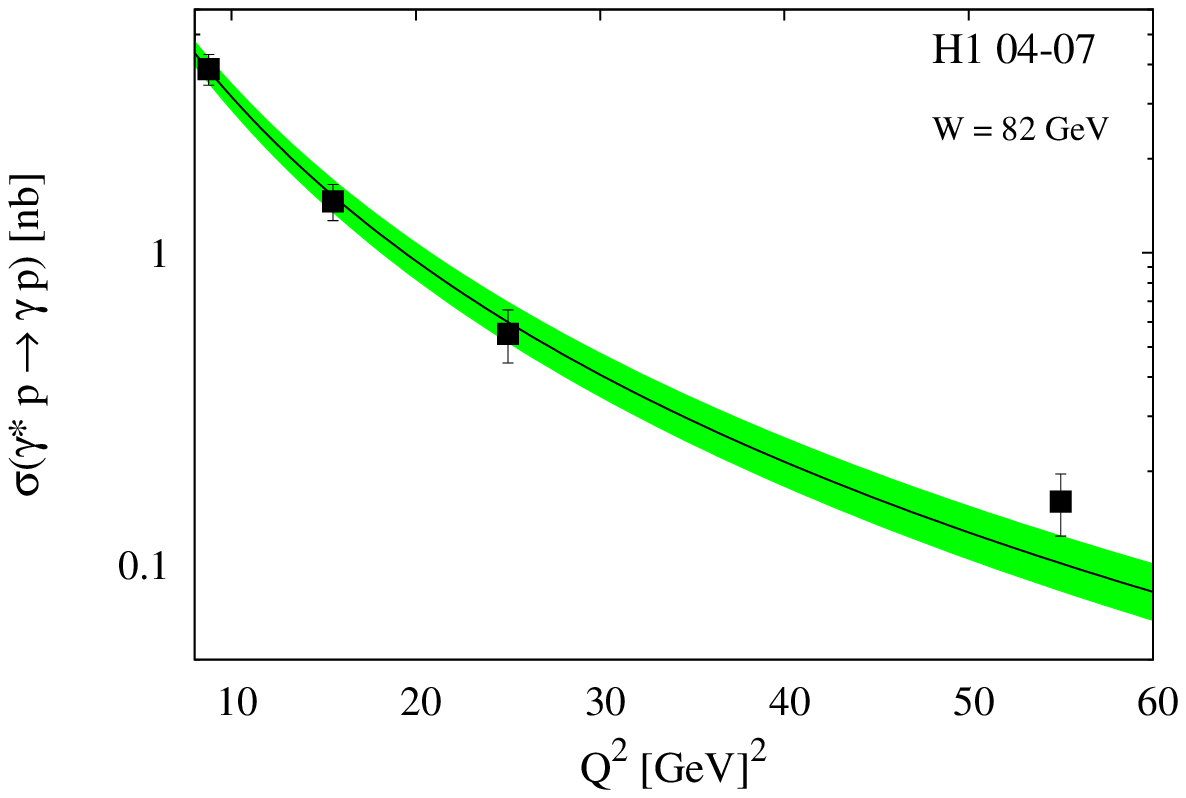}
\includegraphics[clip,scale=0.45]{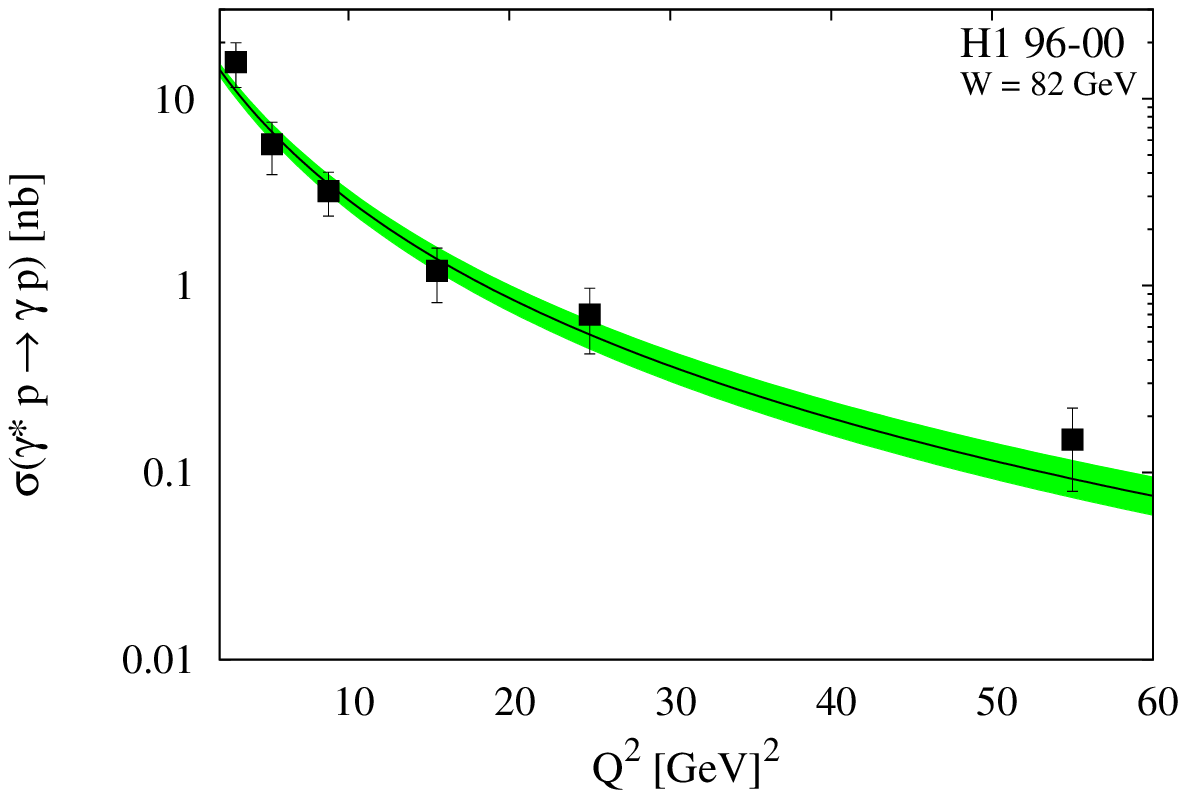}
\includegraphics[clip,scale=0.45]{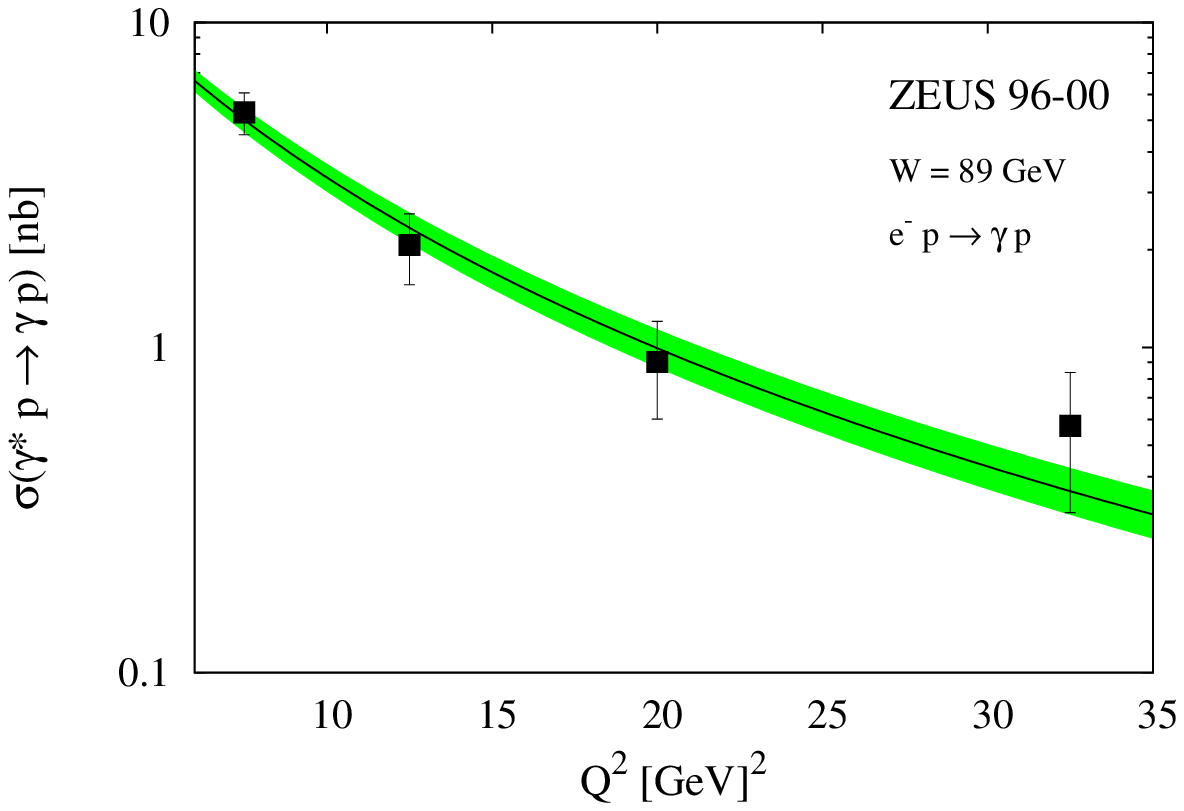}
\includegraphics[clip,scale=0.45]{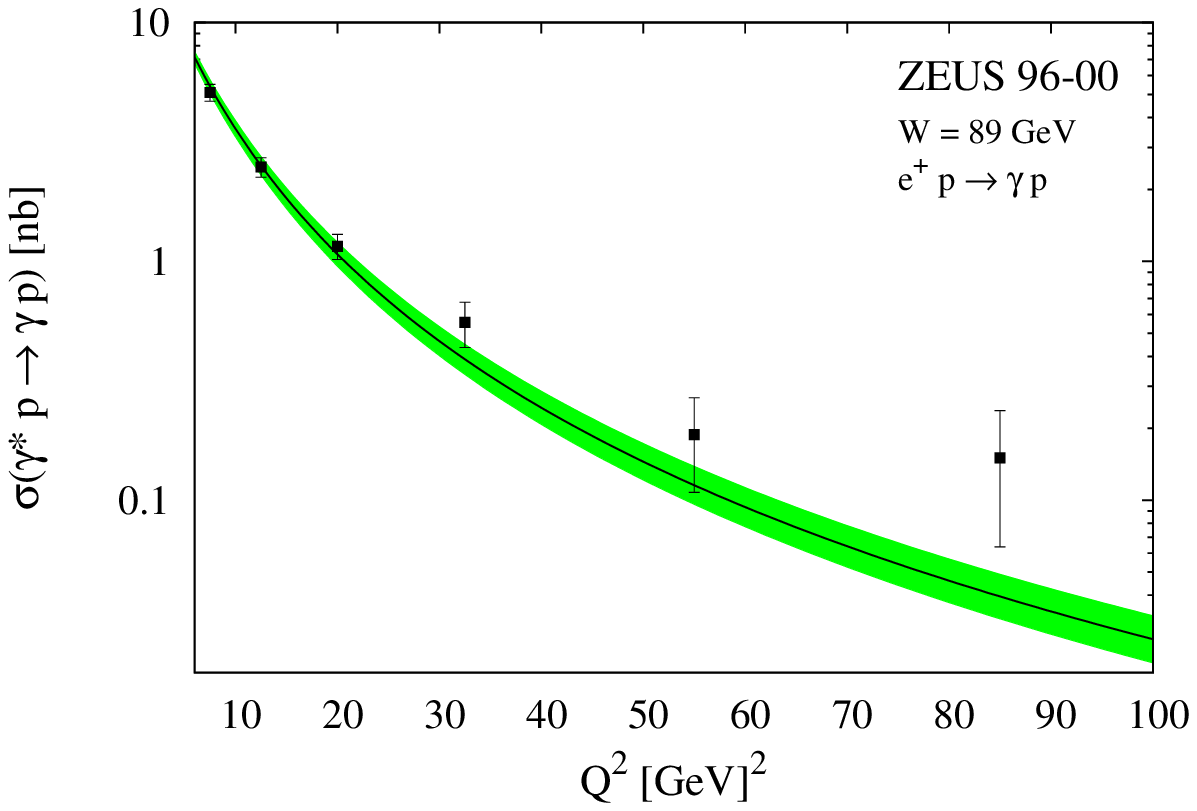}
\includegraphics[clip,scale=0.45]{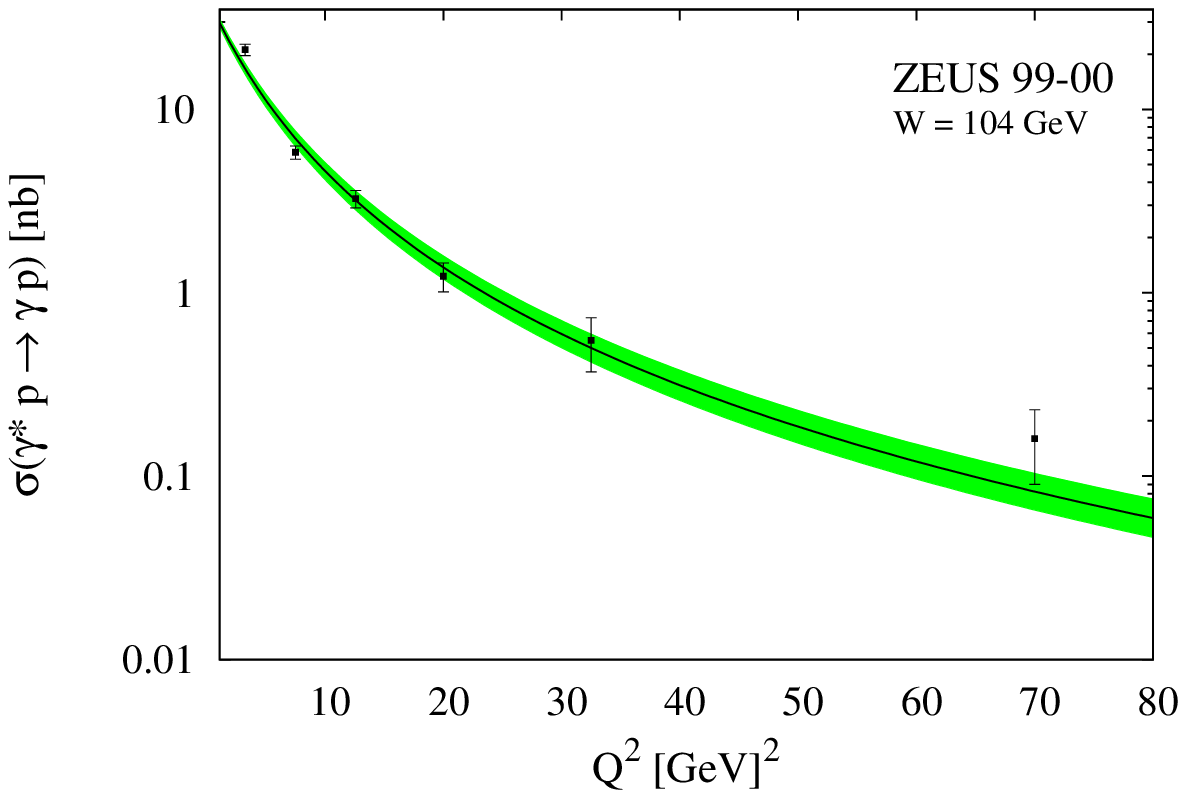}
\caption{\label{fig:dvcsq2}The behaviour according to our model of $\gamma^*p\rightarrow\gamma p$ total cross section as function of $Q^2$ is compared with data from Refs.~\cite{d1,d3,d4} measured by the H1 and ZEUS Collaborations for several values of $W$. The green bands are calculated accordingly with the uncertainties on the free parameter $|A_0|$.}
\end{figure*}

% DVCS sigma(W)

\begin{figure*}
\includegraphics[clip,scale=0.45]{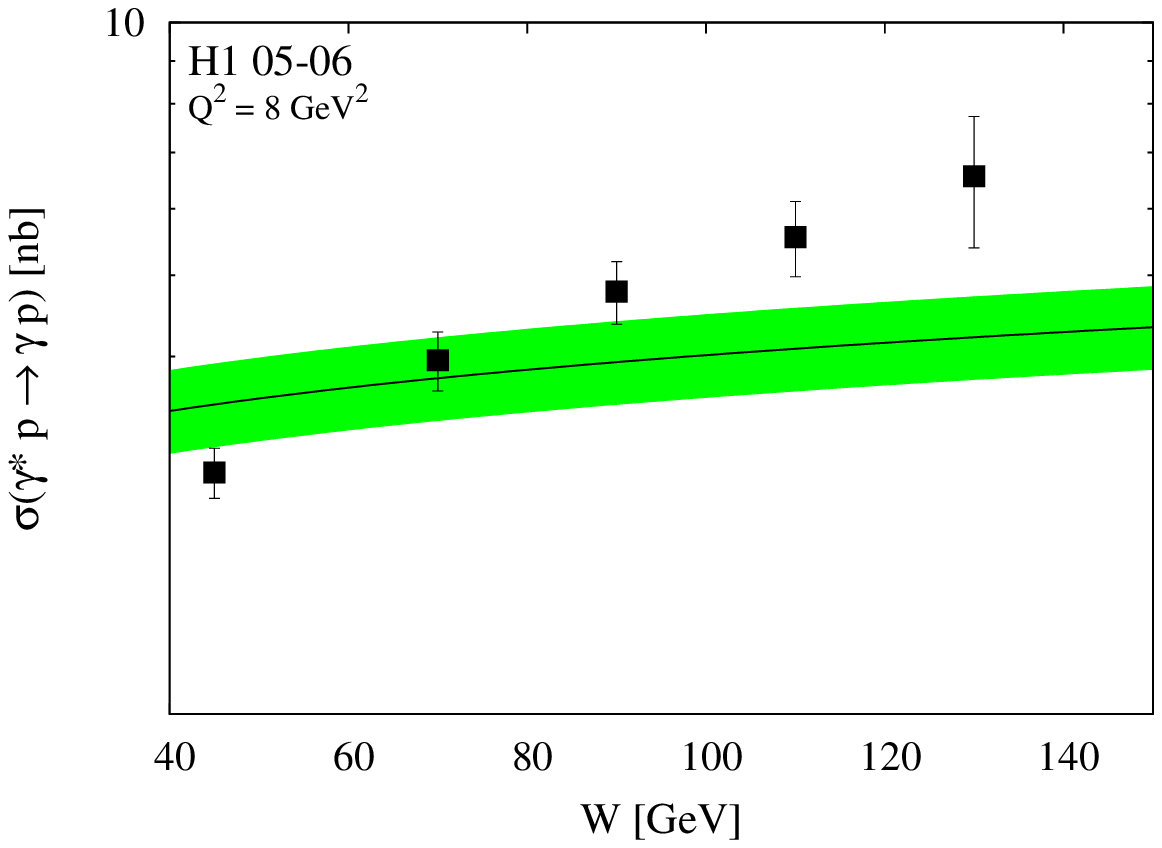}
\includegraphics[clip,scale=0.45]{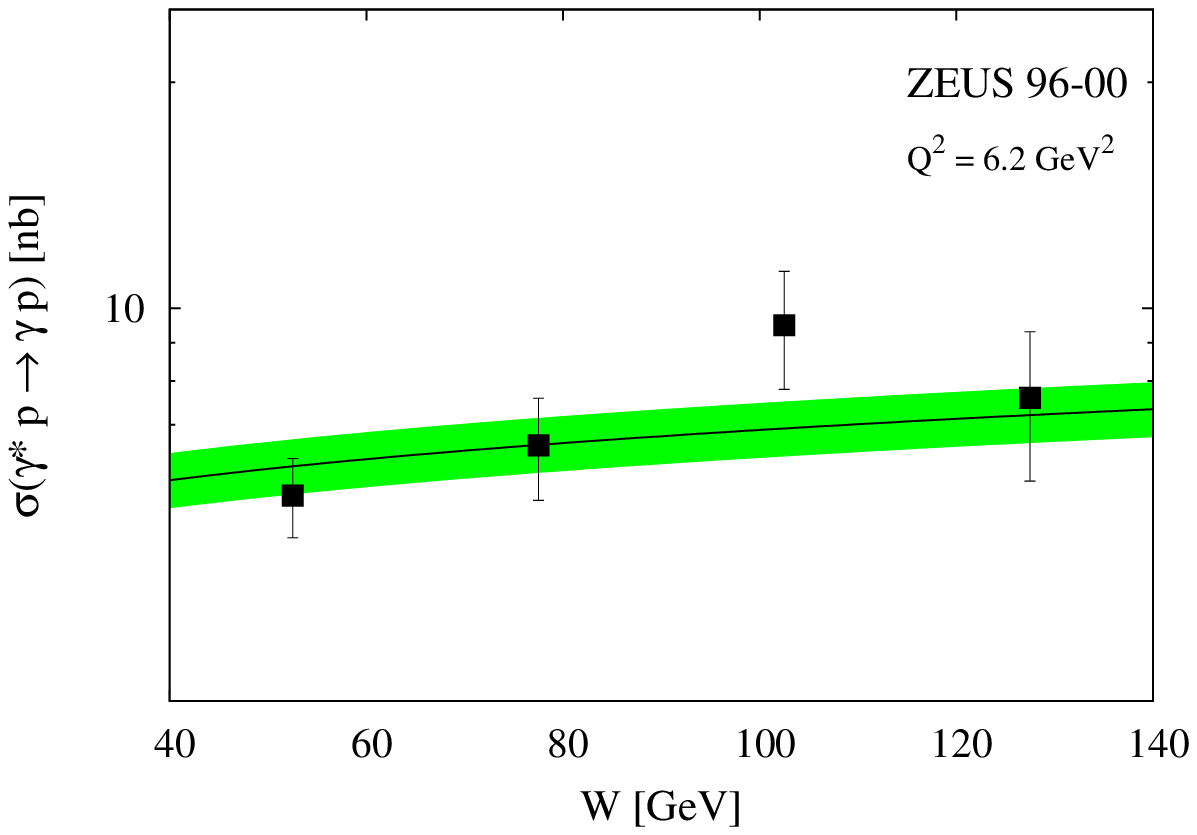}
\includegraphics[clip,scale=0.45]{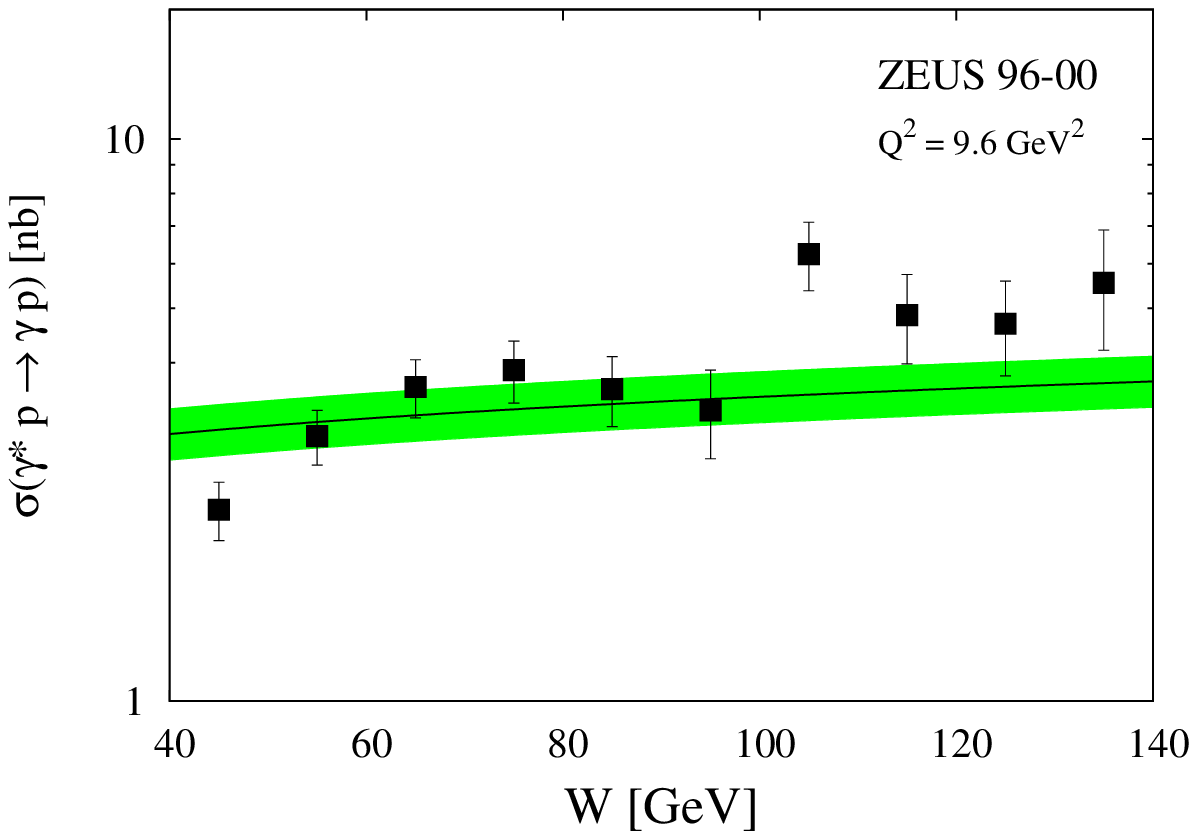}
\includegraphics[clip,scale=0.45]{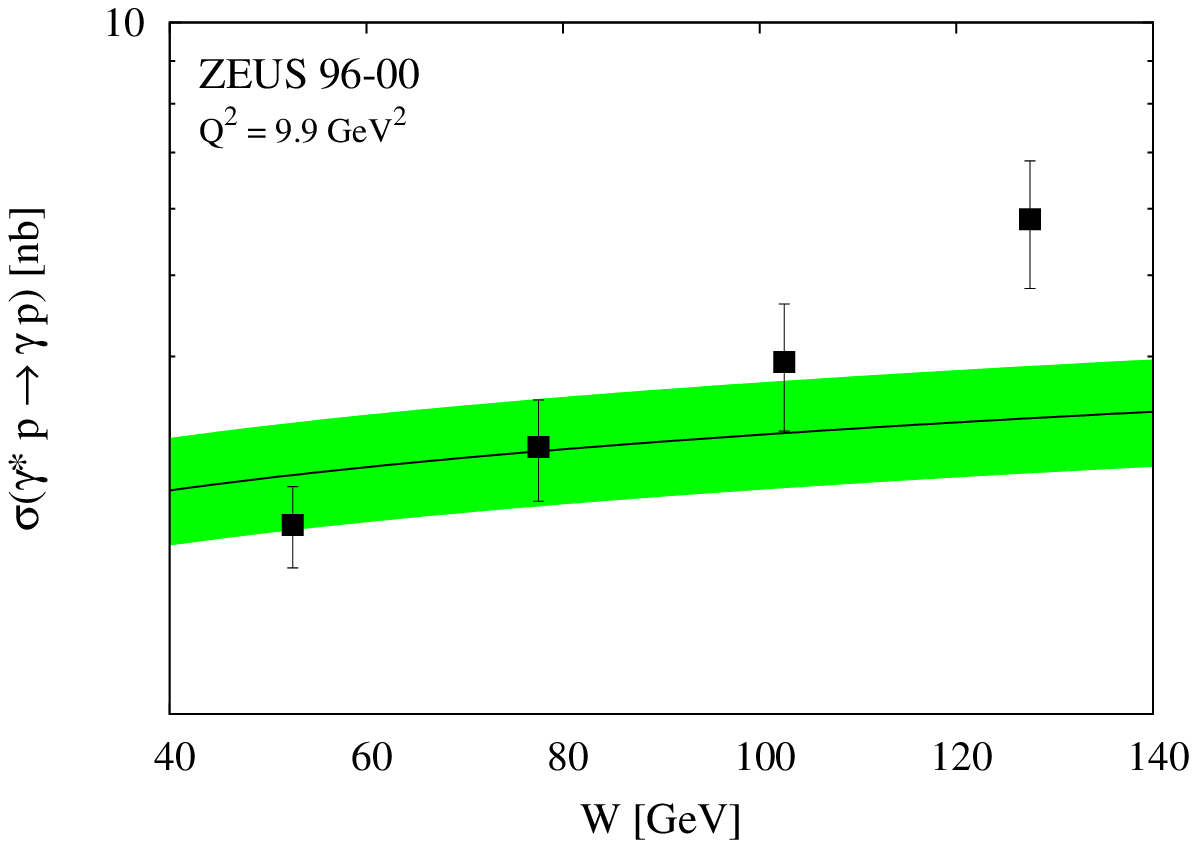}
\includegraphics[clip,scale=0.45]{figure_dvcs_vmp-2011/dvcs/sigma_int_log/dvcs_int_w/sigma_int_log_ZEUS_96-00_w_Q2_9.9_dvcs.eps}
\includegraphics[clip,scale=0.45]{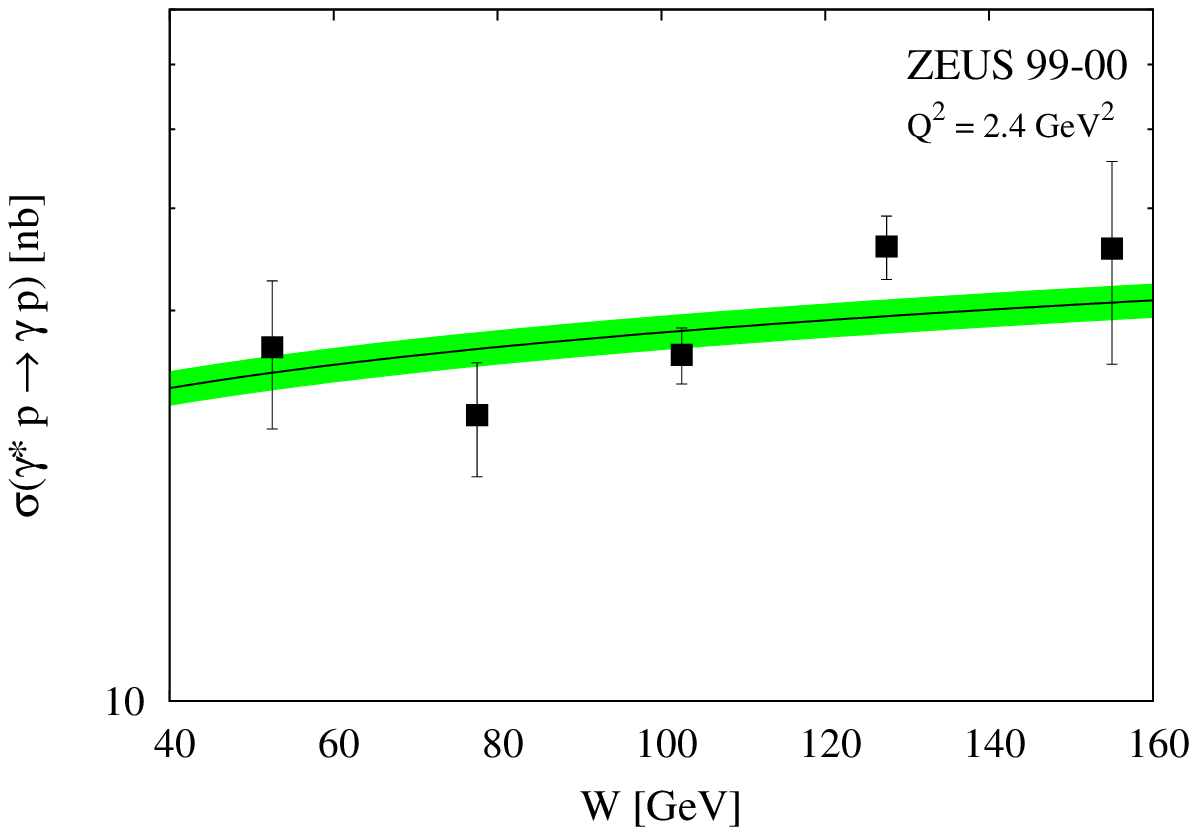}
\includegraphics[clip,scale=0.45]{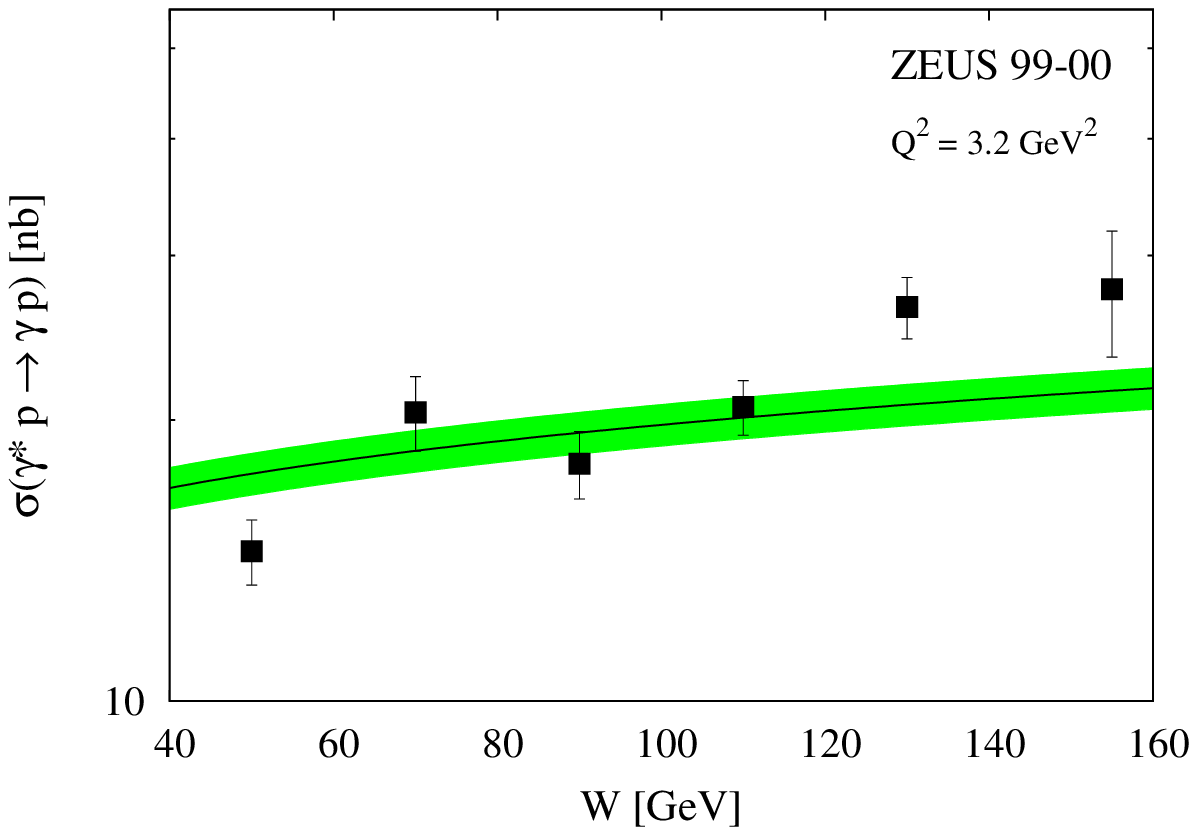}
\caption{\label{fig:dvcsw}The behaviour according to our model of $\gamma^*p\rightarrow\gamma p$ total cross section as function of $W$ is compared with data from Refs.~\cite{d1,d3,d4} measured by the H1 and ZEUS Collaborations for several values of $Q^2$. The green bands are calculated according with the uncertainties on the free parameter $|A_0|$.}
\end{figure*}

% DVCS dsigma/dt H1 04-07

\begin{figure*}
\includegraphics[clip,scale=0.427]{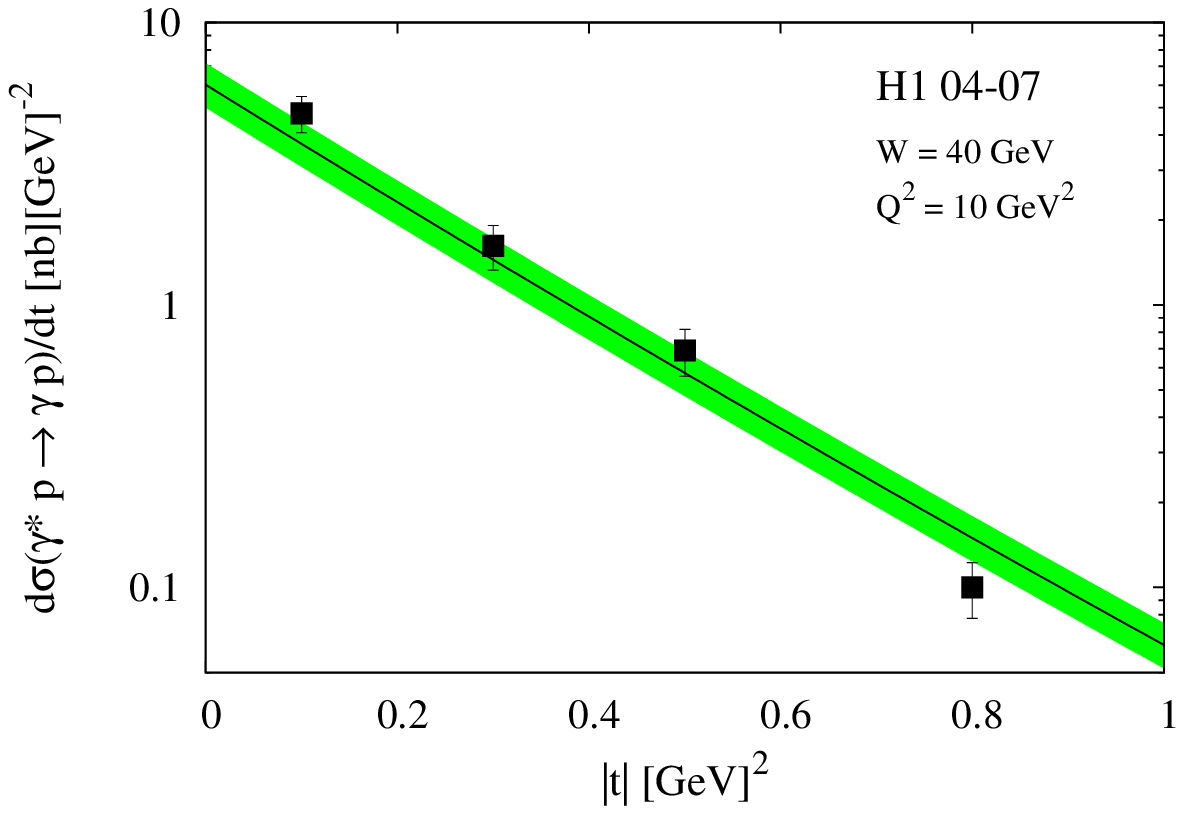}
\includegraphics[clip,scale=0.427]{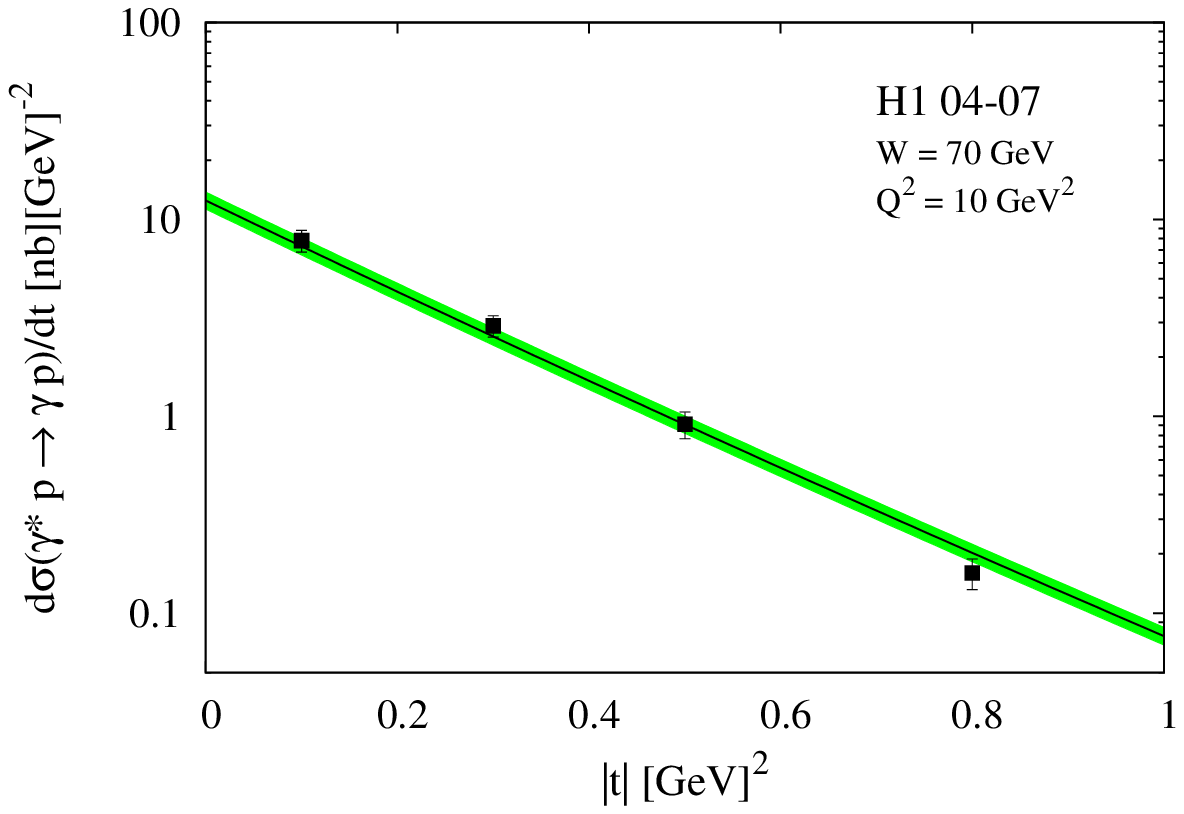}
\includegraphics[clip,scale=0.427]{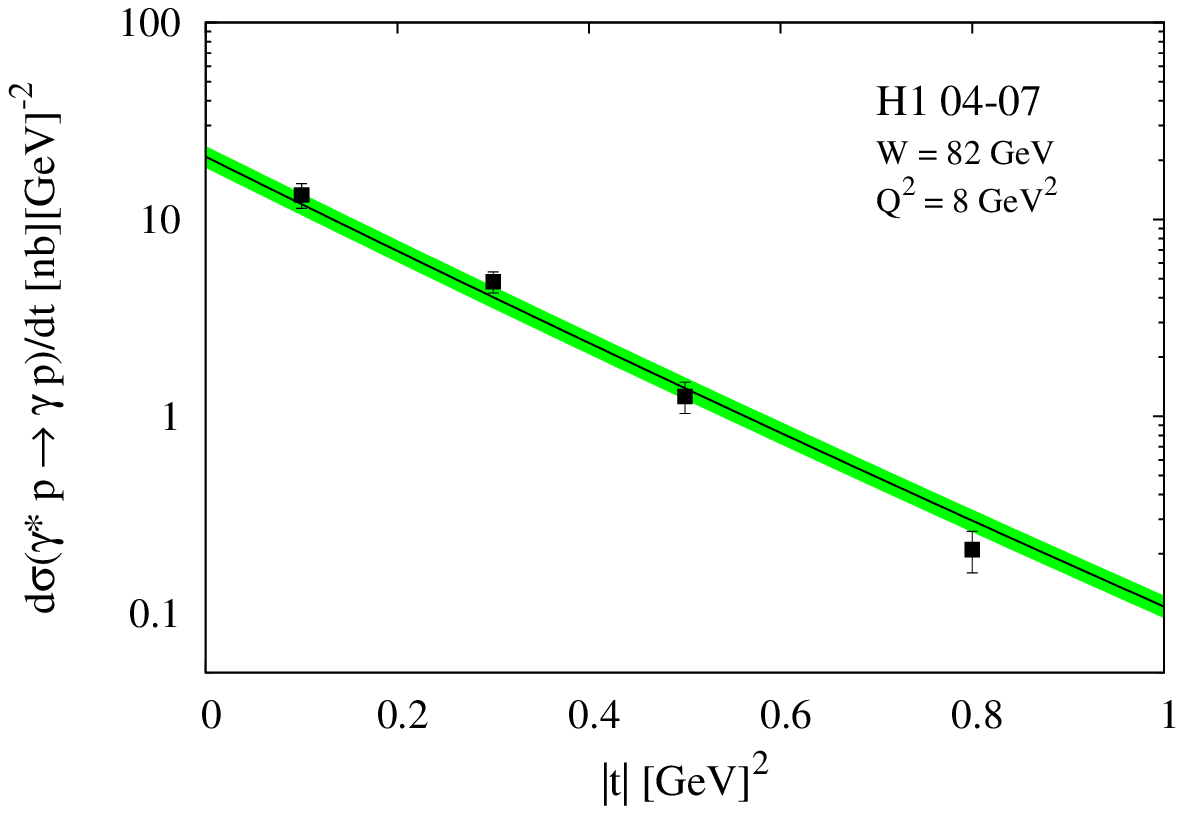}
\includegraphics[clip,scale=0.427]{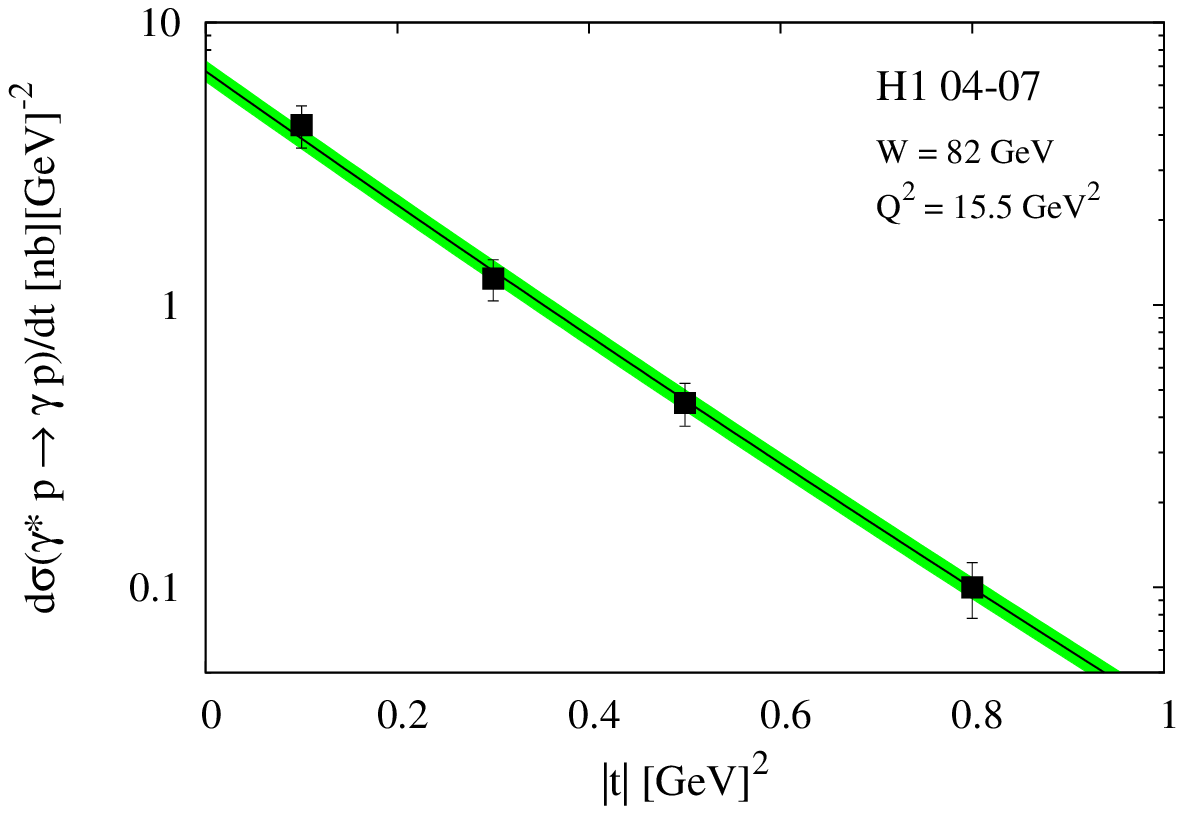}
\includegraphics[clip,scale=0.427]{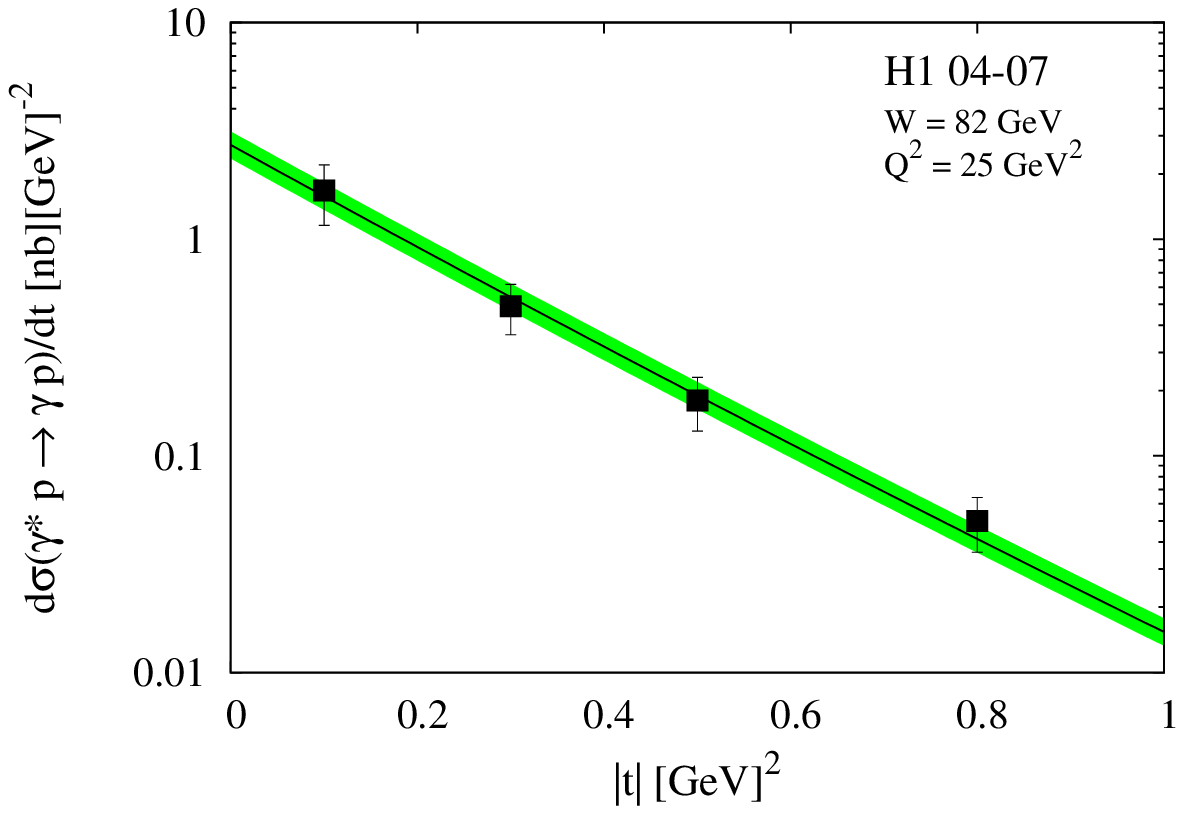}
\includegraphics[clip,scale=0.427]{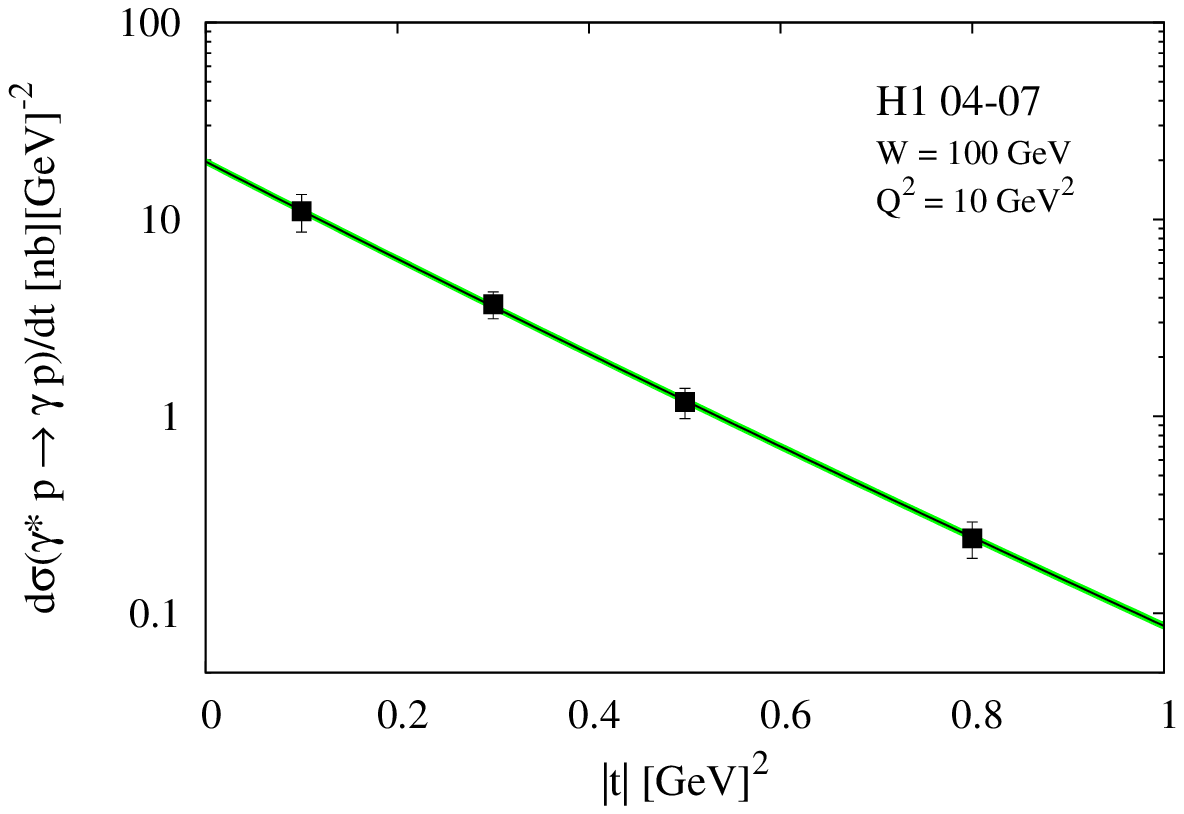}
\includegraphics[clip,scale=0.427]{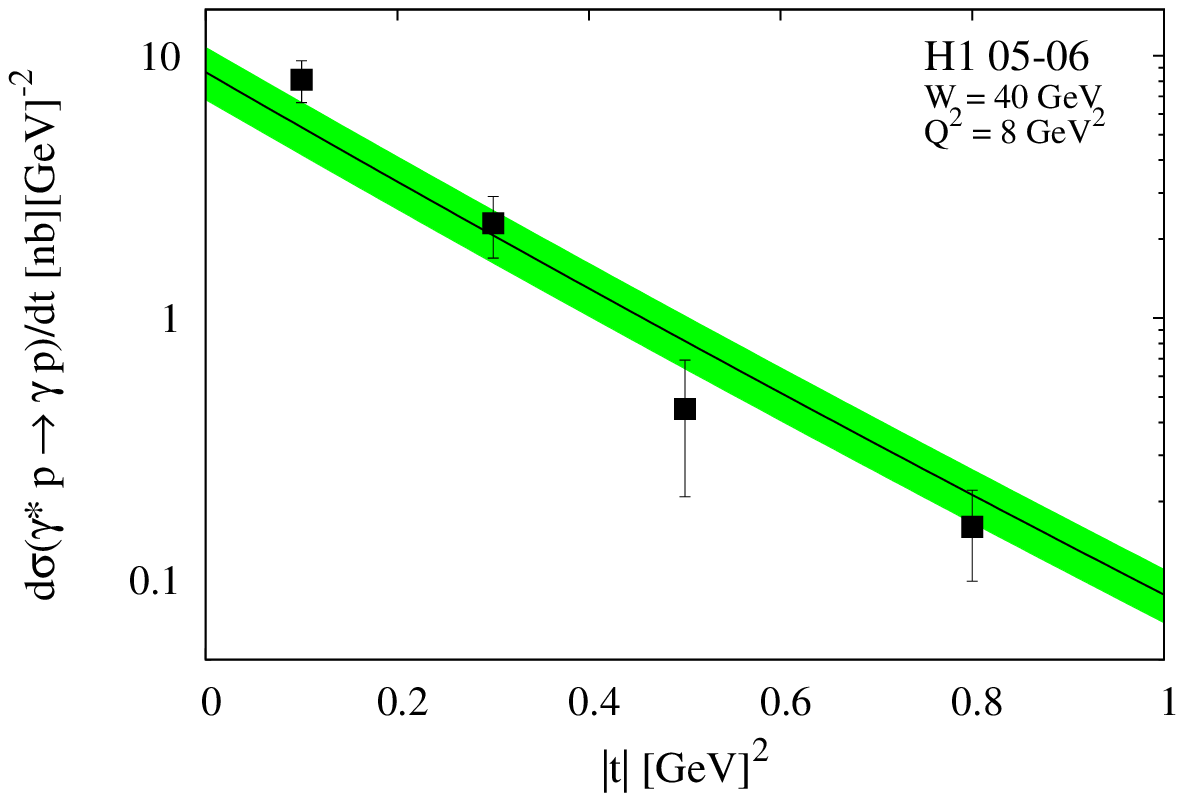}
\includegraphics[clip,scale=0.427]{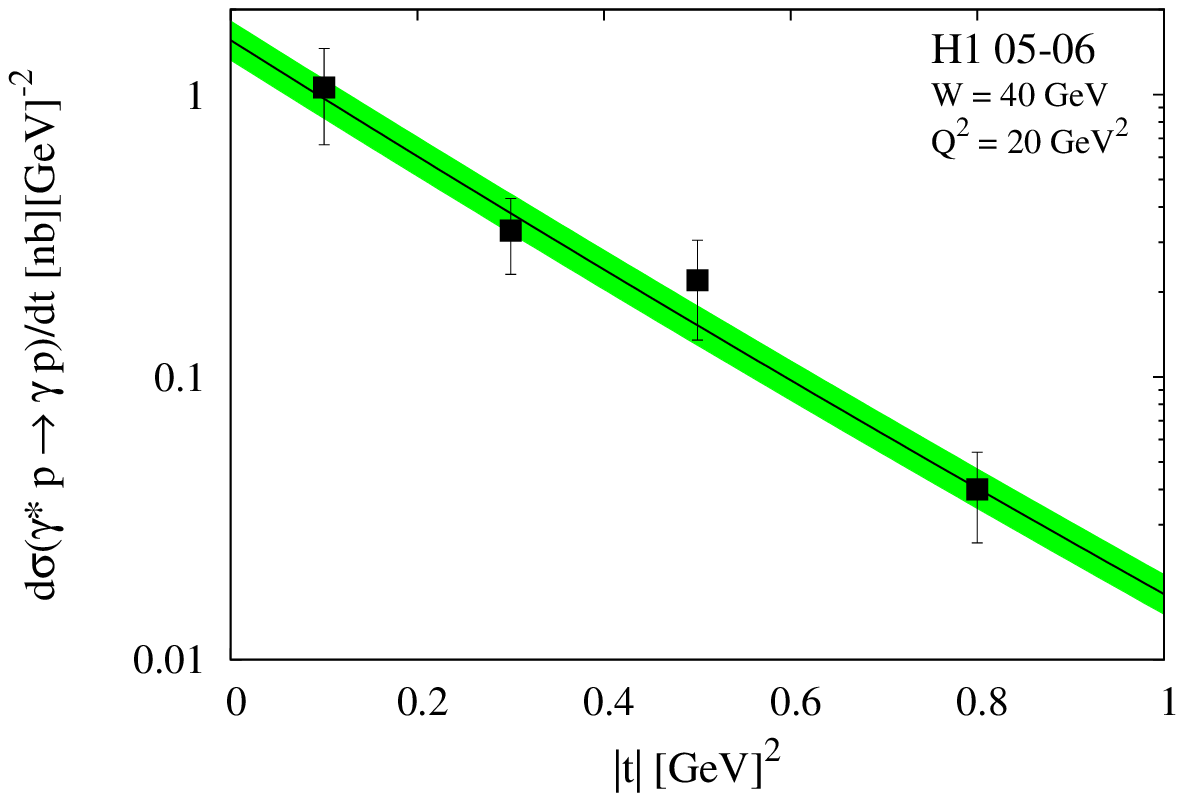}
\includegraphics[clip,scale=0.427]{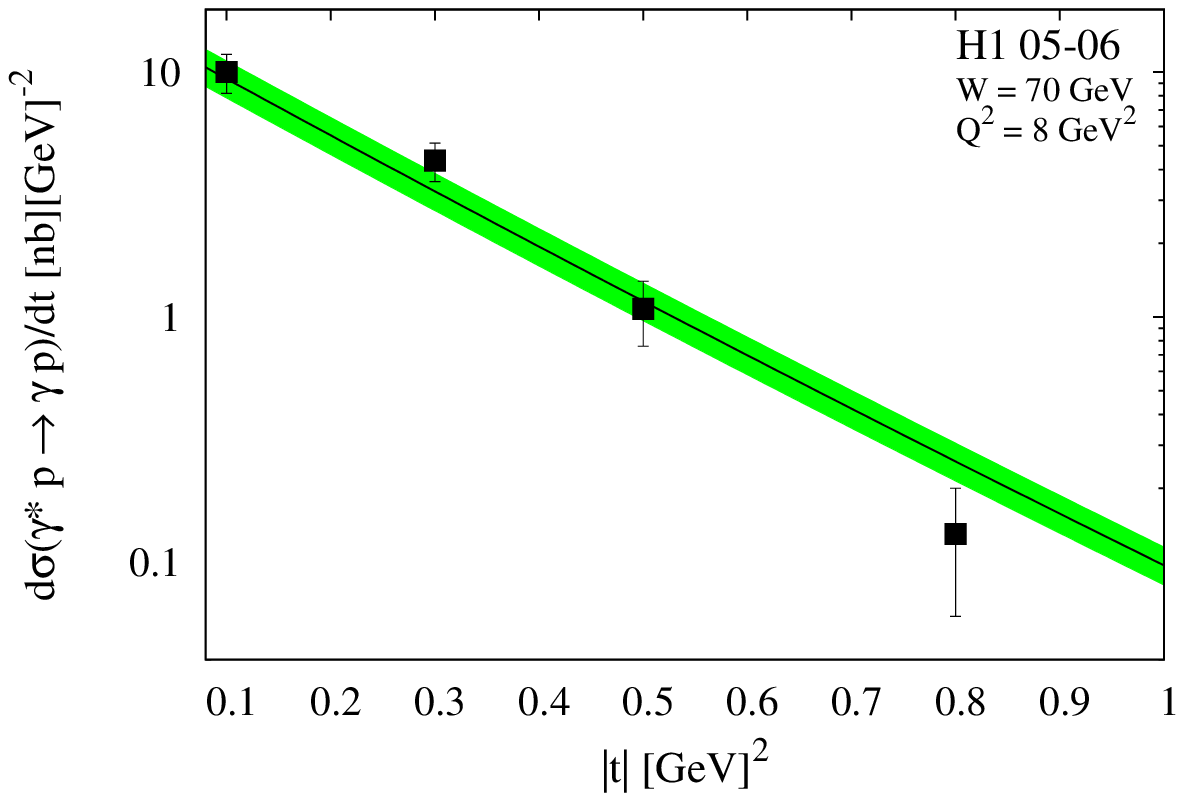}
\includegraphics[clip,scale=0.427]{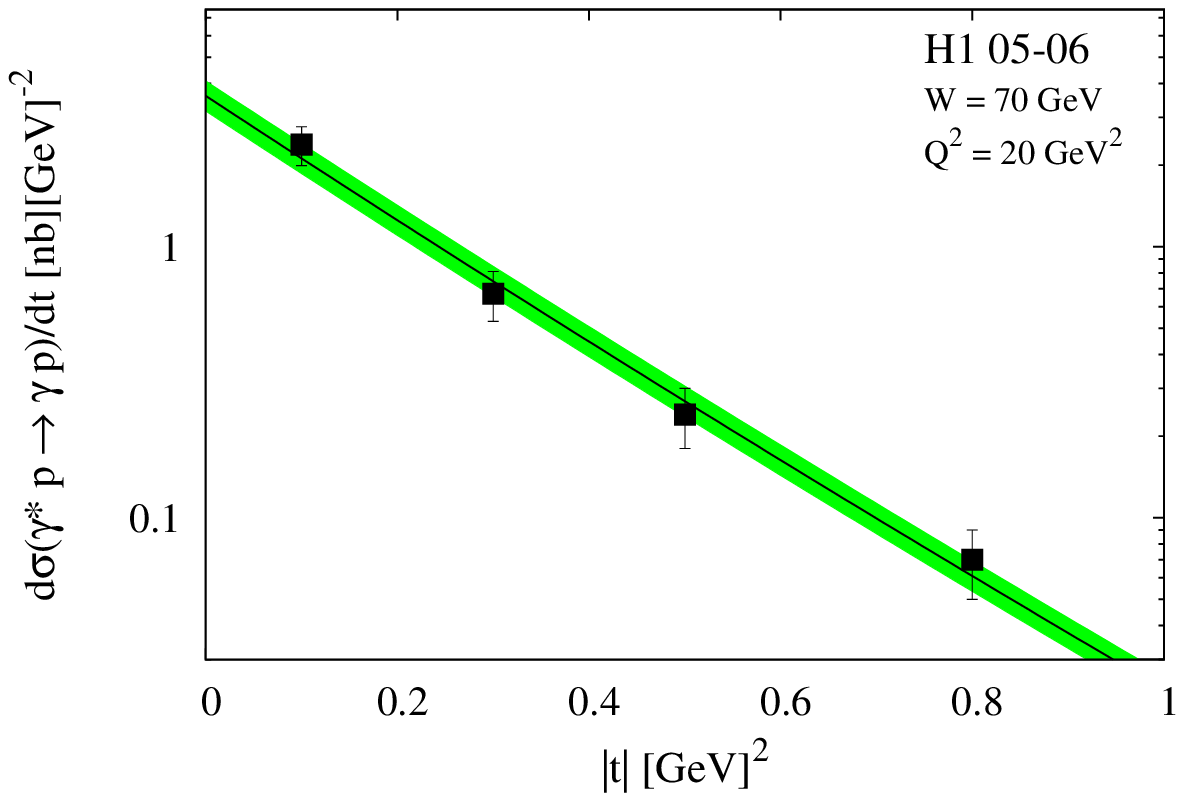}
\includegraphics[clip,scale=0.427]{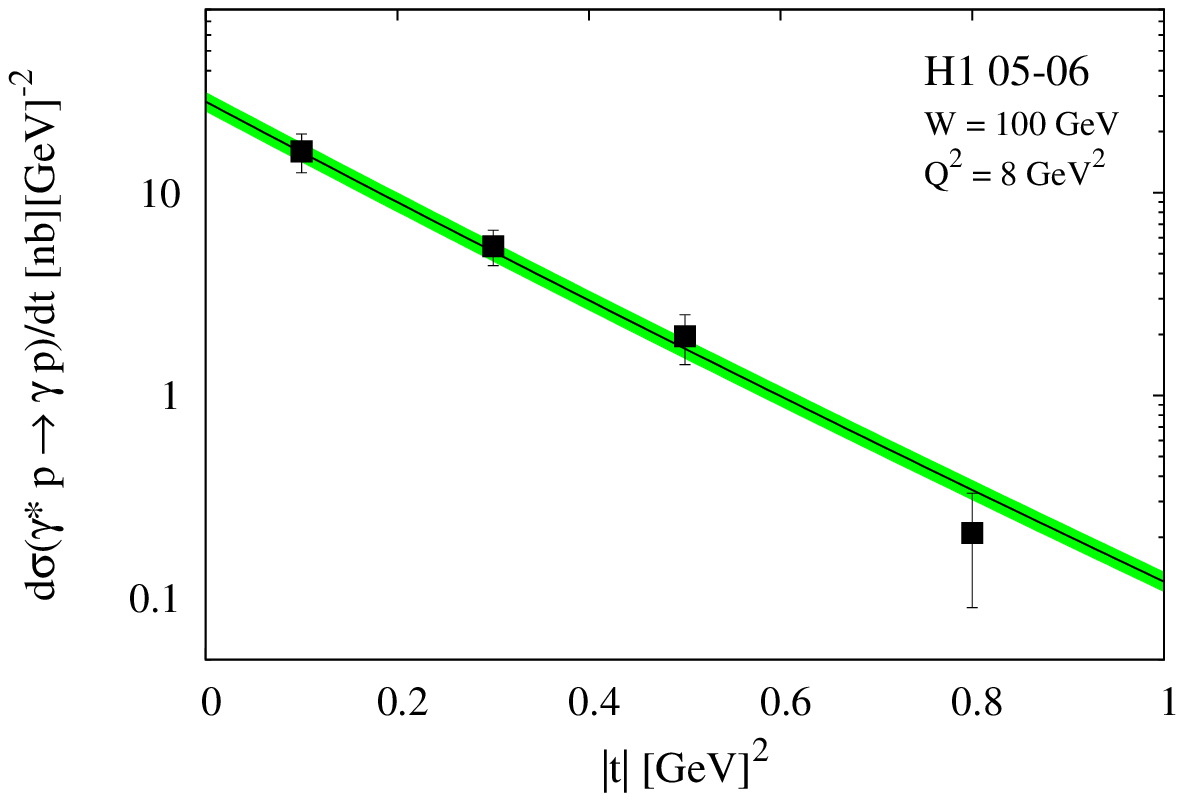}
\includegraphics[clip,scale=0.427]{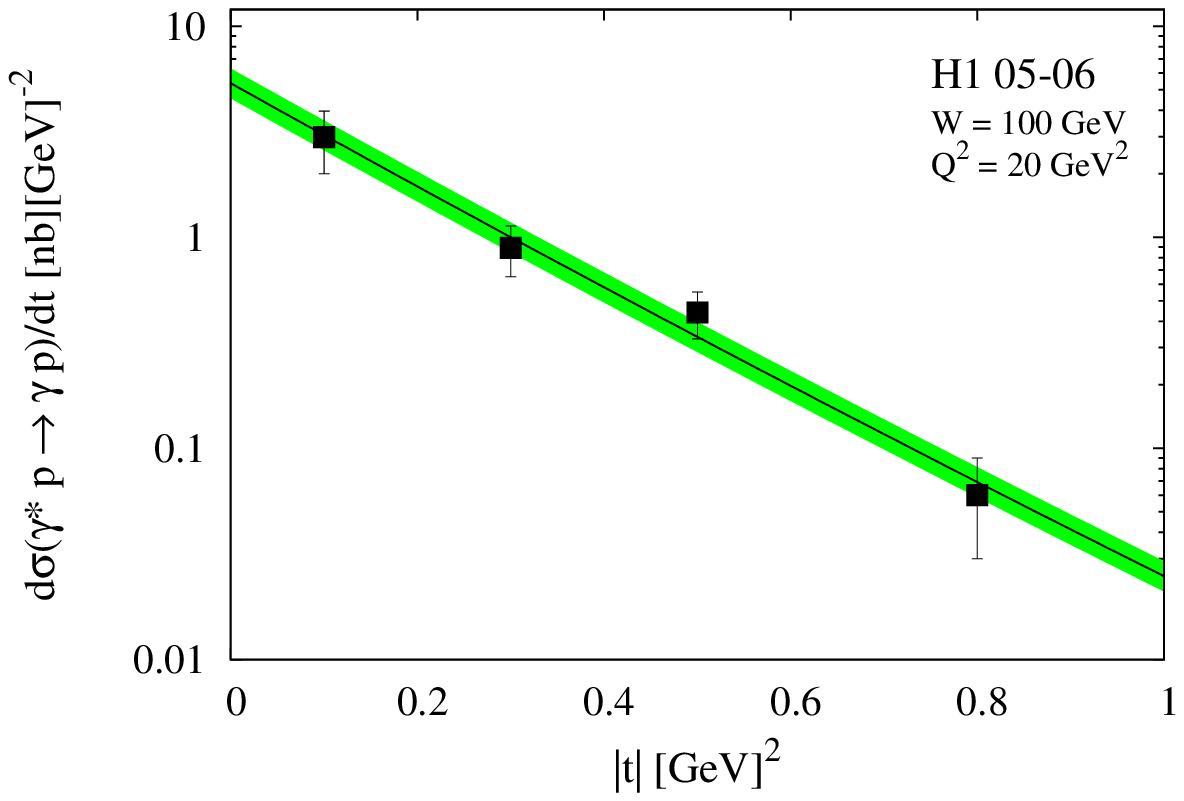}
\includegraphics[clip,scale=0.427]{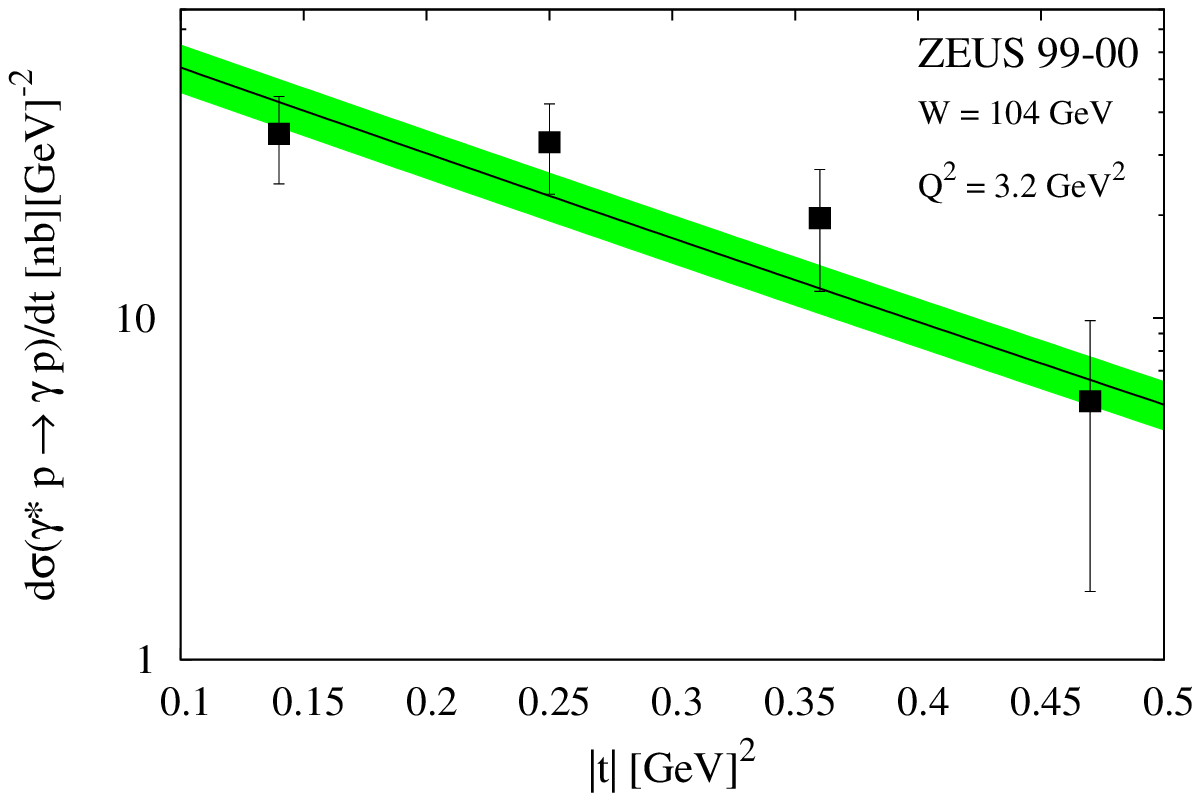}
\caption{\label{fig:dvcsdsdth1}The behaviour according to our model of $\gamma^*p\rightarrow\gamma p$ differential cross section as function of $t$ is compared with data from Refs.~\cite{d2,d3,d4} measured by the H1 and ZEUS Collaborations for several values of $Q^2$ and $W$. The green bands are calculated accordingly with the uncertainties on the free parameter $|A_0|$.}
\end{figure*}

\begin{figure*}[htb]
\includegraphics[clip,scale=0.5]{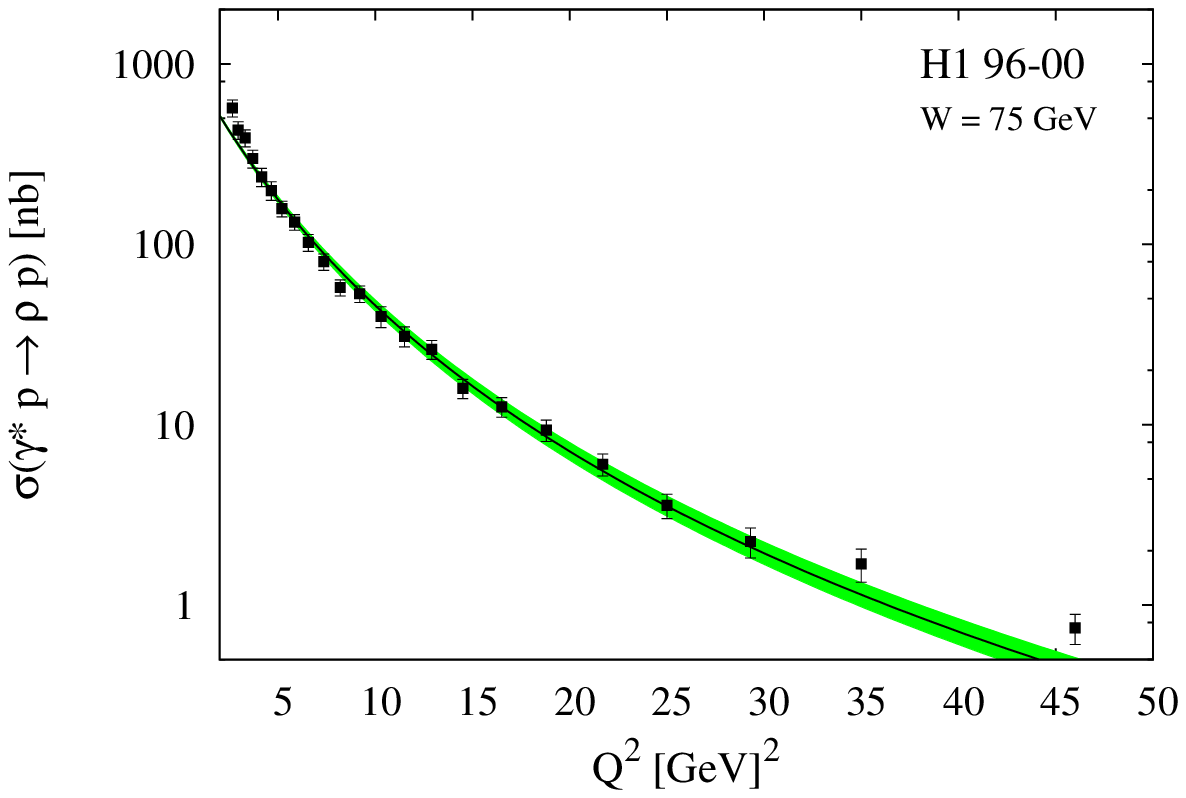}
\includegraphics[clip,scale=0.5]{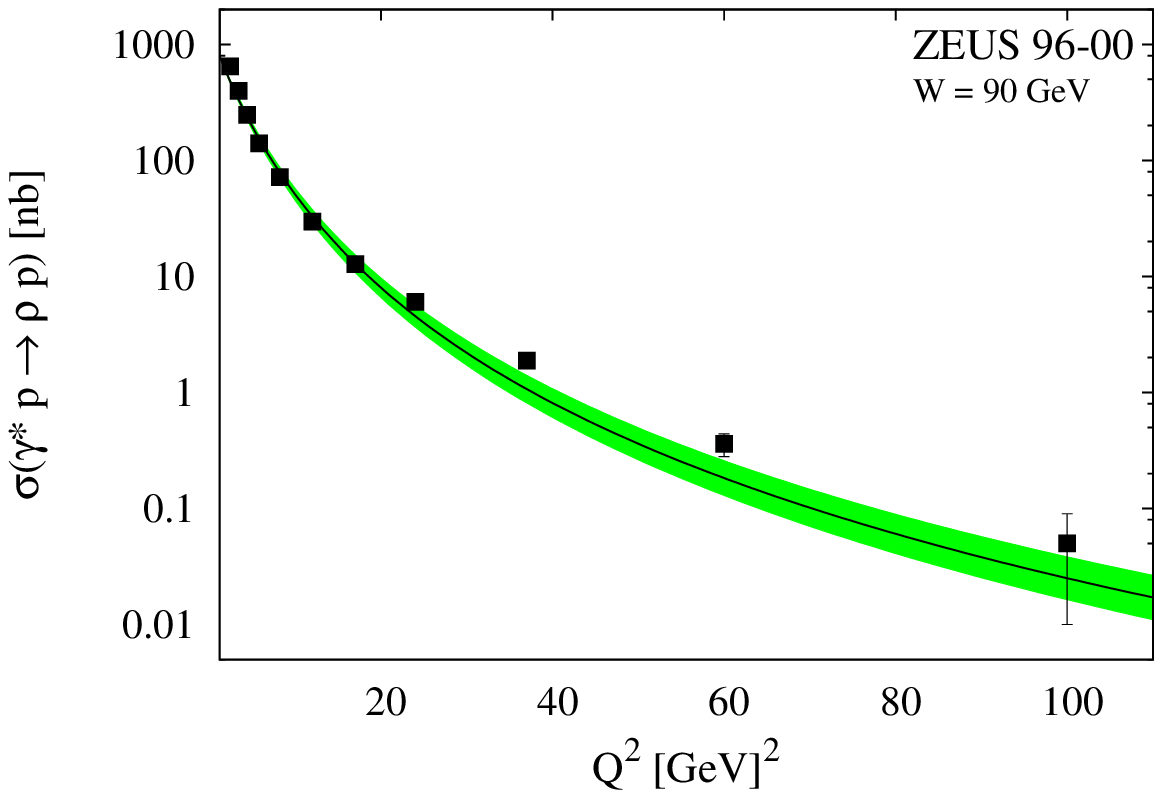}
\caption{\label{fig:rhoq2} The behaviour according to our model of the $\gamma^*p\rightarrow\rho^0 p$ total cross section as function of $Q^2$ is compared with data from Refs.~\cite{r1,r2} measured by the H1 and ZEUS Collaborations for fixed values of $W$. The green bands are calculated accordingly with the uncertainties on the free parameter $|A_0|$.}
\end{figure*}

% rho sigma(W) H1 96-00

\begin{figure*}[htb]
\includegraphics[clip,scale=0.43]{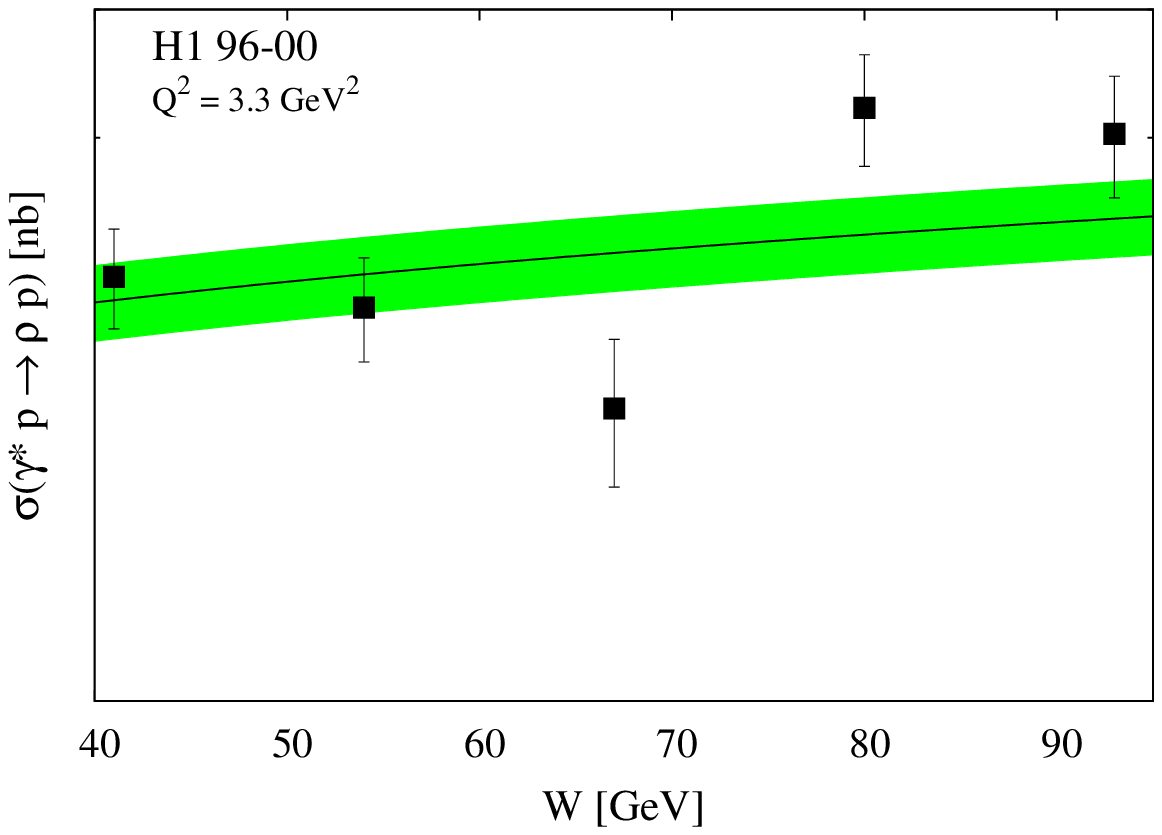}
\includegraphics[clip,scale=0.43]{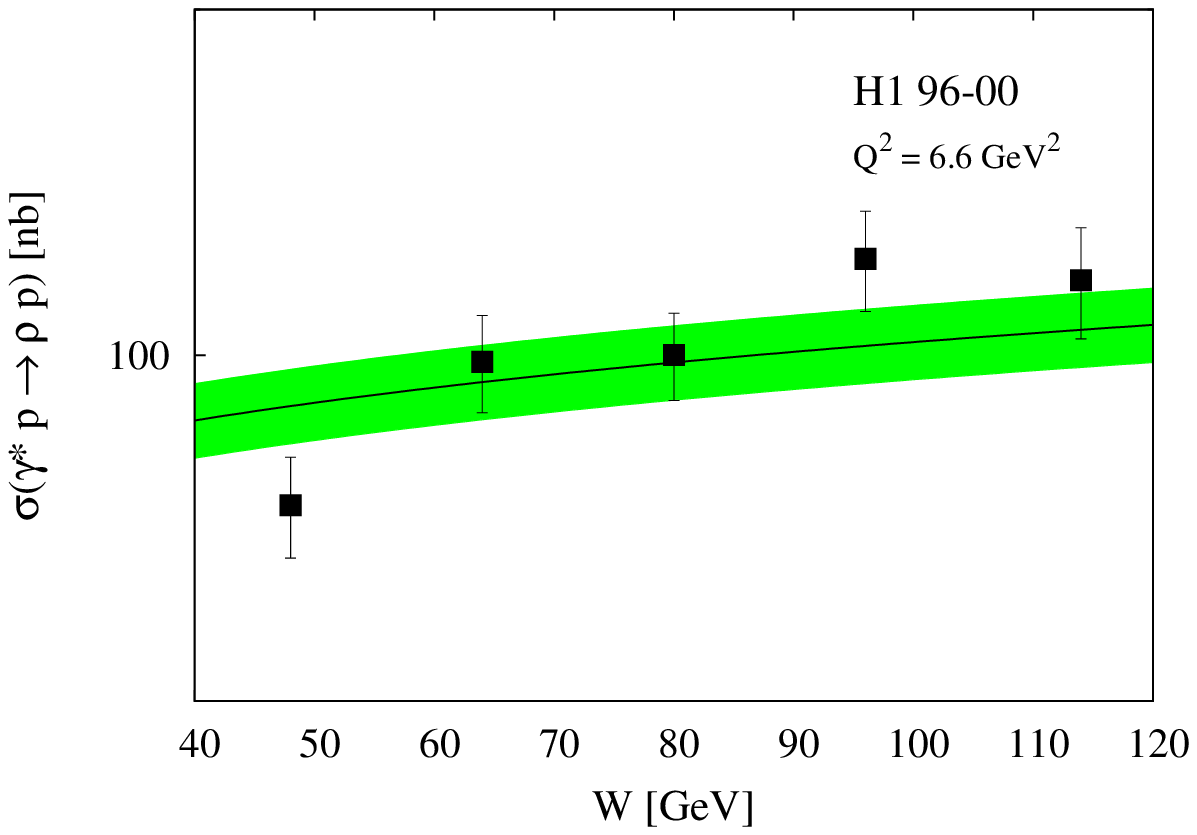}
\includegraphics[clip,scale=0.43]{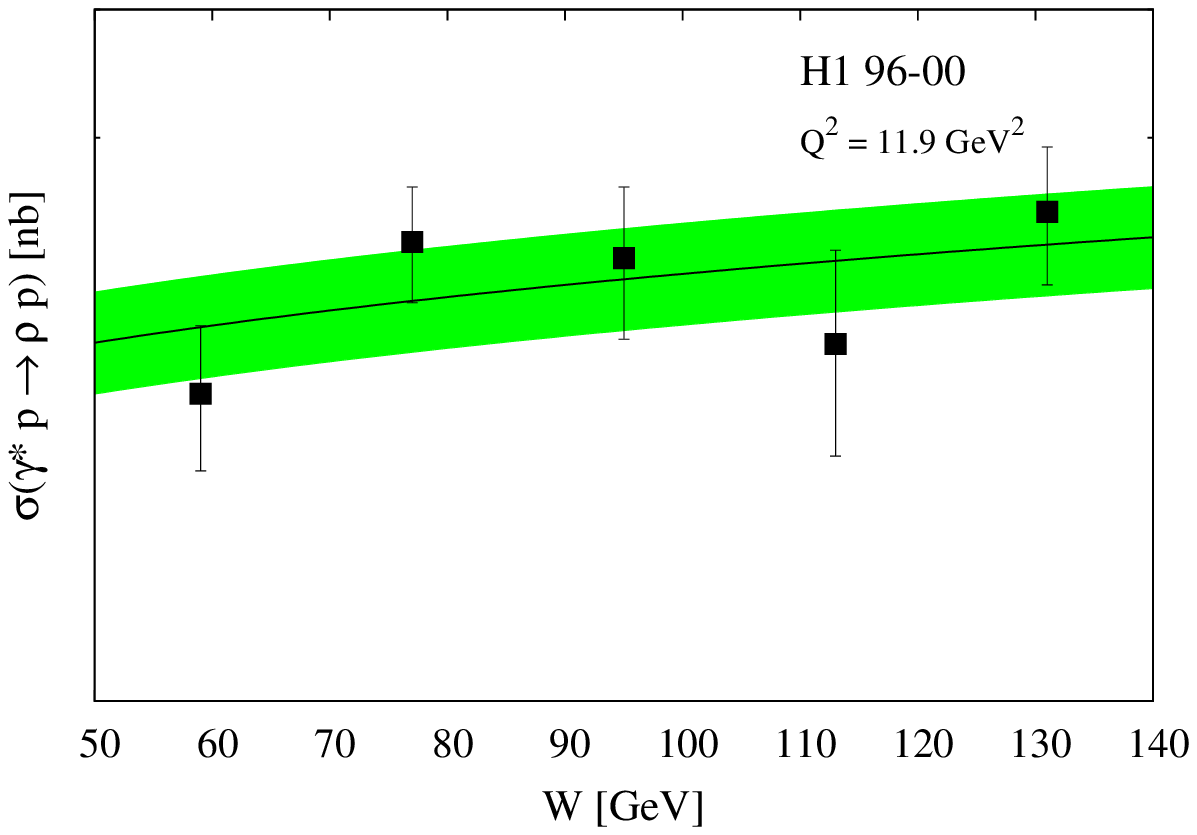}
\includegraphics[clip,scale=0.43]{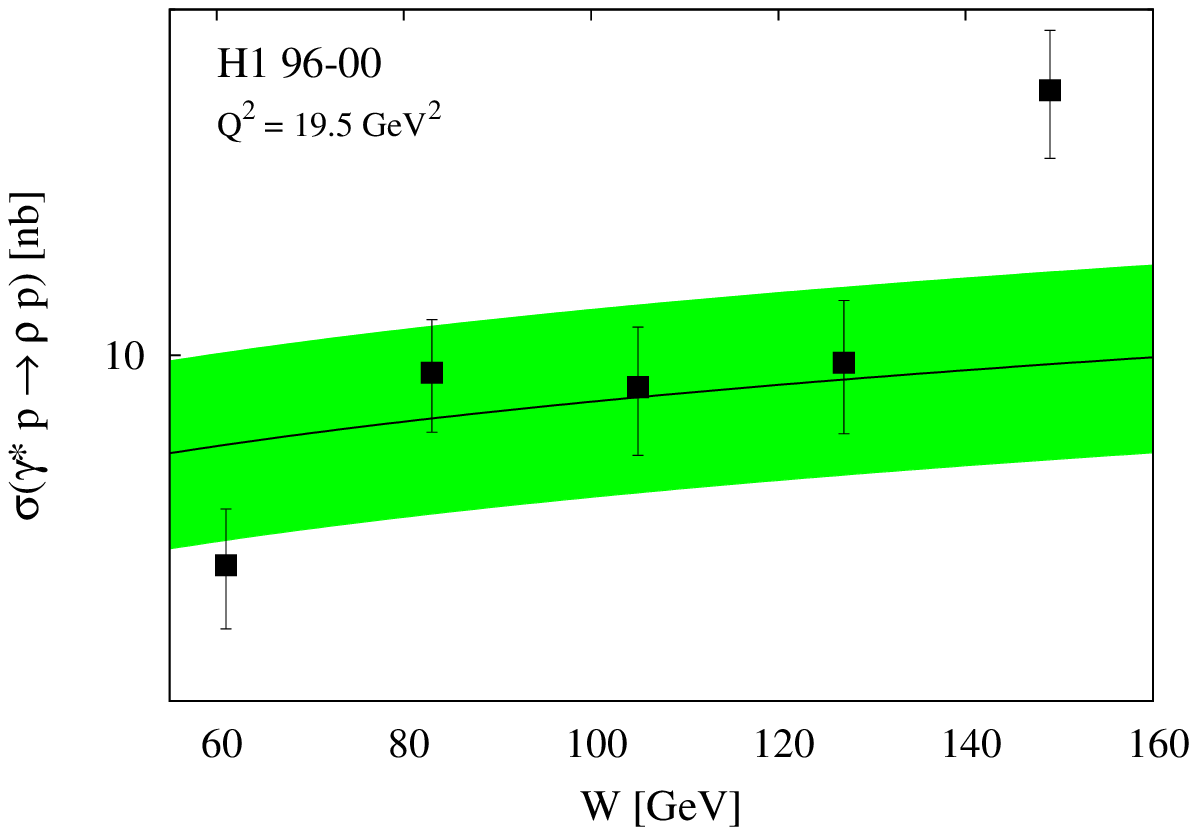}
\includegraphics[clip,scale=0.43]{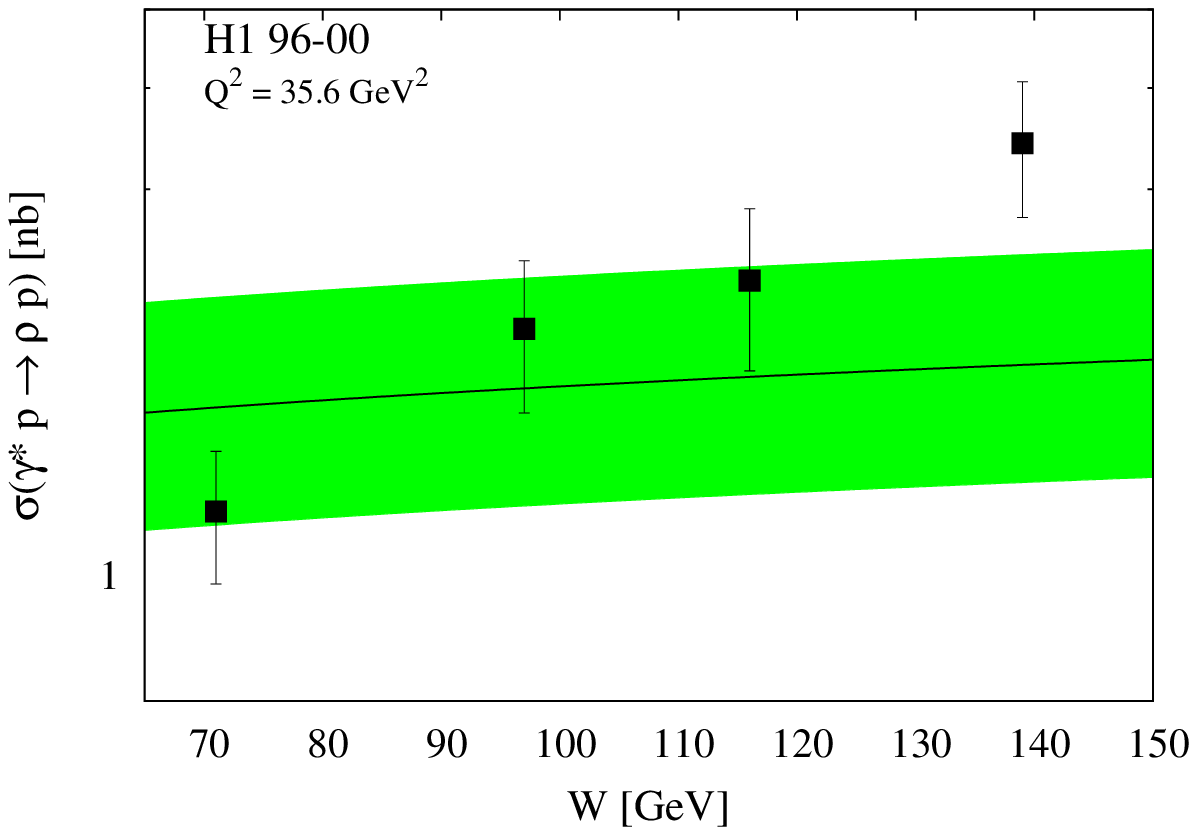}
\includegraphics[clip,scale=0.43]{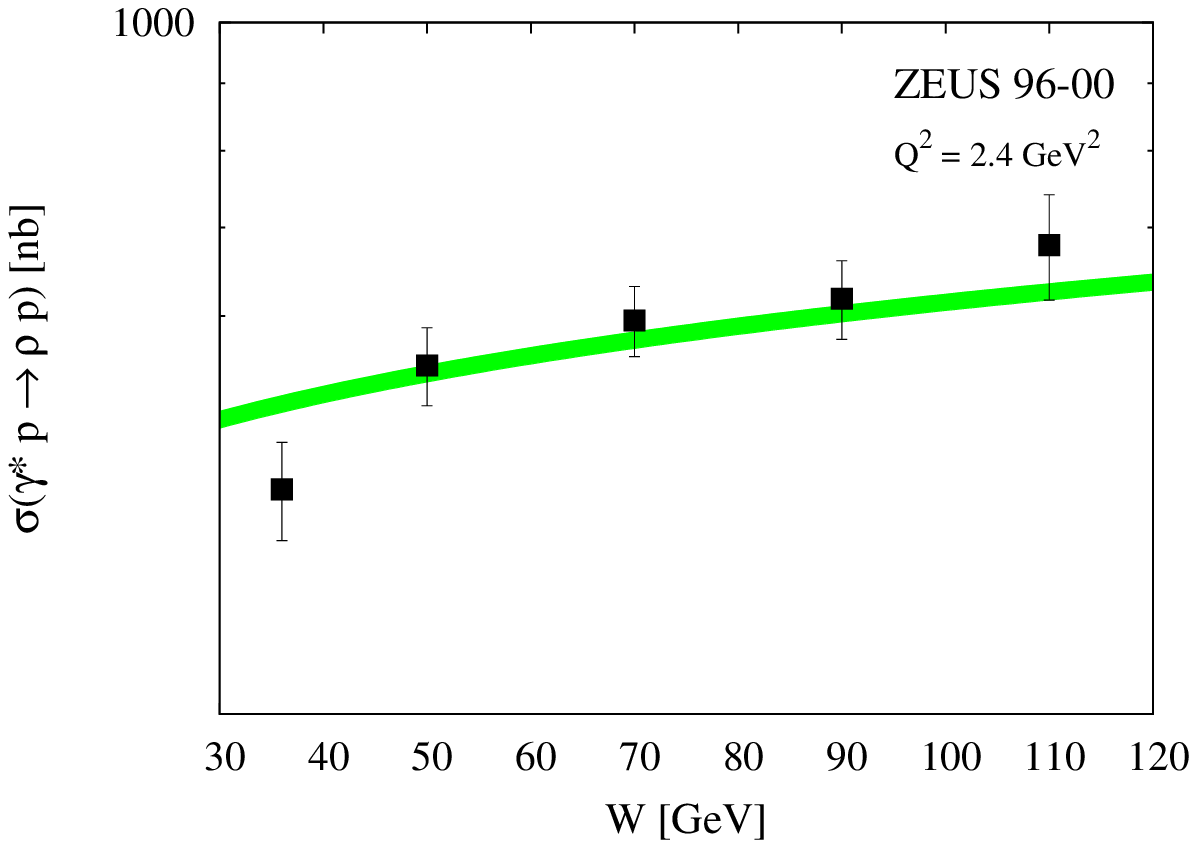}
\includegraphics[clip,scale=0.43]{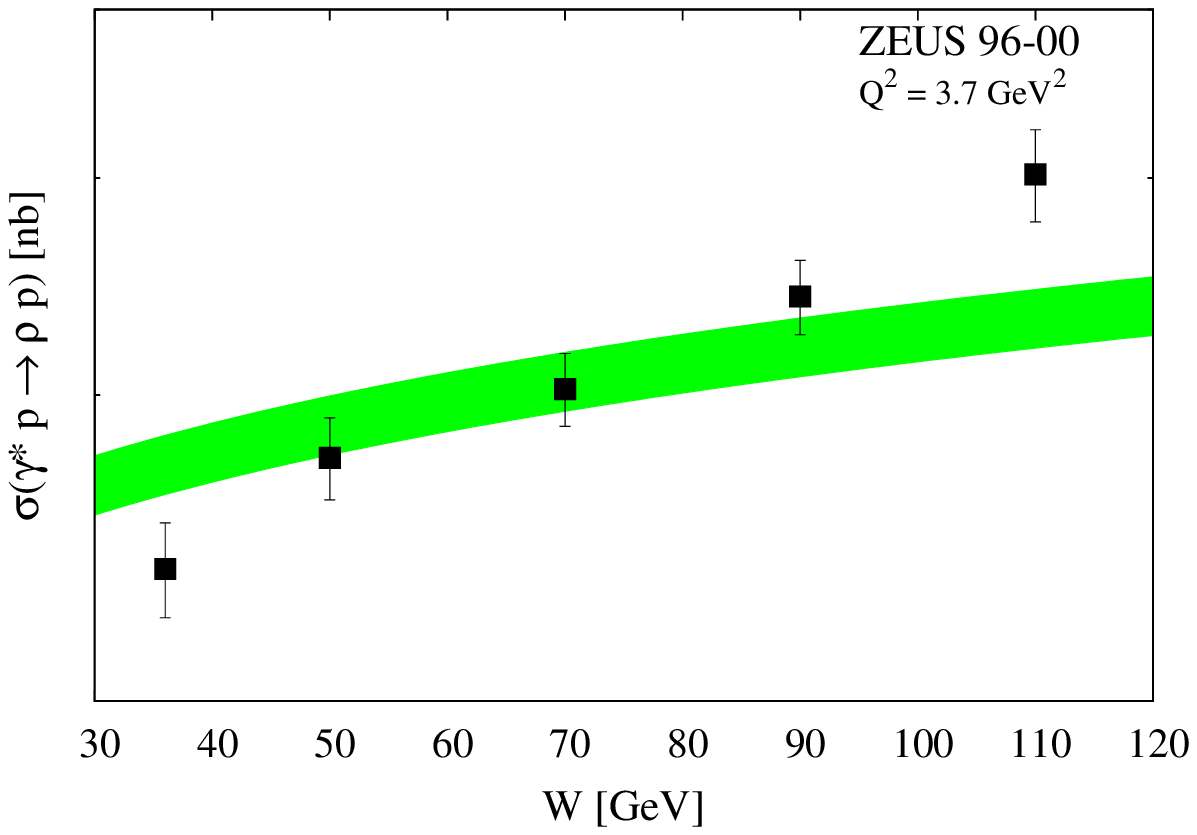}
\includegraphics[clip,scale=0.43]{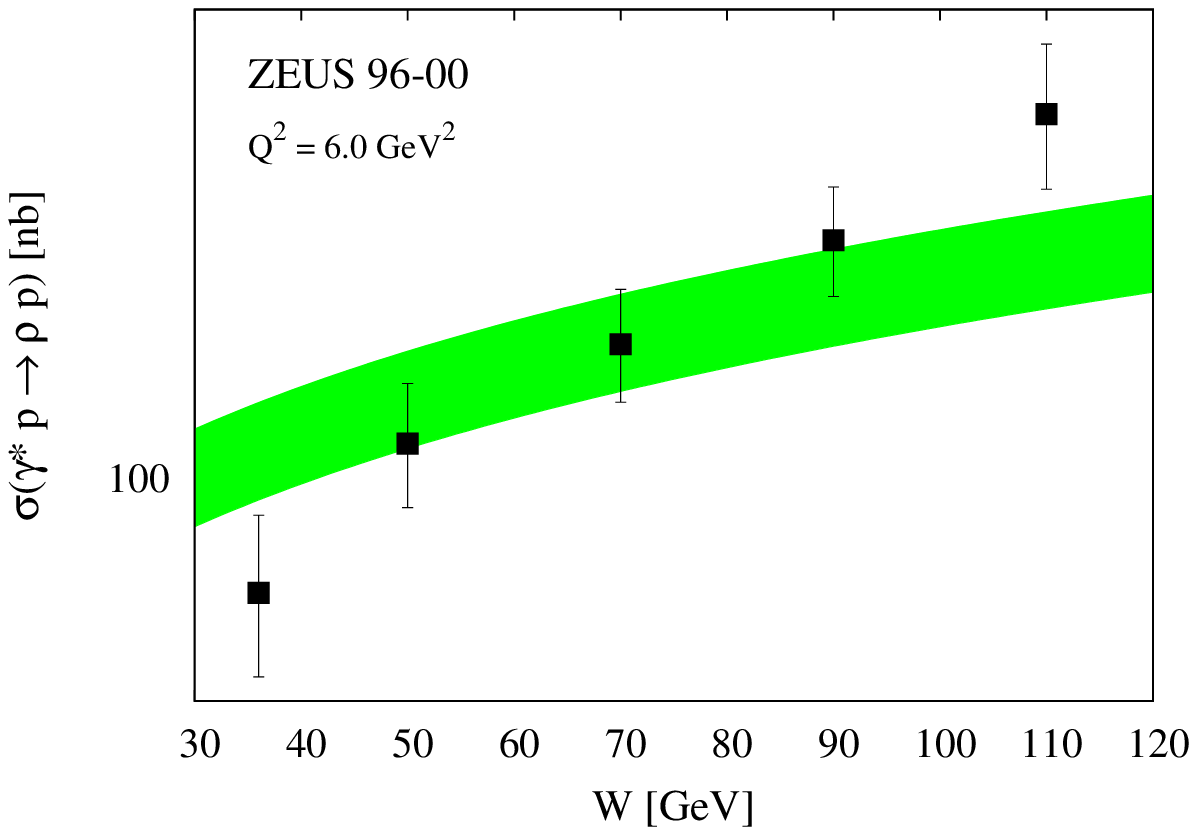}
\includegraphics[clip,scale=0.43]{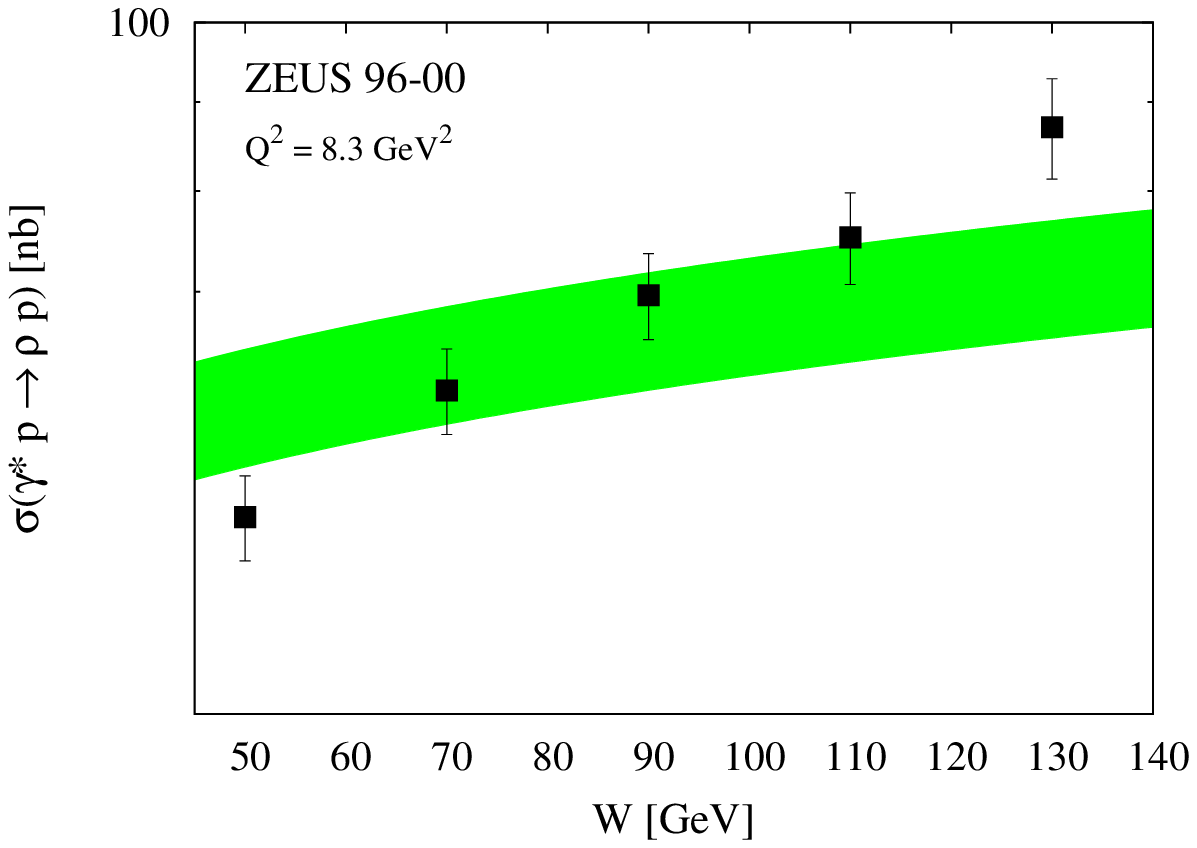}
\includegraphics[clip,scale=0.43]{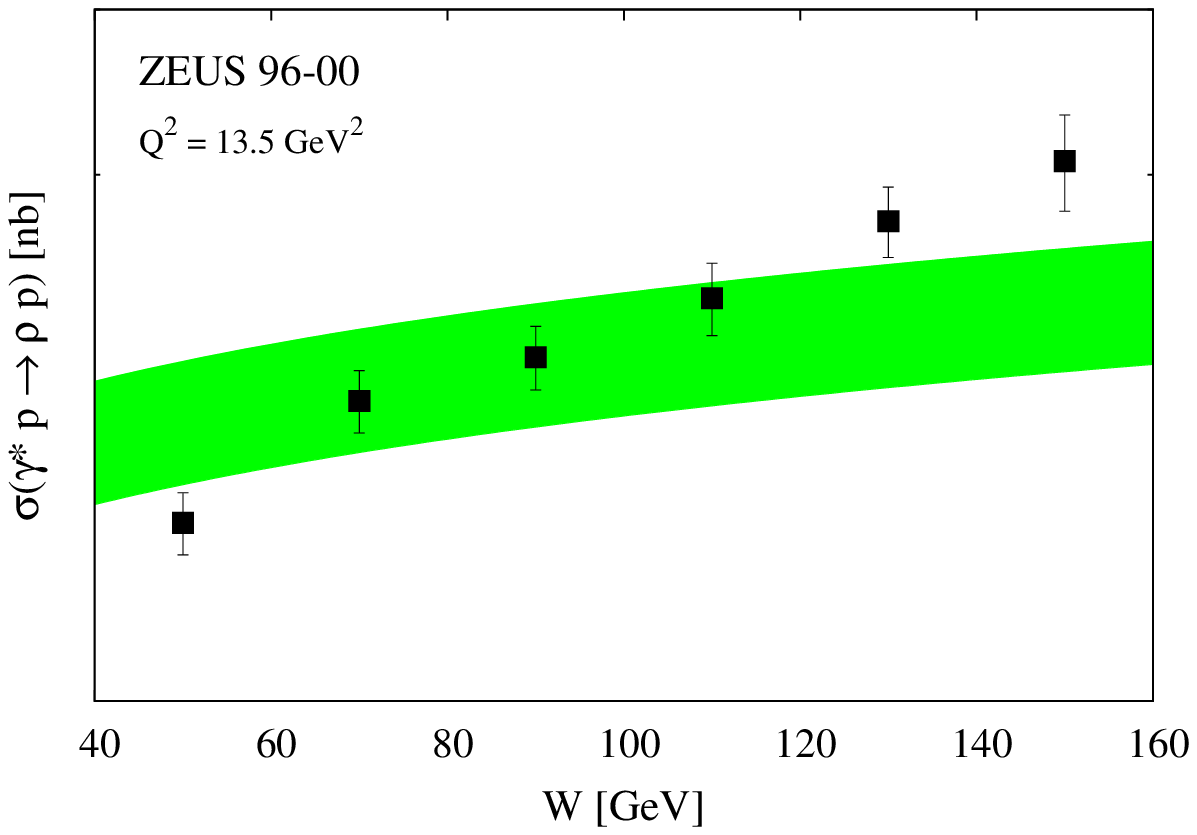}
\caption{\label{fig:rhow}The behaviour according to our model of the $\gamma^*p\rightarrow\rho^0 p$ total cross section as function of $W$ is compared with data from Ref.~\cite{r1,r2} measured by the H1 and ZEUS Collaborations for several values of $Q^2$. The green bands are calculated according with the uncertainties on the free parameter $|A_0|$.}
\end{figure*}

%\begin{figure*}
%\includegraphics[clip,scale=0.45]{figure_dvcs_vmp-2011/rho/int/rho_int_w/rho_int_w_ZEUS_96-00/sigmael_ZEUS_96-00_w_Q2=2-4_rho.eps}
%\includegraphics[clip,scale=0.45]{figure_dvcs_vmp-2011/rho/int/rho_int_w/rho_int_w_ZEUS_96-00/sigmael_ZEUS_96-00_w_Q2=3-7_rho.eps}
%\includegraphics[clip,scale=0.45]{figure_dvcs_vmp-2011/rho/int/rho_int_w/rho_int_w_ZEUS_96-00/sigmael_ZEUS_96-00_w_Q2=6-0_rho.eps}
%\includegraphics[clip,scale=0.45]{figure_dvcs_vmp-2011/rho/int/rho_int_w/rho_int_w_ZEUS_96-00/sigmael_ZEUS_96-00_w_Q2=8-3_rho.eps}
%\includegraphics[clip,scale=0.45]{figure_dvcs_vmp-2011/rho/int/rho_int_w/rho_int_w_ZEUS_96-00/sigmael_ZEUS_96-00_w_Q2=13-5_rho.eps}
%\caption{\label{fig:rhow_zeus}The behaviour according to our model of $\gamma^*p\rightarrow\rho^0 p$ total cross section as function of $W$ is compared with data from Ref.~\cite{r2} measured by the ZEUS Collaboration for several values of $Q^2$. The green bands are calculated according with the uncertainties on the free parameter $|A_0|$.}
%\end{figure*}

% rho dsigma/dt 

\begin{figure*}[htb]
\includegraphics[clip,scale=0.427]{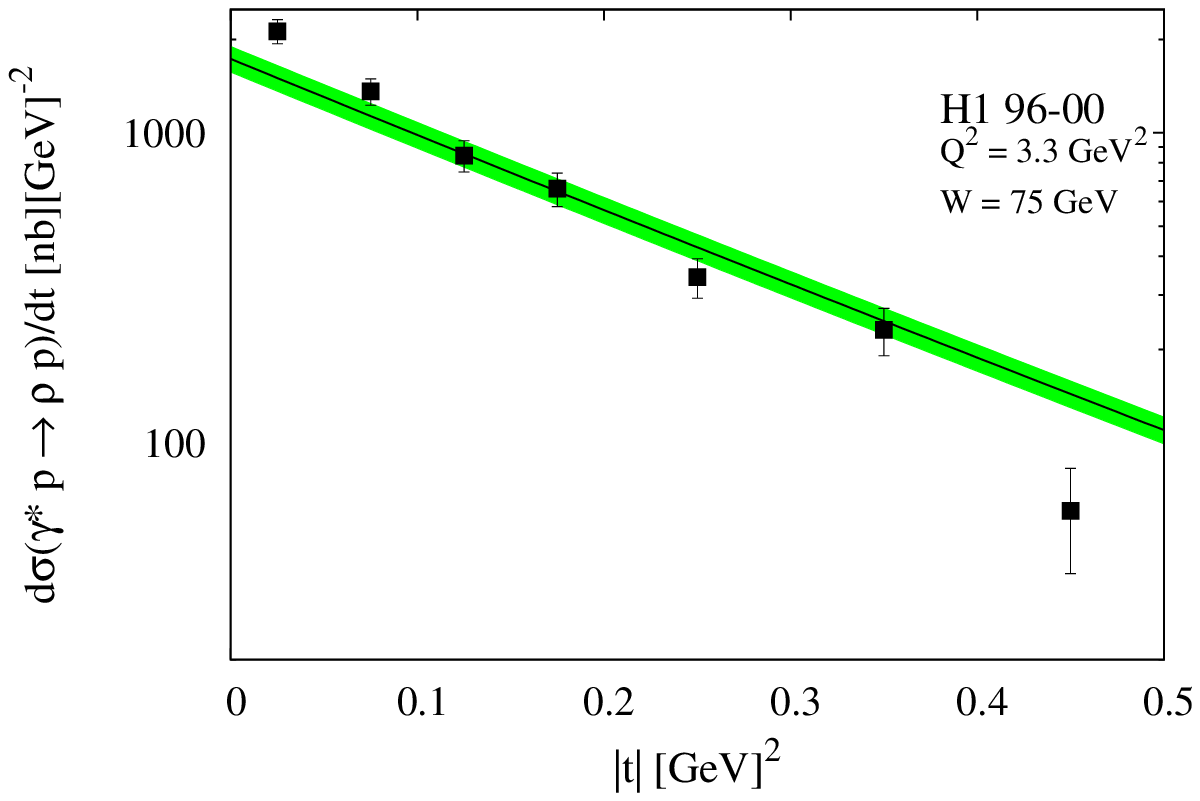}
\includegraphics[clip,scale=0.427]{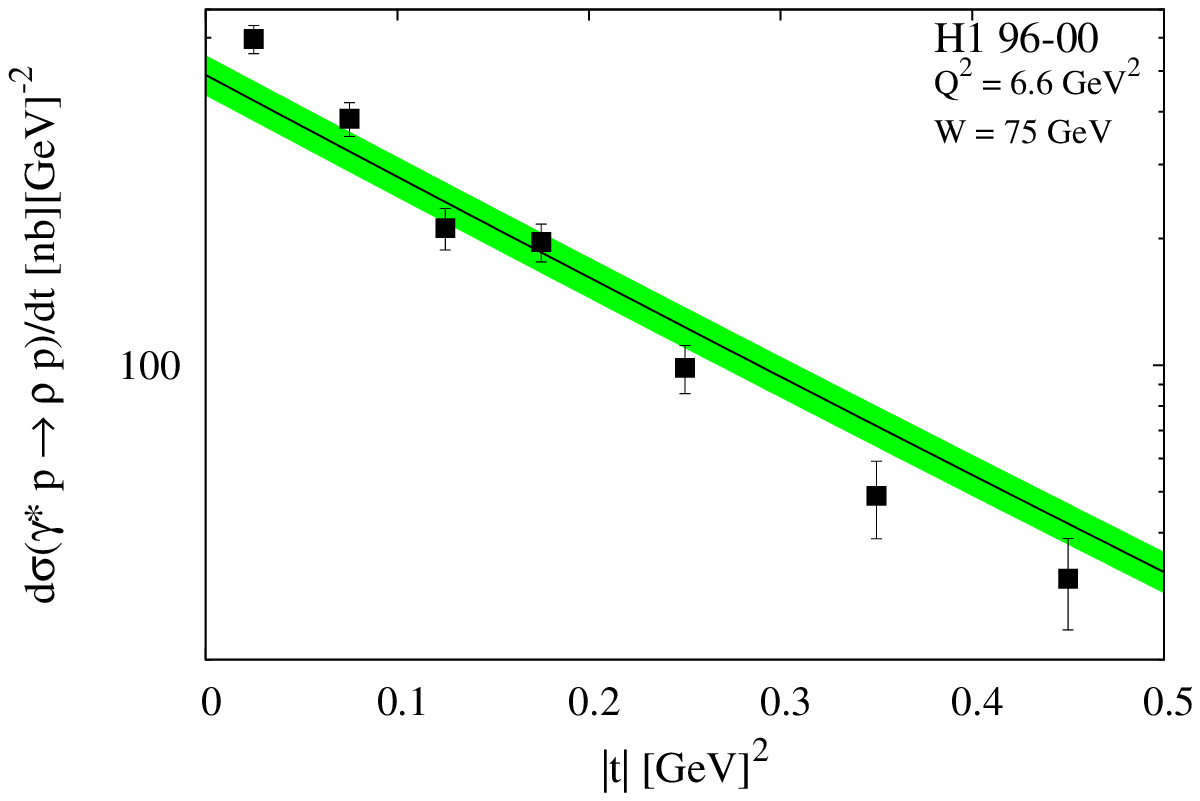}
\includegraphics[clip,scale=0.427]{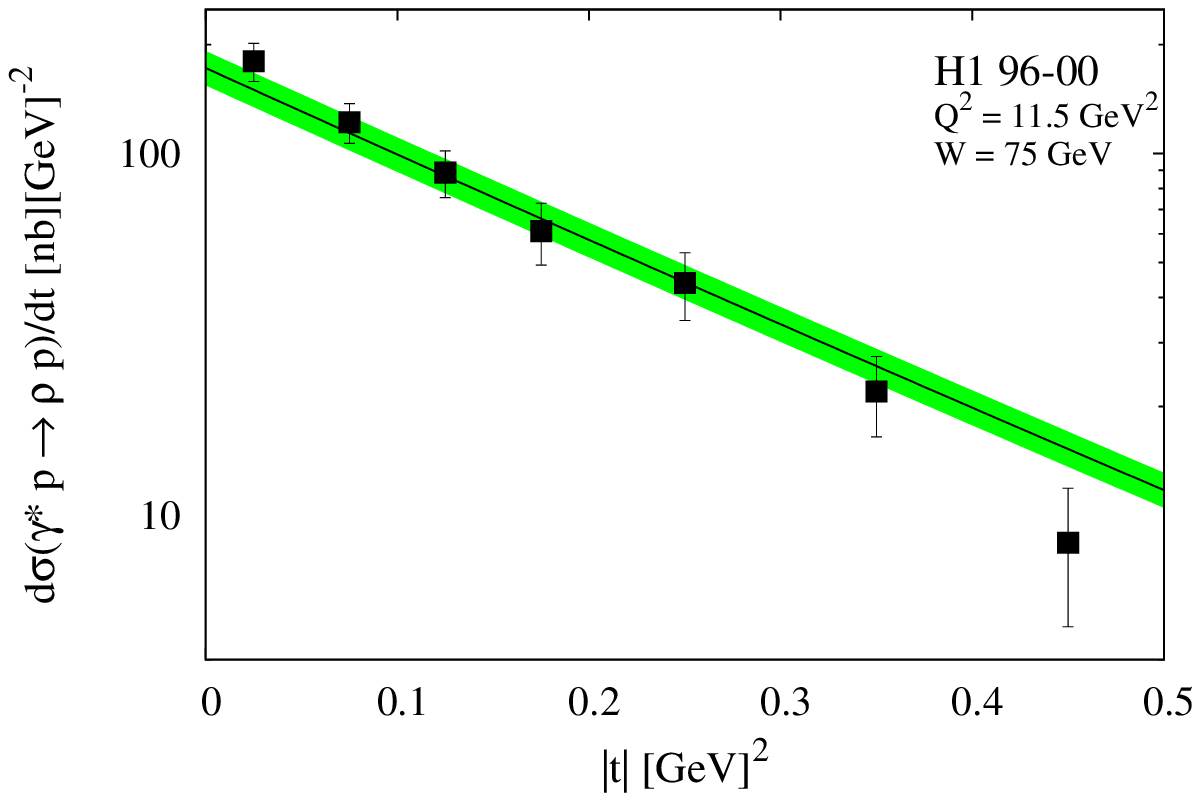}
\includegraphics[clip,scale=0.427]{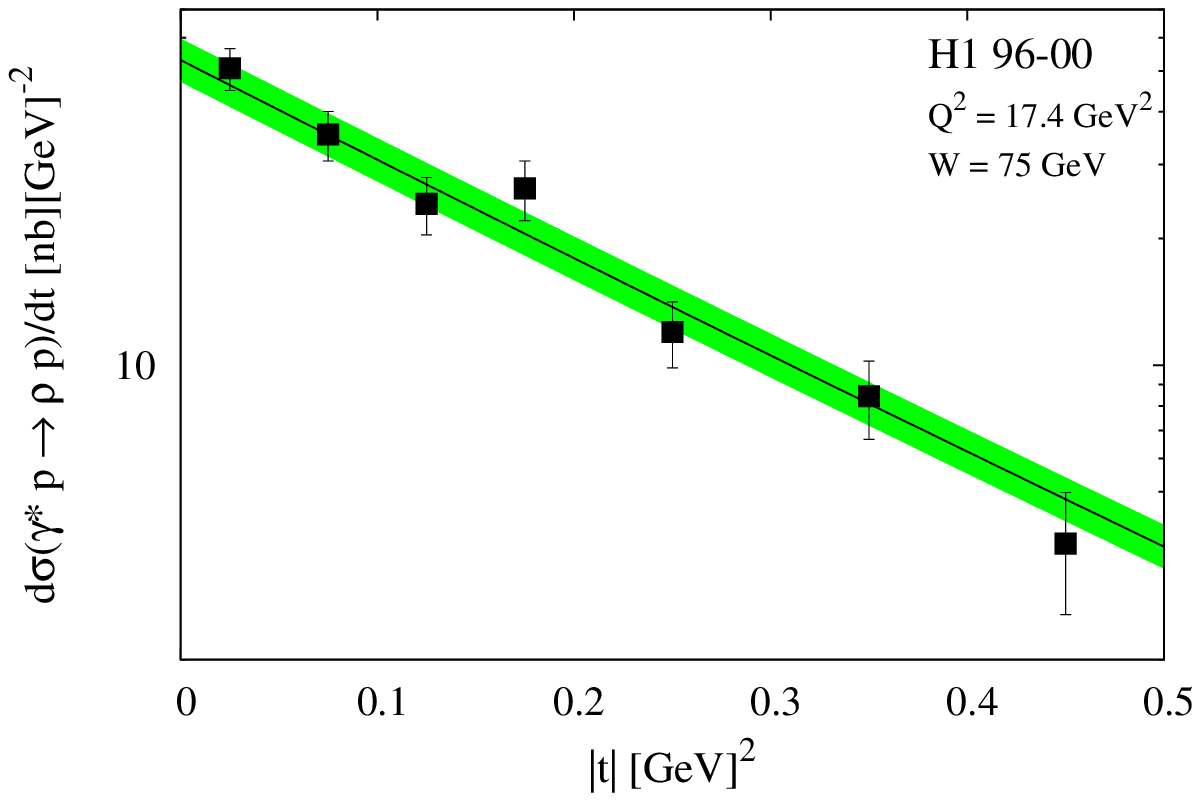}
\includegraphics[clip,scale=0.427]{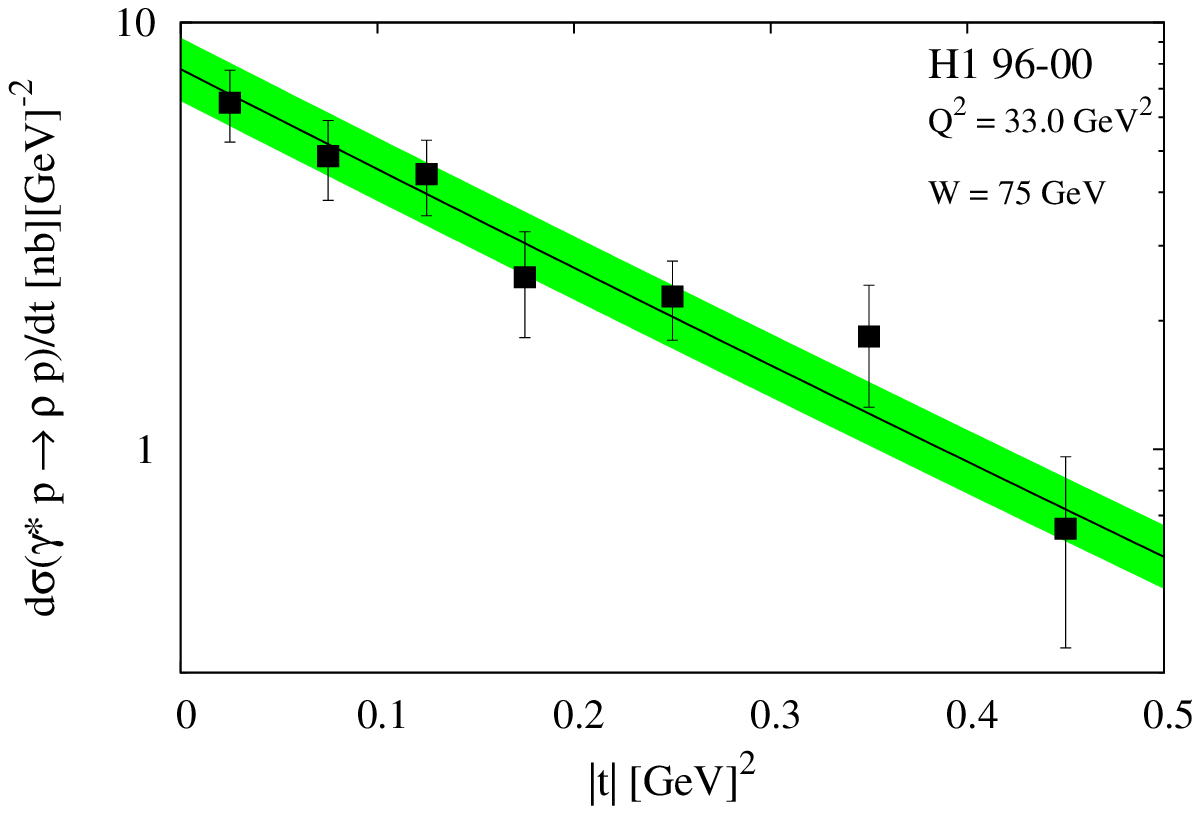}
\includegraphics[clip,scale=0.427]{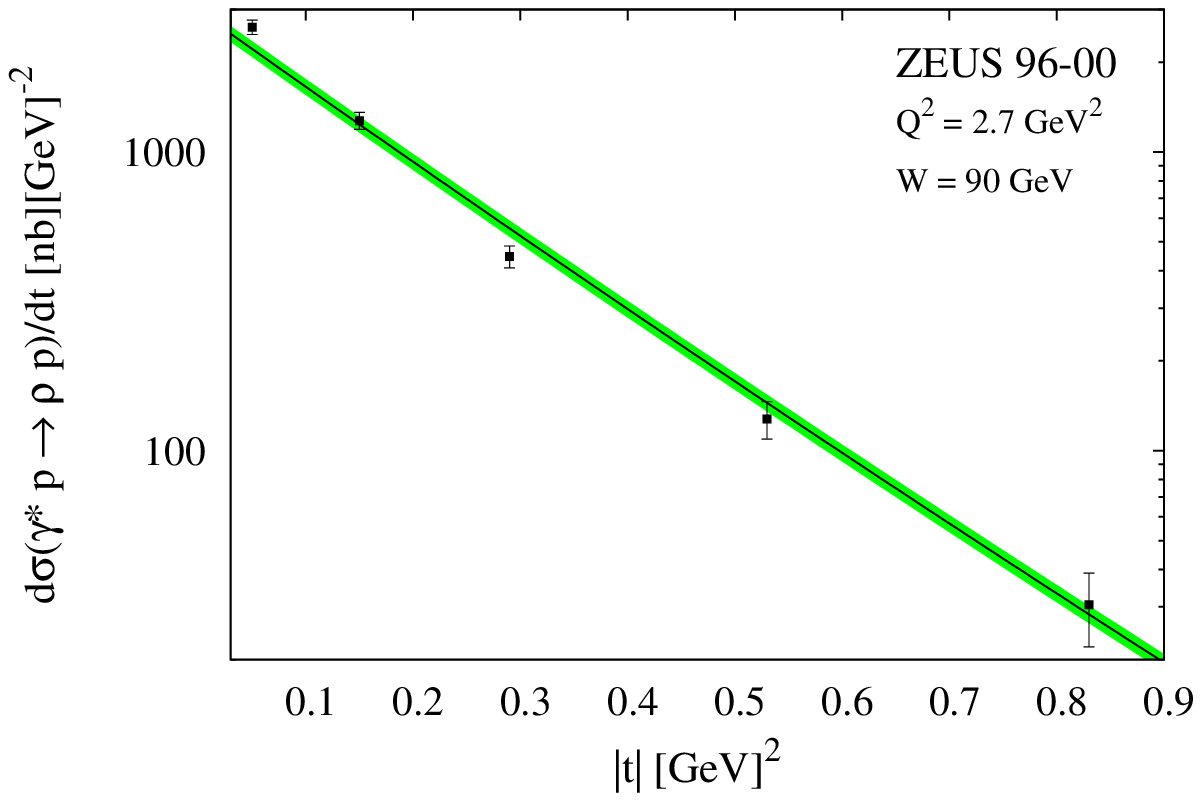}
\includegraphics[clip,scale=0.427]{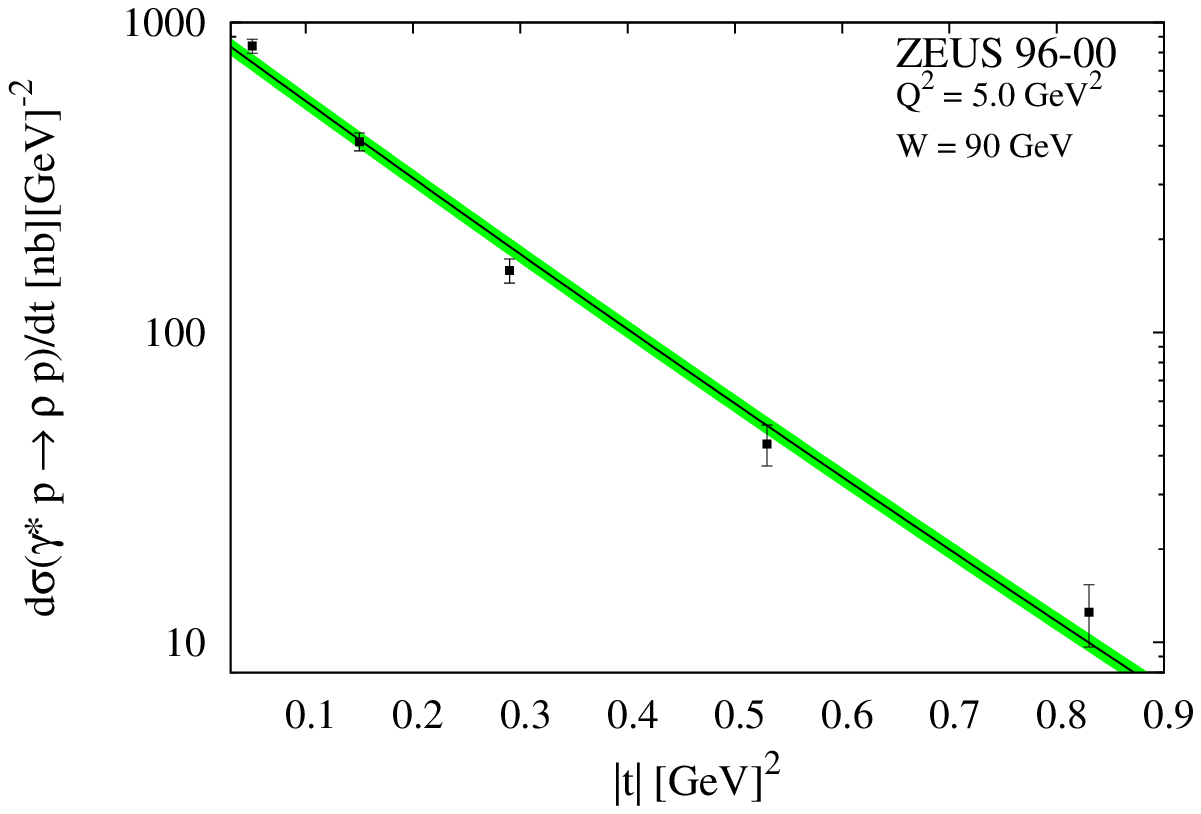}
\includegraphics[clip,scale=0.427]{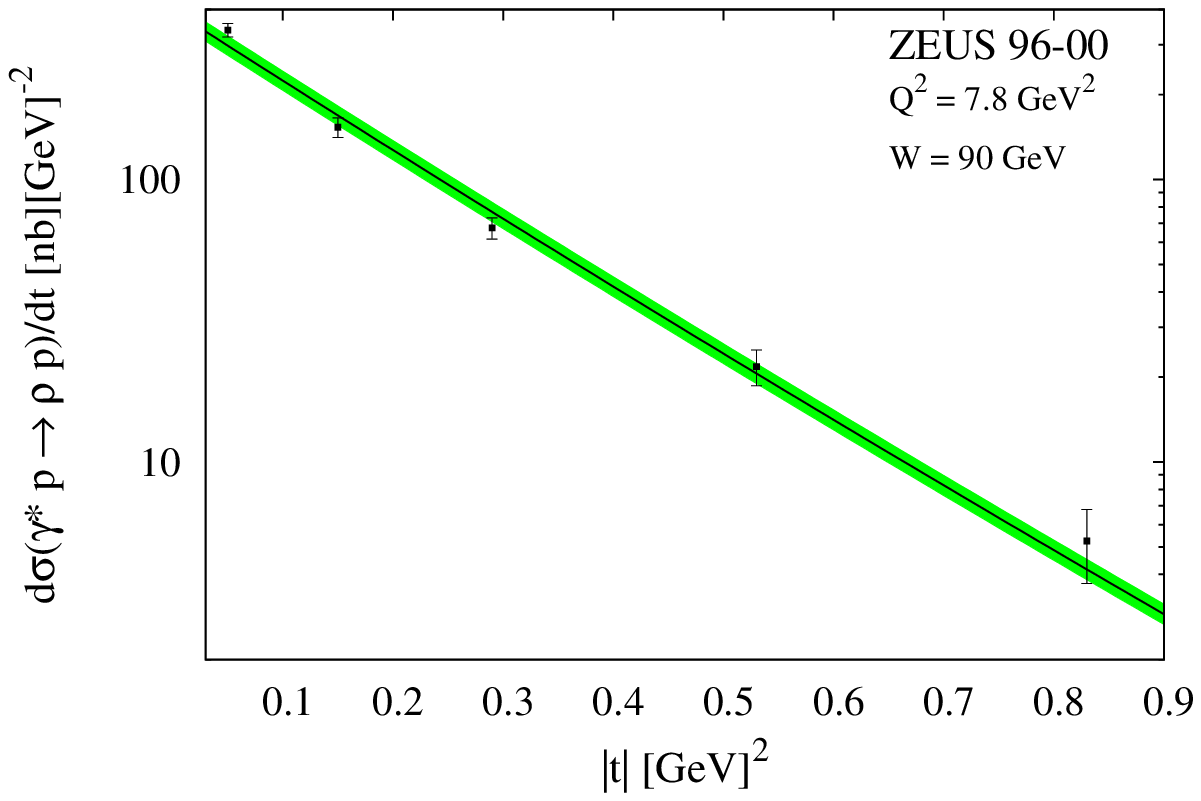}
\includegraphics[clip,scale=0.427]{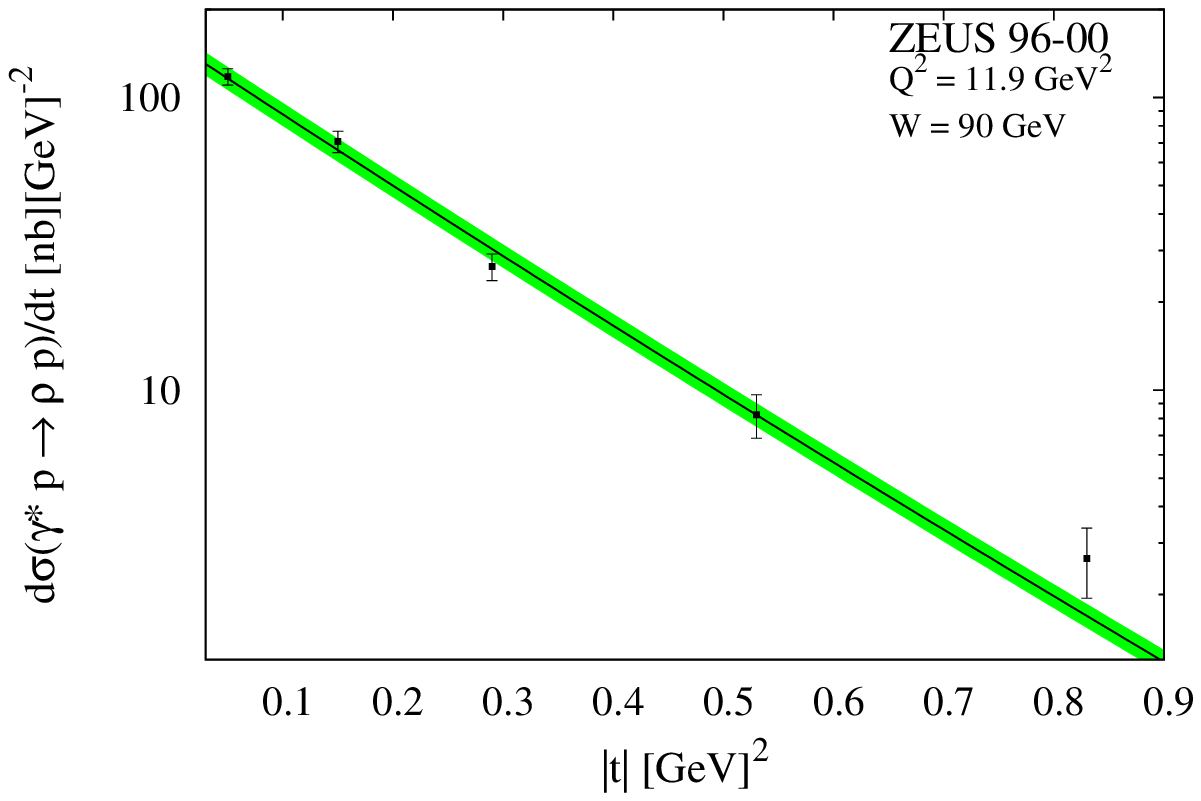}
\includegraphics[clip,scale=0.427]{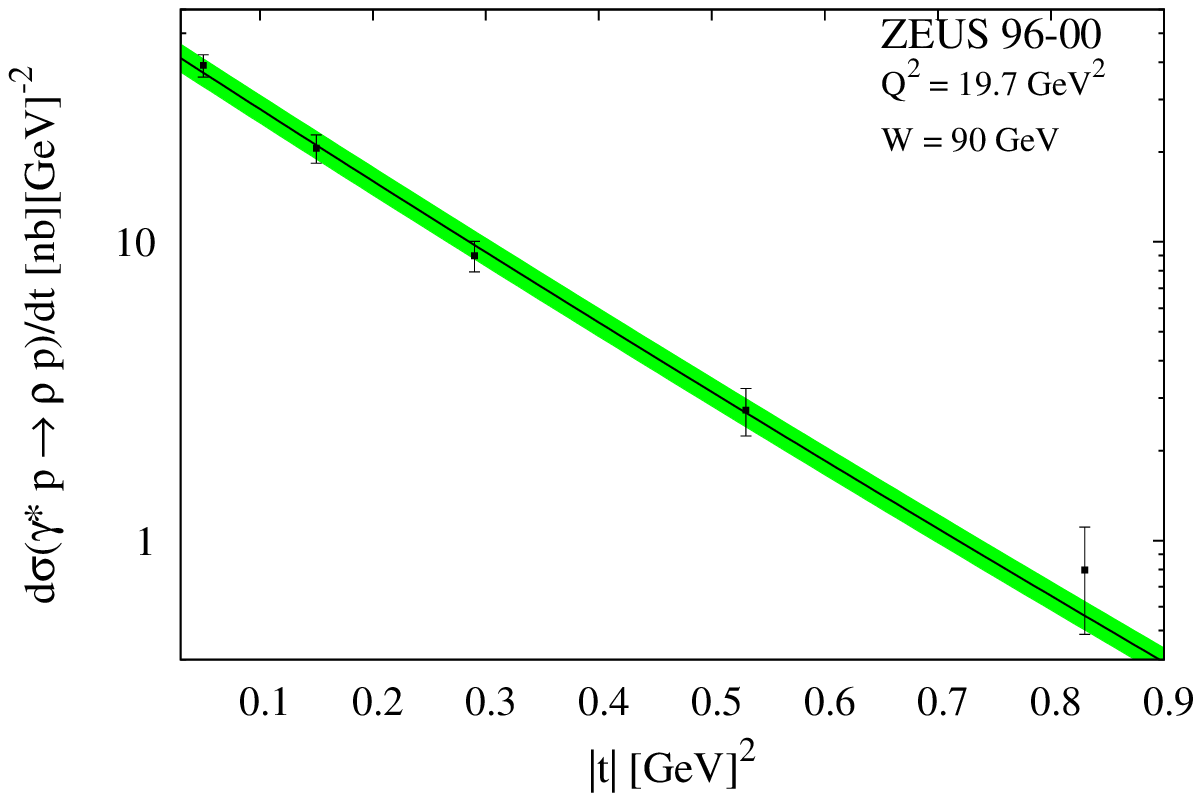}
\includegraphics[clip,scale=0.427]{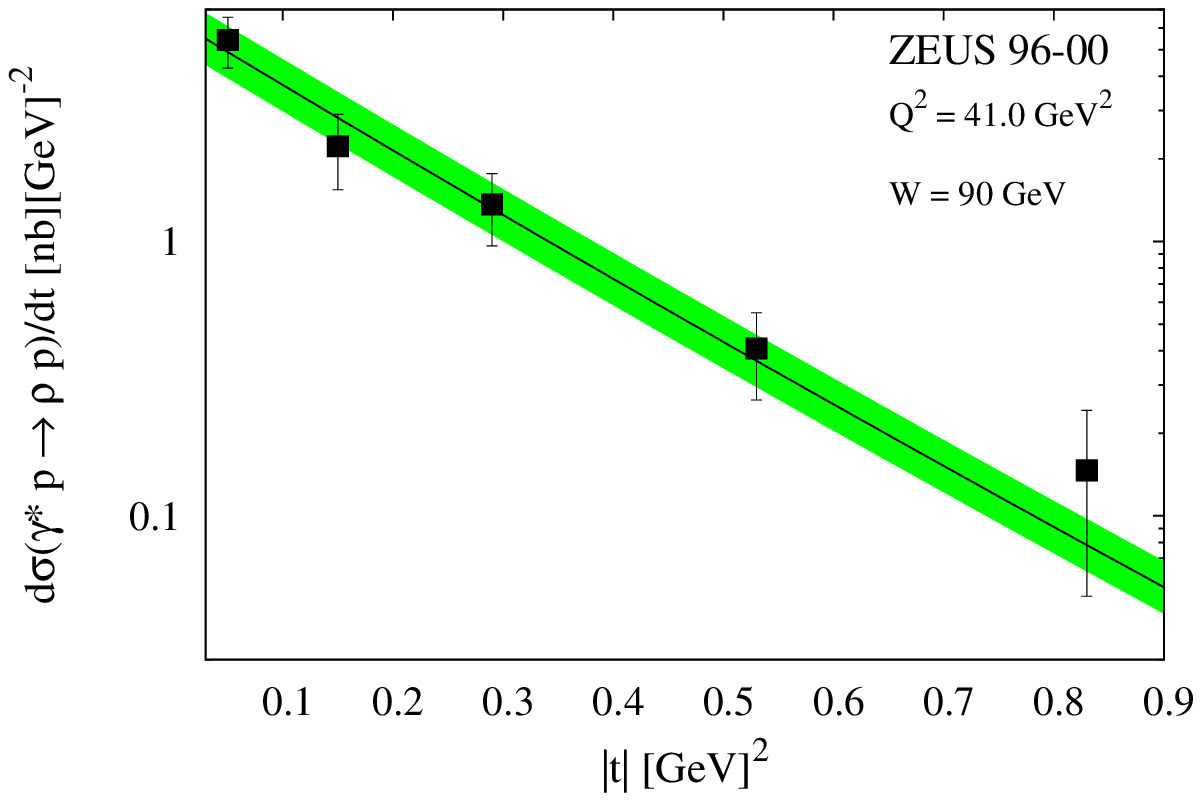}
\caption{\label{fig:rhodsdt}The behaviour of $\gamma^*p\rightarrow\rho^0 p$ differential cross section as function of $t$ is compared with data from Ref.~\cite{r1,r2} measured by the H1 and ZEUS Collaborations for several values of $Q^2$ and $W$. The green bands are calculated accordingly with the uncertainties on the free parameter $|A_0|$.}
\end{figure*}

% rho dsigma/dt ZEUS 96-00

%\begin{figure*}
%\includegraphics[clip,scale=0.427]{figure_dvcs_vmp-2011/rho/diff_b2_int/rho_diff_ZEUS_96-00/dsigmadt_ZEUS_96-00_W90_Q2_2-7_rho.eps}
%\includegraphics[clip,scale=0.427]{figure_dvcs_vmp-2011/rho/diff_b2_int/rho_diff_ZEUS_96-00/dsigmadt_ZEUS_96-00_W90_Q2_5-0_rho.eps}
%\includegraphics[clip,scale=0.427]{figure_dvcs_vmp-2011/rho/diff_b2_int/rho_diff_ZEUS_96-00/dsigmadt_ZEUS_96-00_W90_Q2_7-8_rho.eps}
%\includegraphics[clip,scale=0.427]{figure_dvcs_vmp-2011/rho/diff_b2_int/rho_diff_ZEUS_96-00/dsigmadt_ZEUS_96-00_W90_Q2_11-9_rho.eps}
%\includegraphics[clip,scale=0.427]{figure_dvcs_vmp-2011/rho/diff_b2_int/rho_diff_ZEUS_96-00/dsigmadt_ZEUS_96-00_W90_Q2_19-7_rho.eps}
%\includegraphics[clip,scale=0.427]{figure_dvcs_vmp-2011/rho/diff_b2_int/rho_diff_ZEUS_96-00/dsigmadt_ZEUS_96-00_W90_Q2_41-0_rho.eps}
%\caption{\label{fig:rhodsdt_zeus}The behaviour of $\gamma^*p\rightarrow\rho^0 p$ differential cross section as function of $t$ from Ref.~\cite{r2} measured by the ZEUS Collaboration for several values of $Q^2$ and $W$. The green bands are calculated accordingly with the uncertainties on the free parameter $|A_0|$.}
%\end{figure*}

% phi
\begin{figure*}
\includegraphics[clip,scale=0.6]{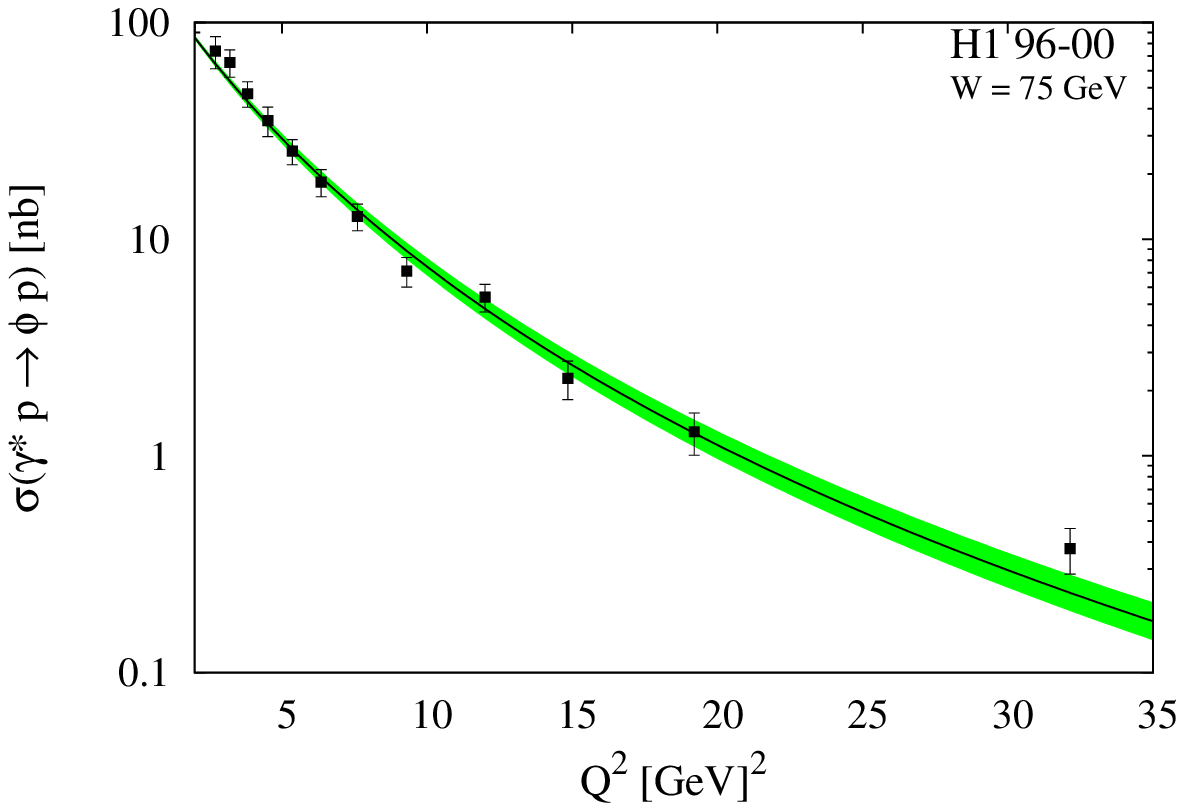}
\includegraphics[clip,scale=0.6]{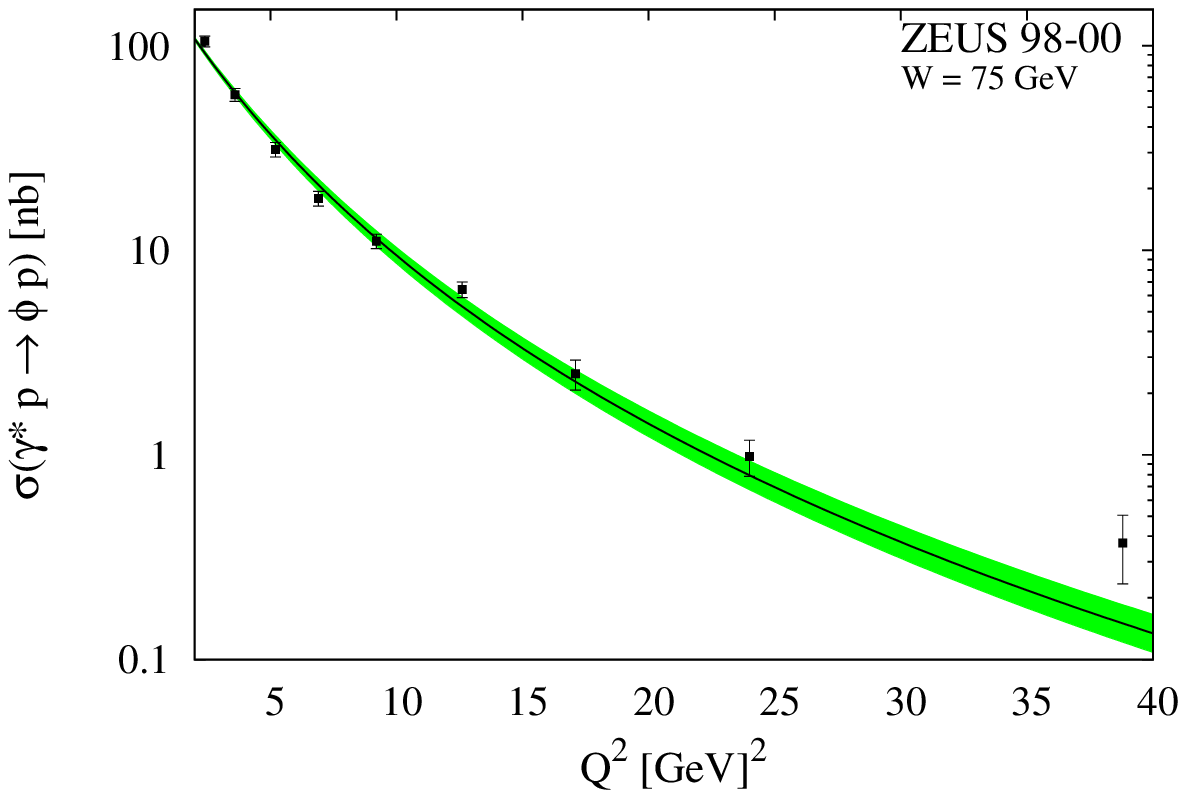}
\caption{\label{fig:phiq2}The behaviour of $\gamma^*p\rightarrow\phi p$ total cross section as function of $Q^2$ is compared to data from Refs.~\cite{r1, phi_zeus} measured by the H1 and ZEUS Collaborations for $W=$ 75 GeV. The green bands are calculated accordingly with the uncertainties on the free parameter $|A_0|$.}
\end{figure*}

% phi sigma(W) H1 96-00

\begin{figure*}
\includegraphics[clip,scale=0.45]{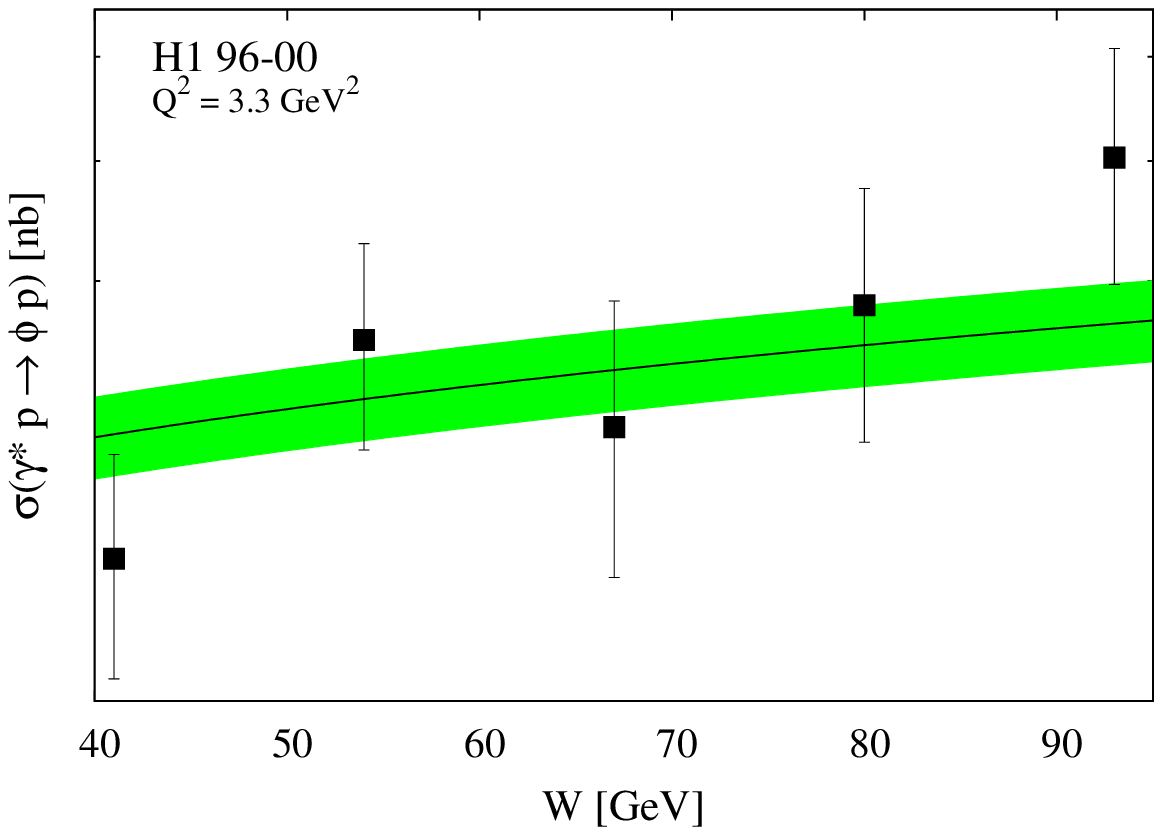}
\includegraphics[clip,scale=0.45]{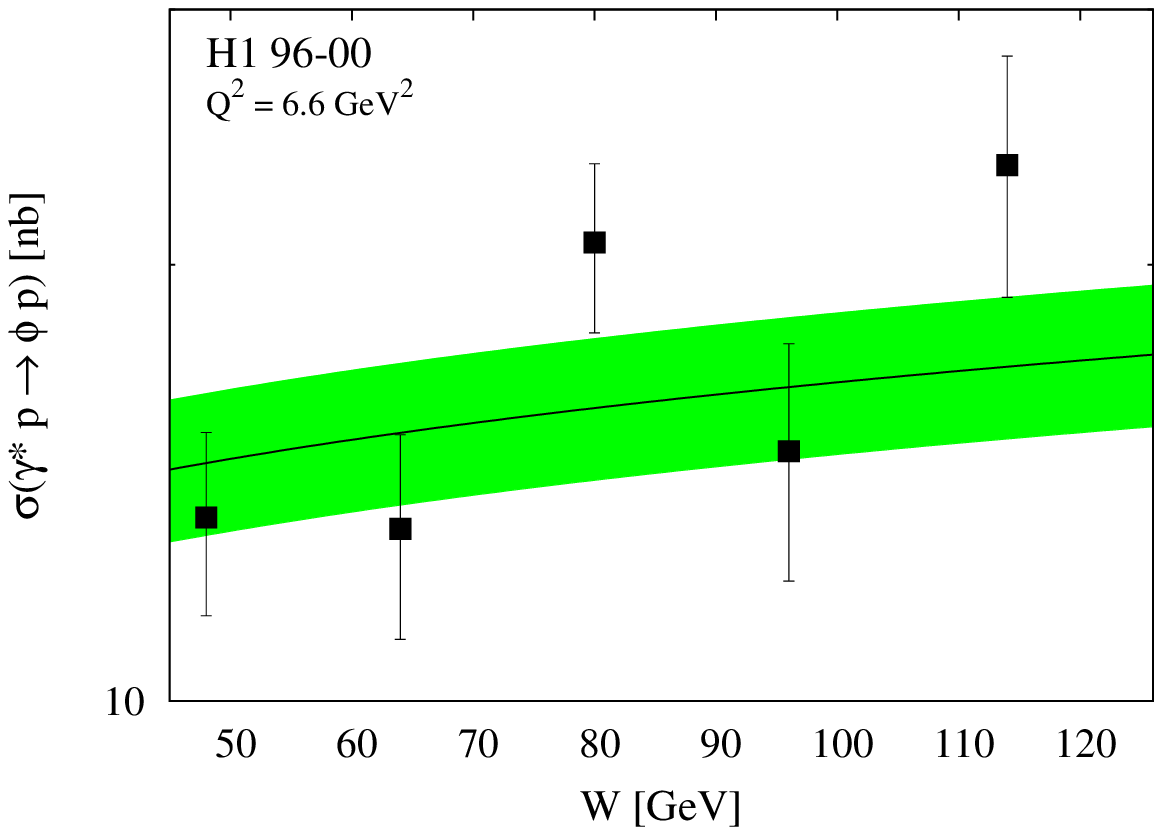}
\includegraphics[clip,scale=0.45]{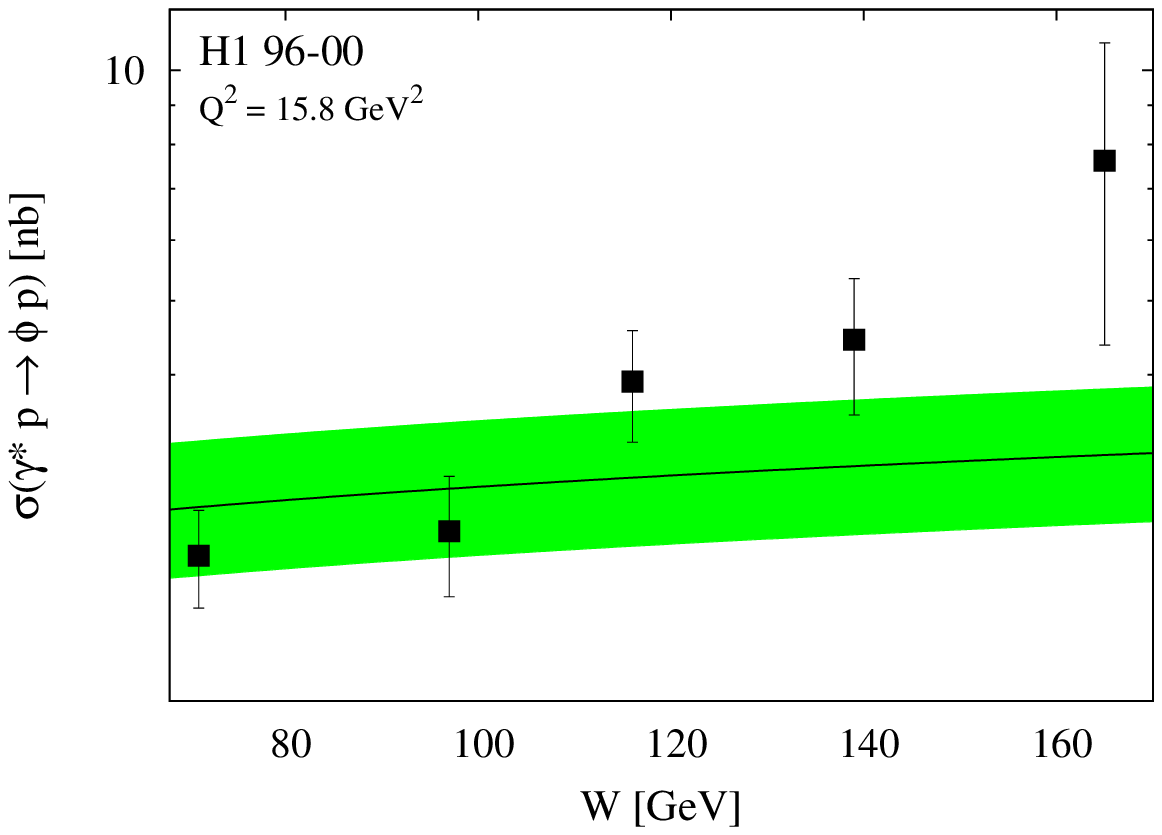}
\includegraphics[clip,scale=0.45]{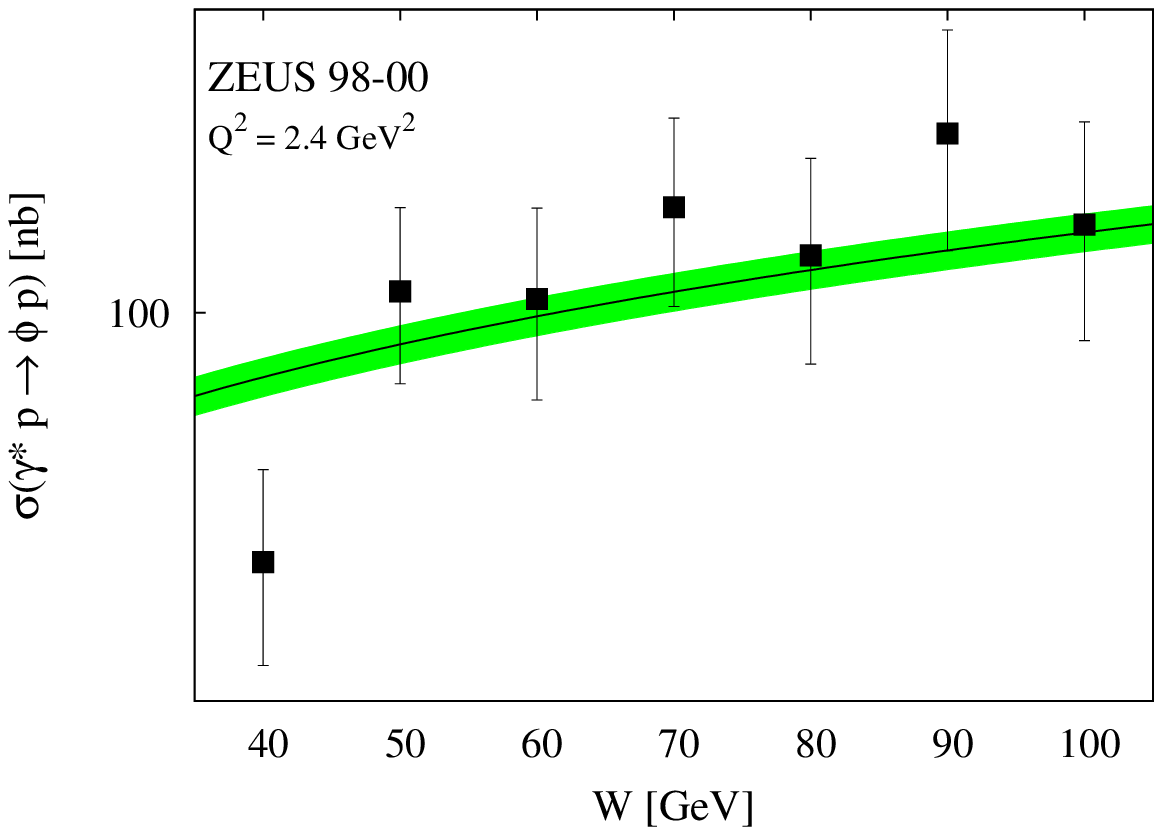}
\includegraphics[clip,scale=0.45]{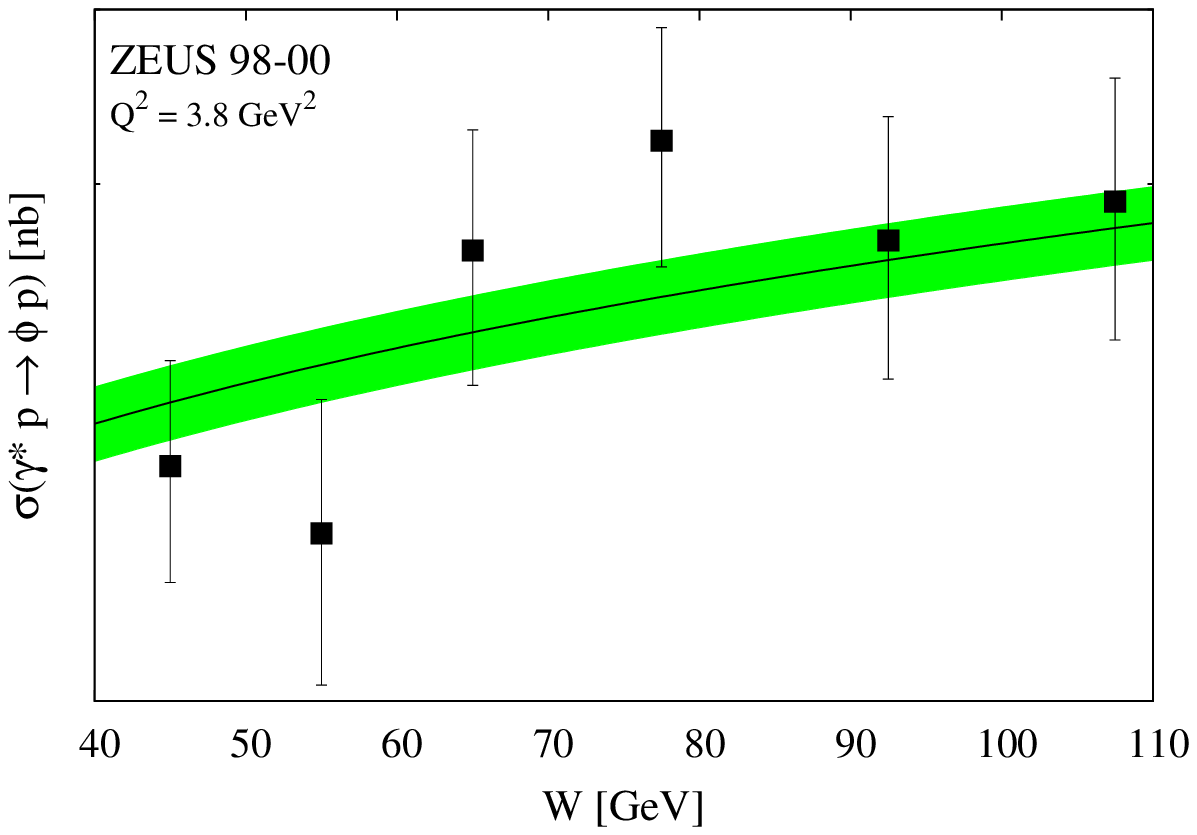}
\includegraphics[clip,scale=0.45]{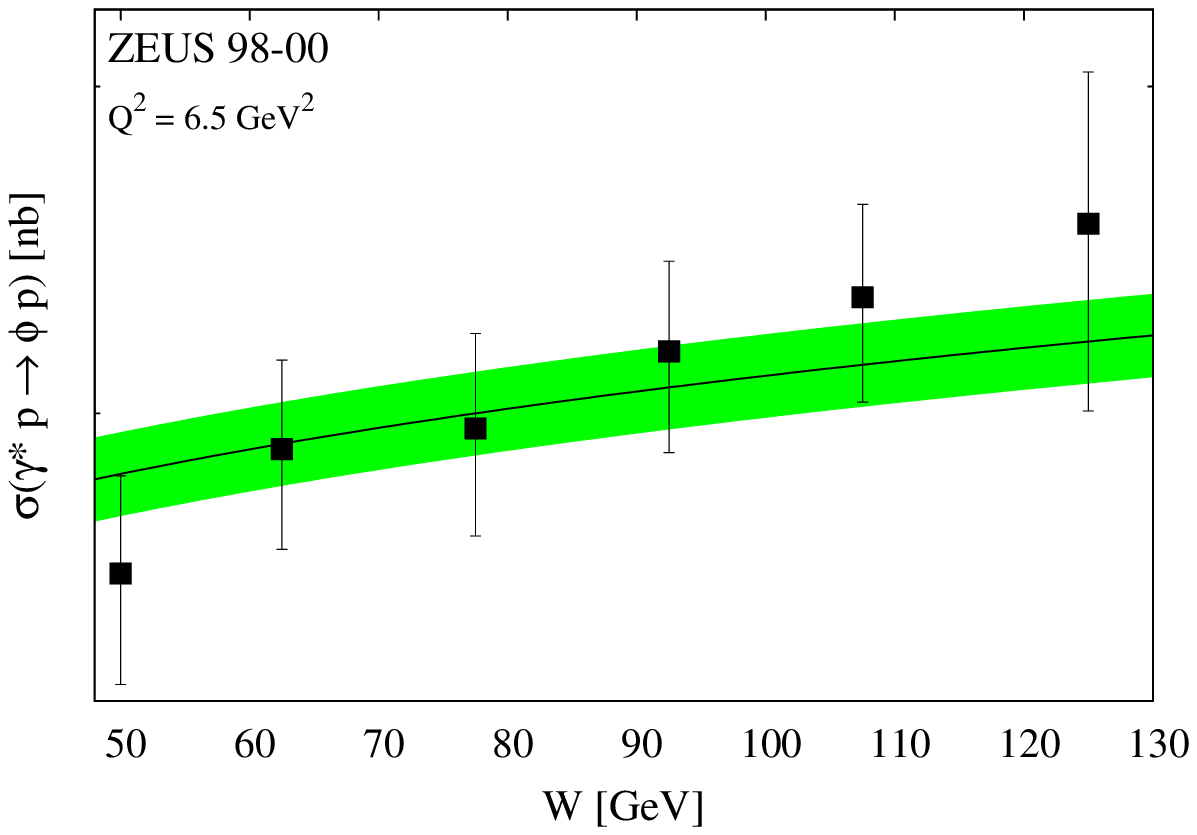}
\includegraphics[clip,scale=0.45]{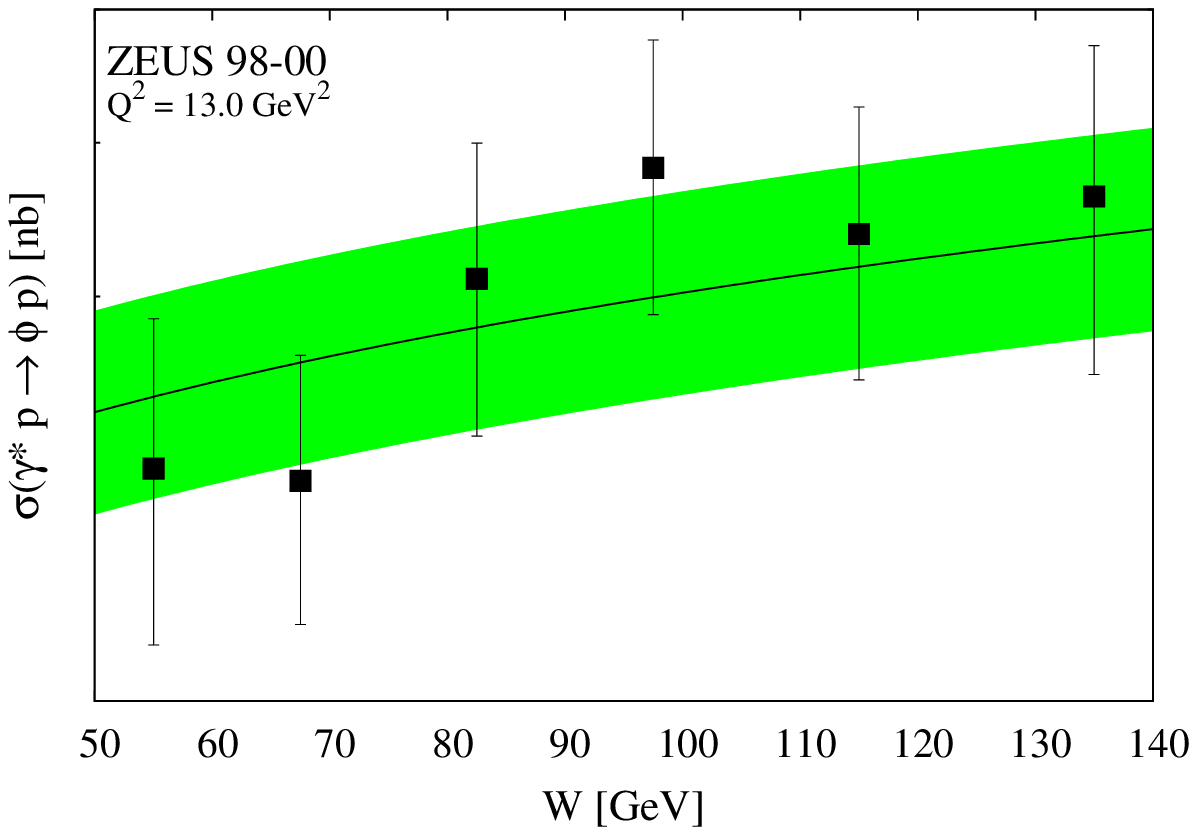}
\caption{\label{fig:phiw}The behaviour of $\gamma^*p\rightarrow\phi p$ total cross section as function of $W$ is compared to data from Ref.~\cite{r1,phi_zeus} measured by the H1 and ZEUS Collaborations for several values of $Q^2$. The green bands are calculated according with the uncertainties on the free parameter $|A_0|$.}
\end{figure*}

% phi sigma(W) ZEUS 96-00

%\begin{figure*}
%\includegraphics[clip,scale=0.45]{figure_dvcs_vmp-2011/phi/int/rho_int_w/phi_int_w_ZEUS_98-00/sigmael_ZEUS_98-00_w_Q2_2-4_phi.eps}
%\includegraphics[clip,scale=0.45]{figure_dvcs_vmp-2011/phi/int/rho_int_w/phi_int_w_ZEUS_98-00/sigmael_ZEUS_98-00_w_Q2_3-8_phi.eps}
%\includegraphics[clip,scale=0.45]{figure_dvcs_vmp-2011/phi/int/rho_int_w/phi_int_w_ZEUS_98-00/sigmael_ZEUS_98-00_w_Q2_6-5_phi.eps}
%\includegraphics[clip,scale=0.45]{figure_dvcs_vmp-2011/phi/int/rho_int_w/phi_int_w_ZEUS_98-00/sigmael_ZEUS_98-00_w_Q2_13-0_phi.eps}
%\caption{\label{fig:phiw_zeus}The behaviour of $\gamma^*p\rightarrow\phi p$ total cross section as function of $W$ is compared with data from Ref.~\cite{phi_zeus} measured by the ZEUS Collaboration for several values of $Q^2$. The green bands are calculated according with the uncertainties on the free parameter $|A_0|$.}
%\end{figure*}

% phi dsigma/dt H1 96-00
\begin{figure*}
\includegraphics[clip,scale=0.427]{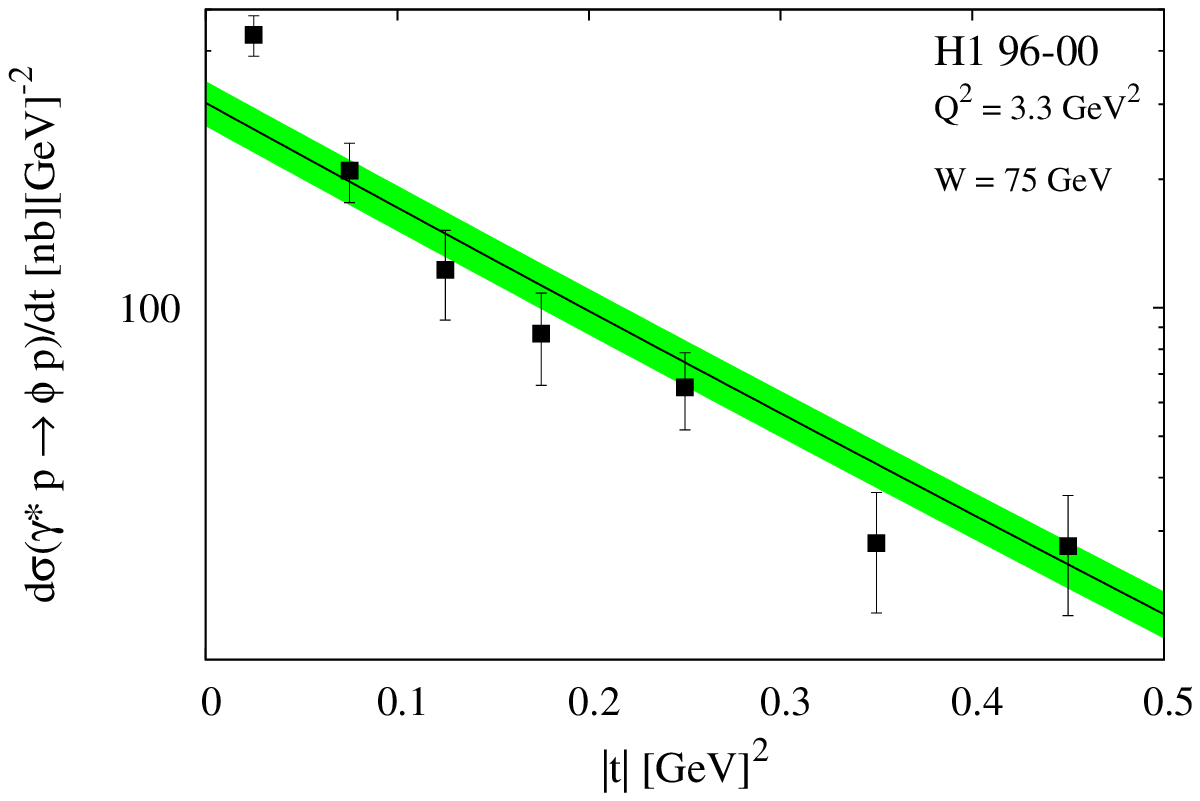}
\includegraphics[clip,scale=0.427]{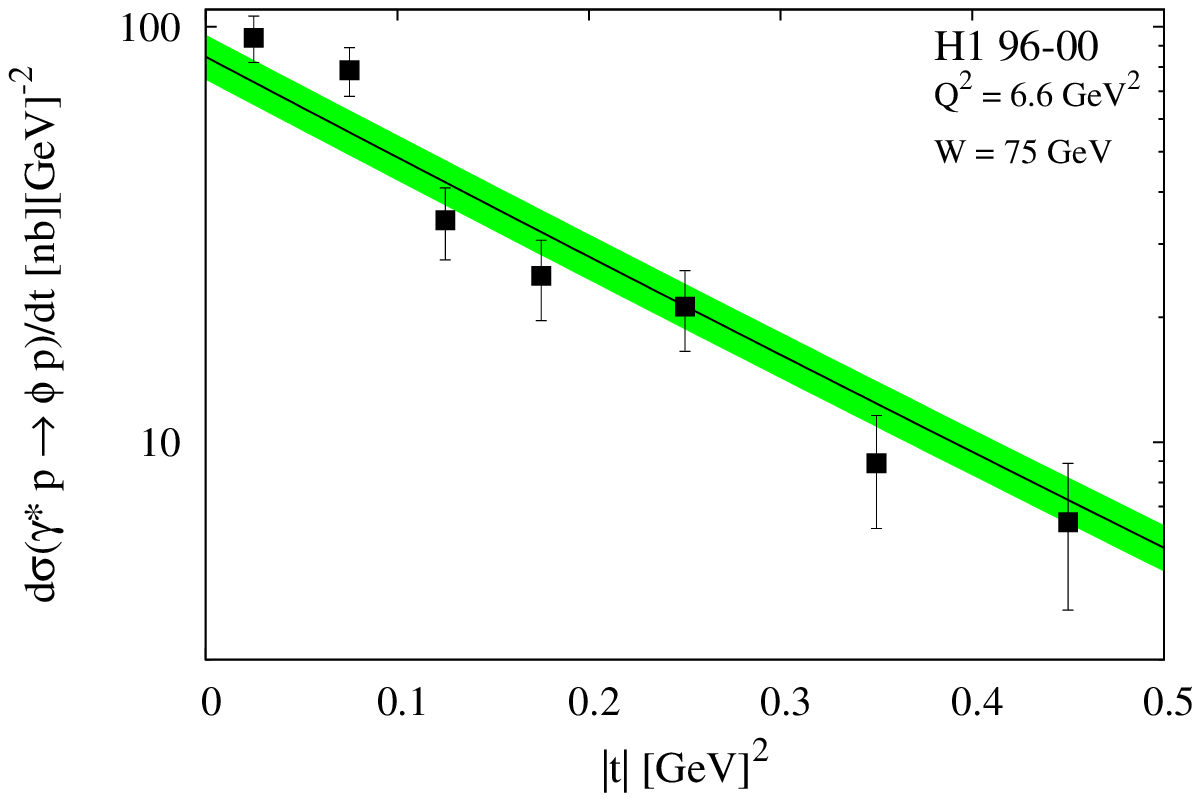}
\includegraphics[clip,scale=0.427]{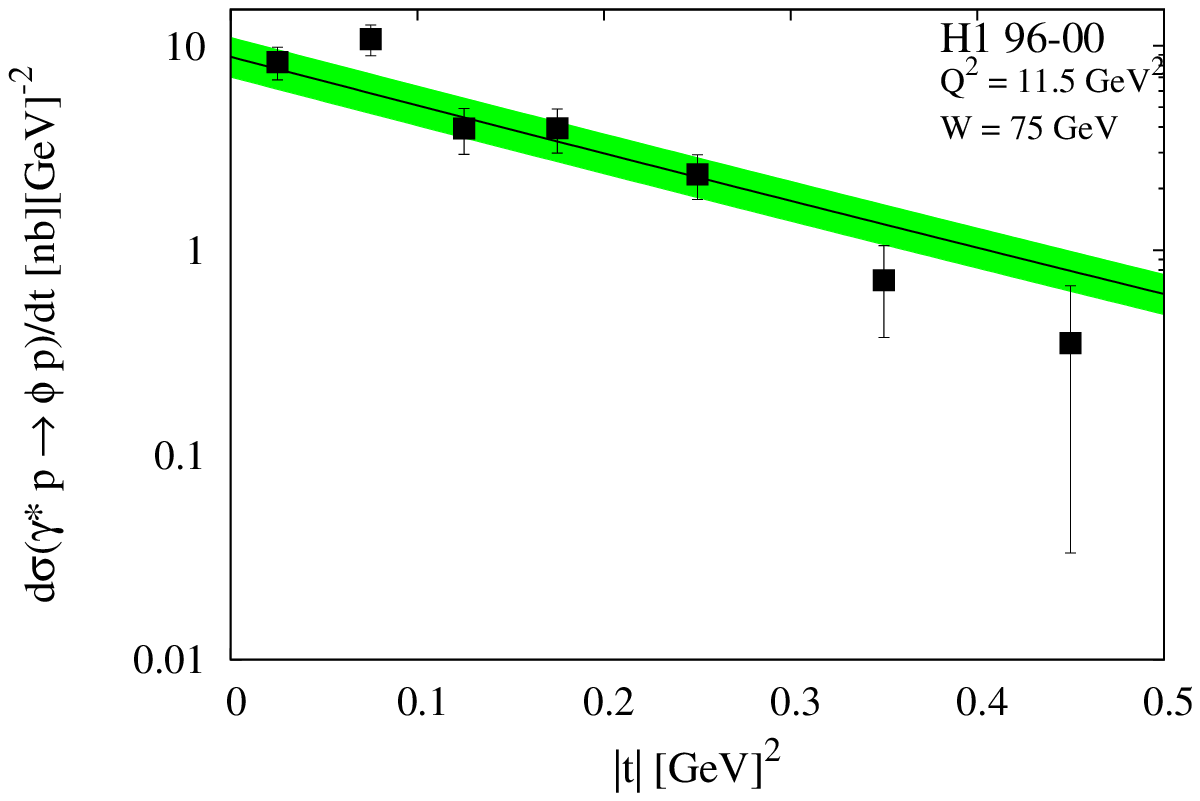}
\caption{\label{fig:phidsdt_h1}The behaviour according to our model of $\gamma^*p\rightarrow\phi p$ differential cross section as function of $t$ is compared with data from Refs.~\cite{r1} measured by the H1 Collaboration for several values of $Q^2$ and $W$. The green bands are calculated accordingly with the uncertainties on the free parameter $|A_0|$.}
\end{figure*}

% phi dsigma/dt ZEUS 98-00
\begin{figure*}
\includegraphics[clip,scale=0.6]{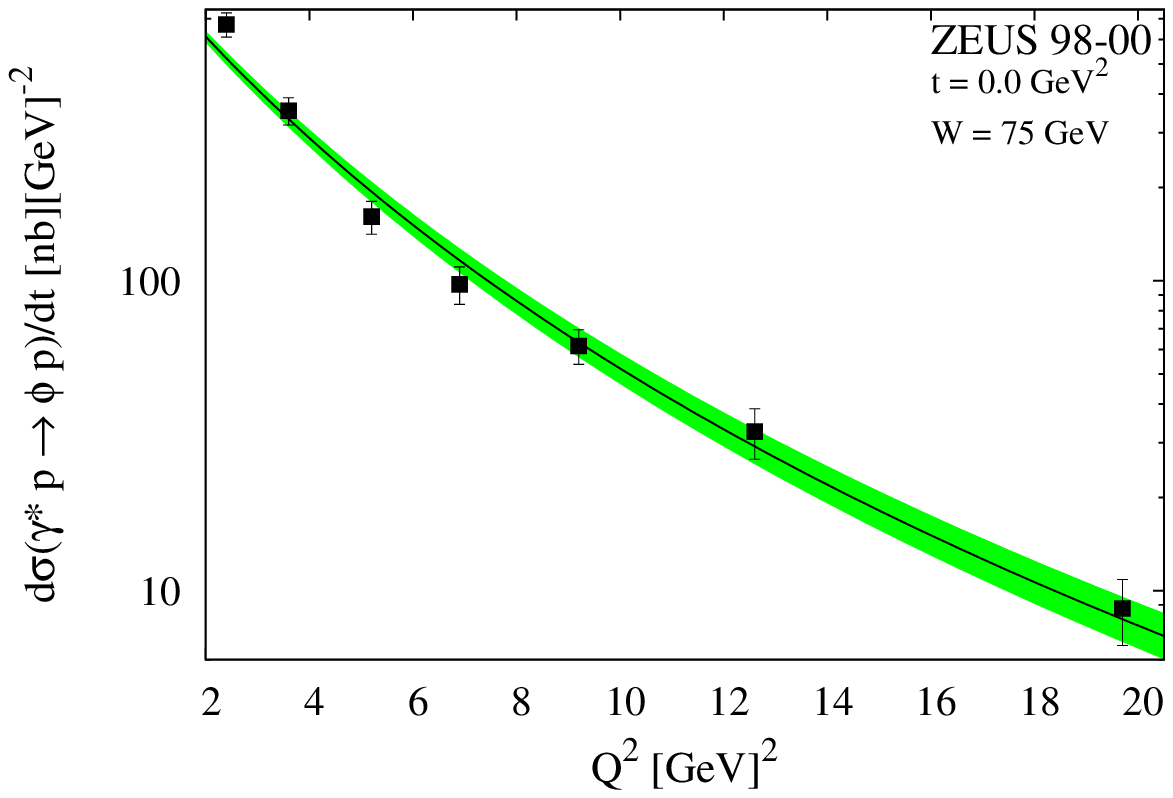}
\caption{\label{fig:phidsdt_zeus}The behaviour according to our model of $\gamma^*p\rightarrow\phi p$ differential cross section as function of Q$^2$ is compared with data from Refs.~\cite{phi_zeus} measured by the ZEUS Collaboration for fixed values $W$ and $t$. The green bands are calculated accordingly with the uncertainties on the free parameter $|A_0|$.}
\end{figure*}

% omega
\begin{figure*}
\includegraphics[clip,scale=0.6]{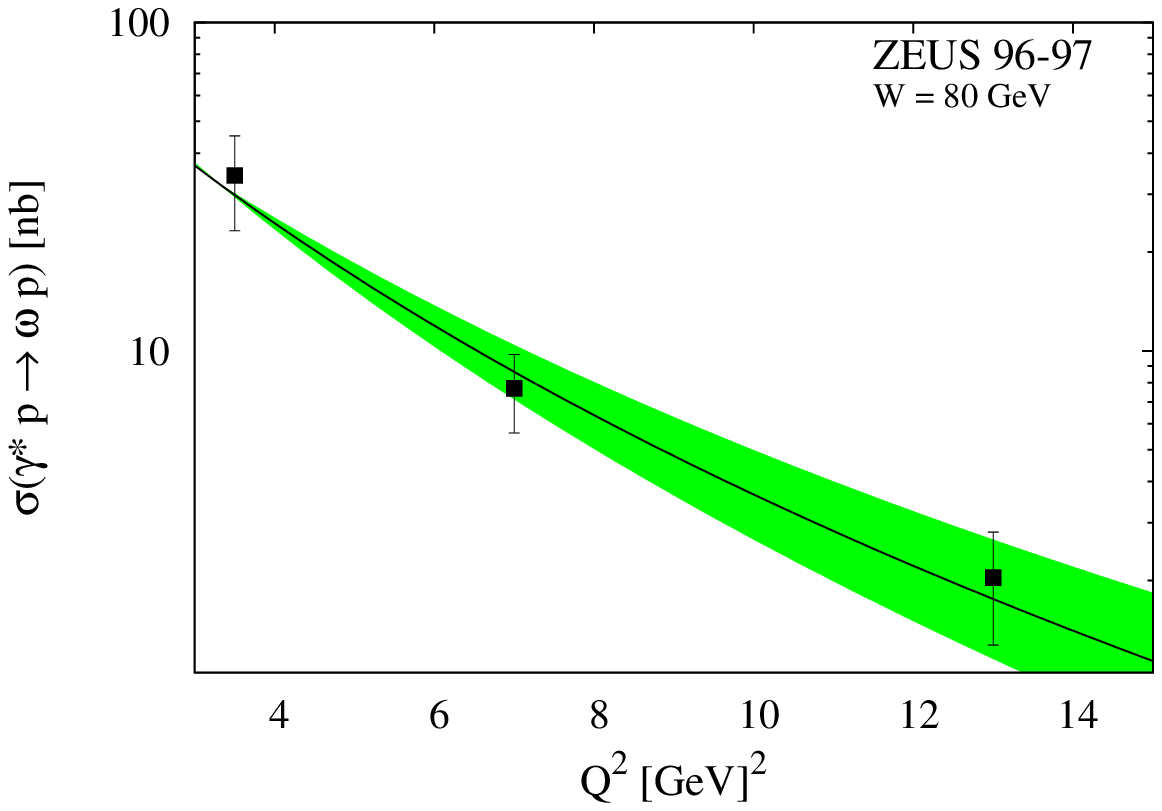}
\includegraphics[clip,scale=0.6]{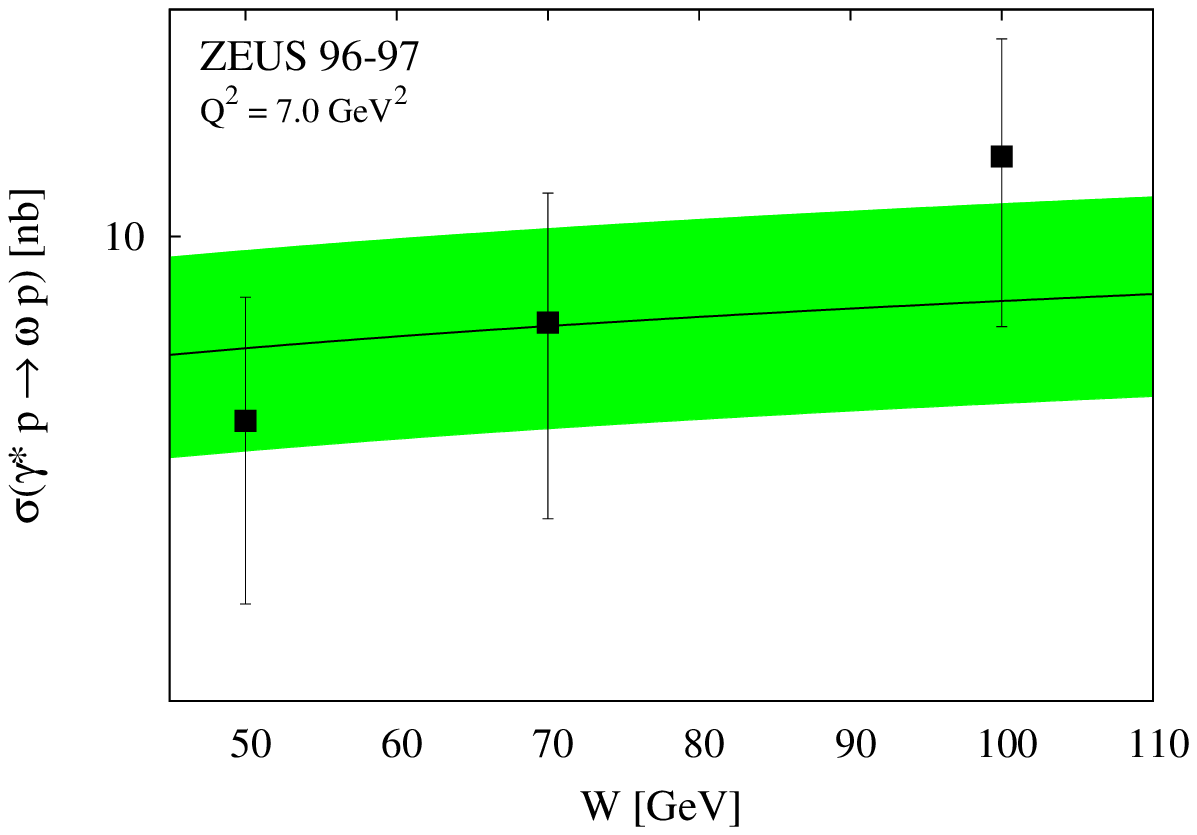}
\caption{\label{fig:omega}The behaviour according to our model of $\gamma^*p\rightarrow\omega p$ total cross section as function of Q$^2$ (left) nad $W$ (right) is compared with data from Ref.~\cite{omega_zeus} measured by the ZEUS Collaboration at fixed value of $W$ nas $Q^2$ respectively. The green bands are calculated accordingly with the uncertainties on the free parameter $|A_0|$.}
\end{figure*}

% omega sigma_tot vs W
%\begin{figure*}
%\includegraphics[clip,scale=0.6]{figure_dvcs_vmp-2011/omega/int/omega_int_w/sigmaint_ZEUS_96-97_w_Q2=7-0_omega.eps}
%\caption{\label{fig:omegaw}The behaviour according to our model of $\gamma^*p\rightarrow\omega p$ total cross section as function of W is compared with data from Ref.~\cite{omega_zeus} measured by the ZEUS Collaboration for fixed value of $Q^2$. The green bands are calculated accordingly with the uncertainties on the free parameter $|A_0|$.}
%\end{figure*}

% j/psi
\begin{figure*}
\includegraphics[clip,scale=0.5]{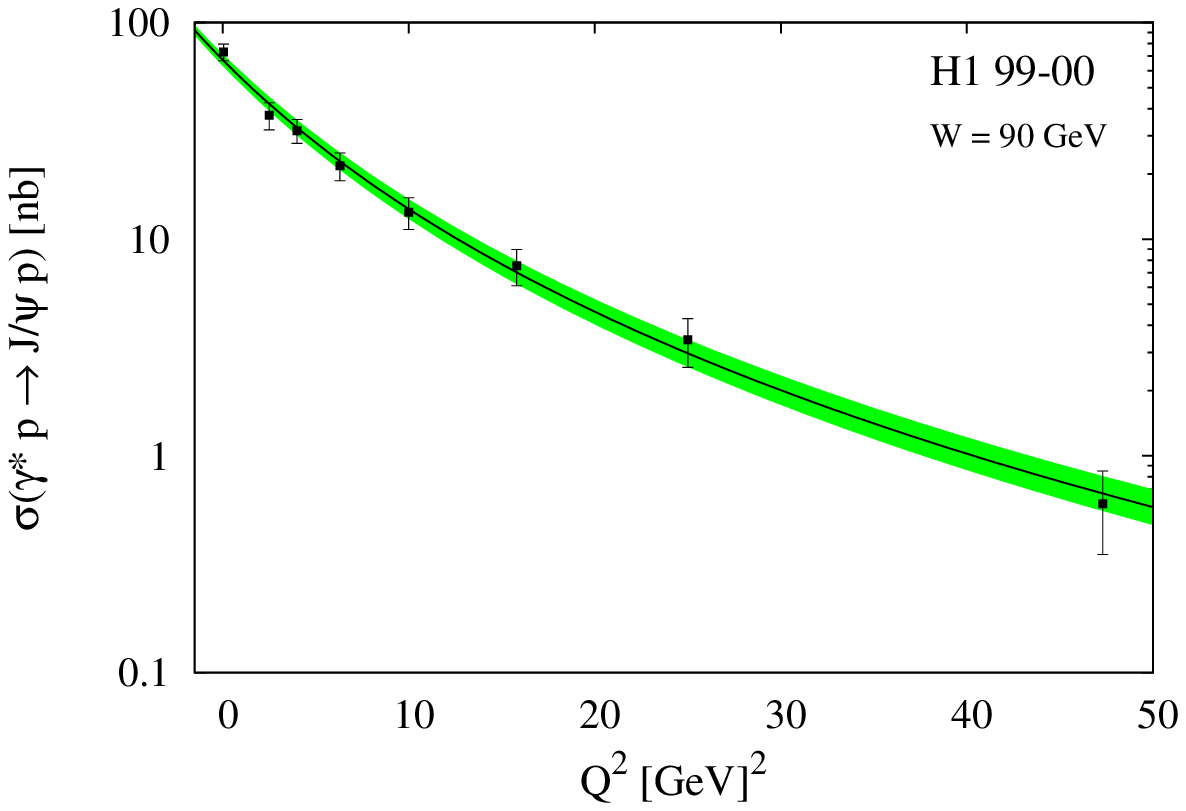}
\includegraphics[clip,scale=0.5]{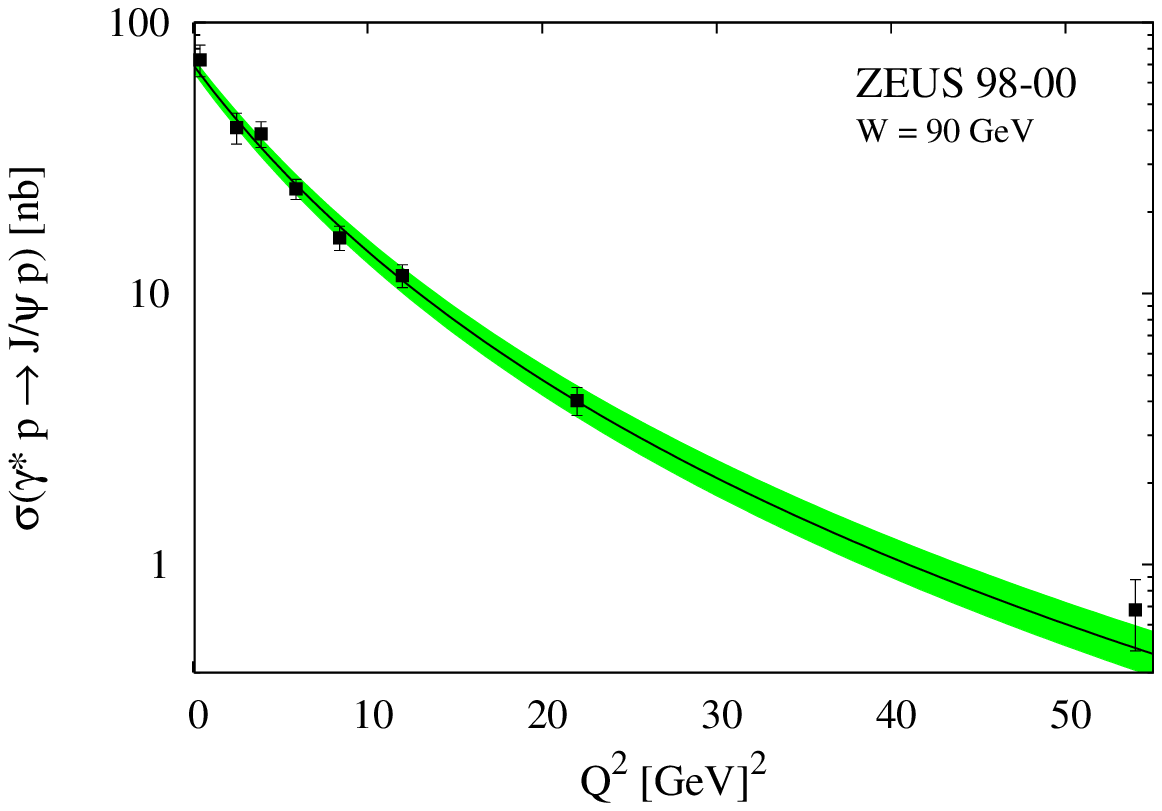}
\caption{\label{fig:jpsiq2}The behaviour according to our model of $\gamma^*p\rightarrow J/\psi p$ total cross section as function of $Q^2$ is compared with data from Refs.~\cite{j1,j2} measured by the H1 and ZEUS Collaborations for $W=90$ GeV. The green bands are calculated accordingly with the uncertainties on the free parameter $|A_0|$.}
\end{figure*}

% j/psi sigma(W) H1 96-00
\begin{figure*}
\includegraphics[clip,scale=0.427]{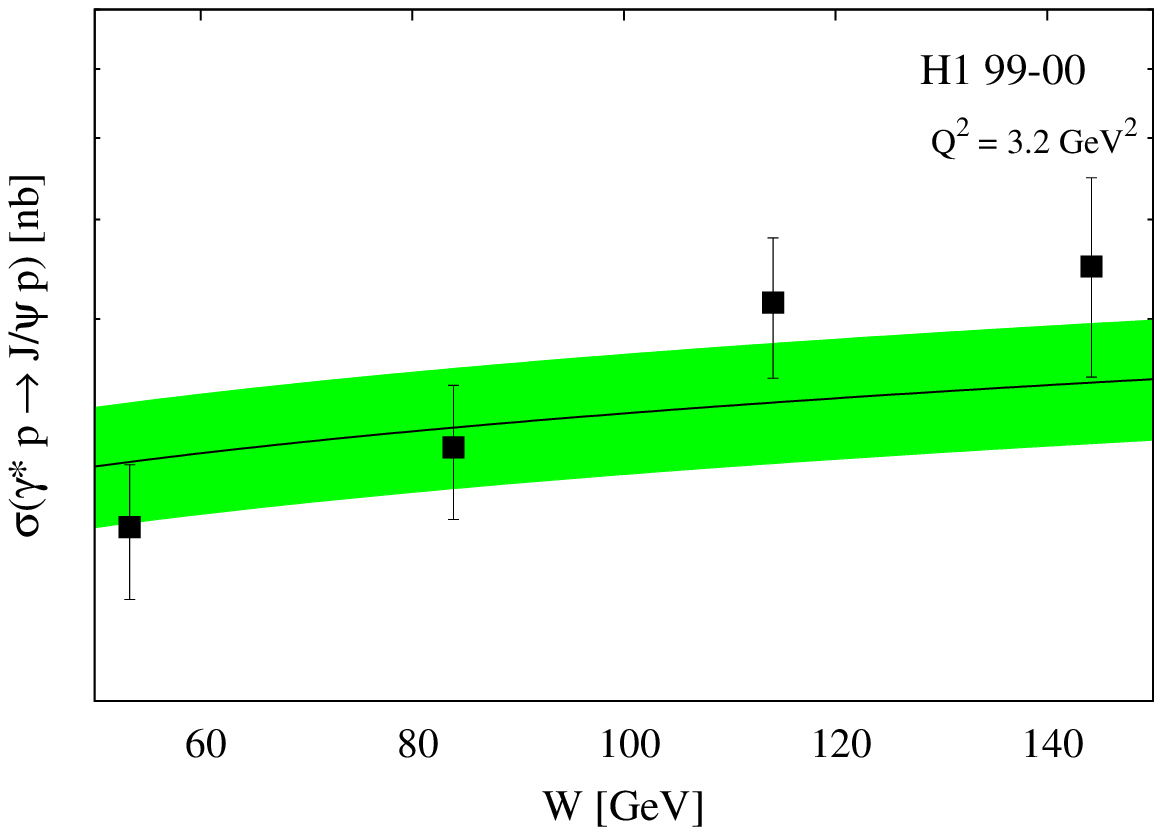}
\includegraphics[clip,scale=0.427]{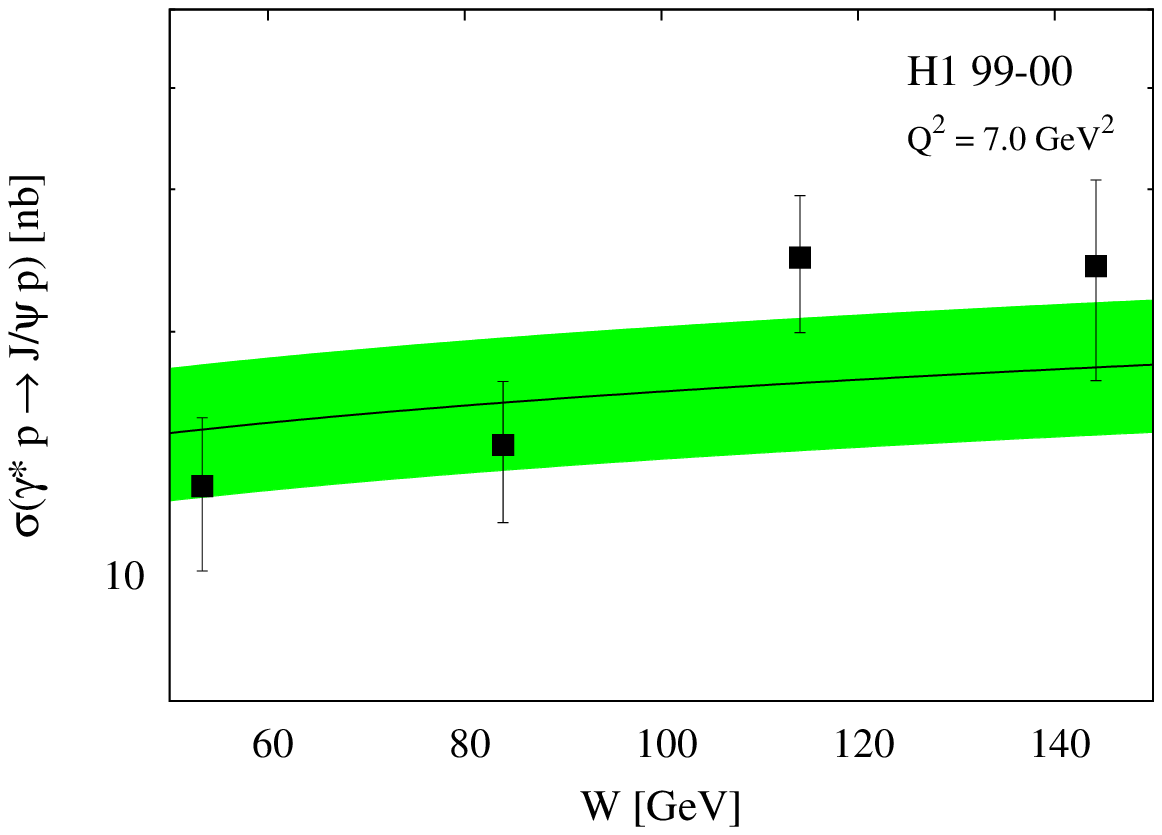}
\includegraphics[clip,scale=0.427]{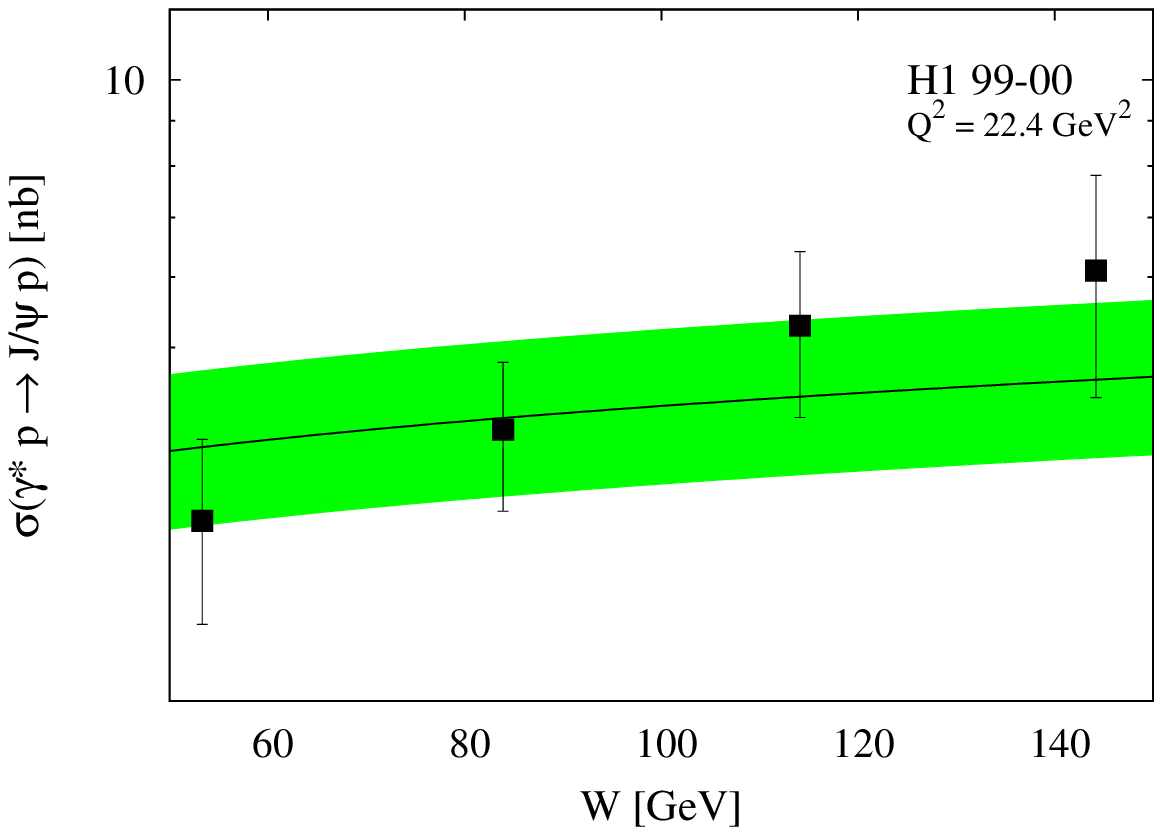}
\includegraphics[clip,scale=0.427]{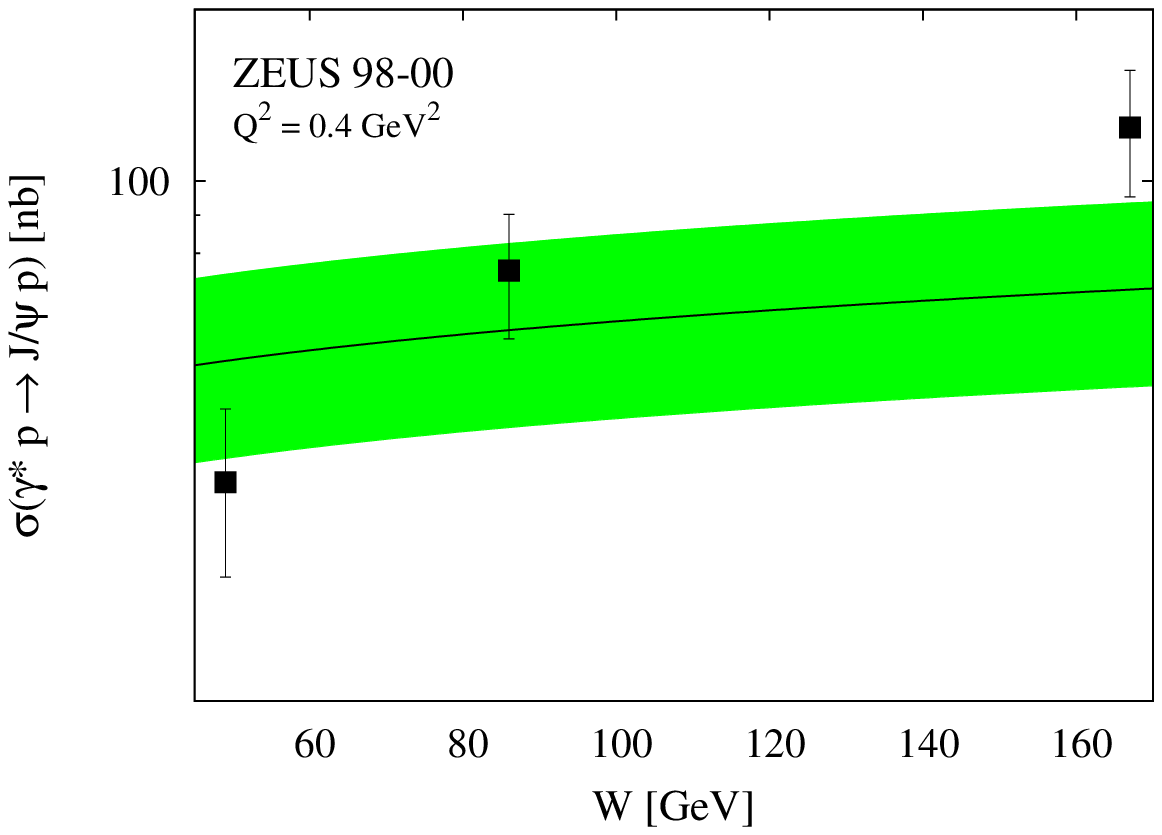}
\includegraphics[clip,scale=0.427]{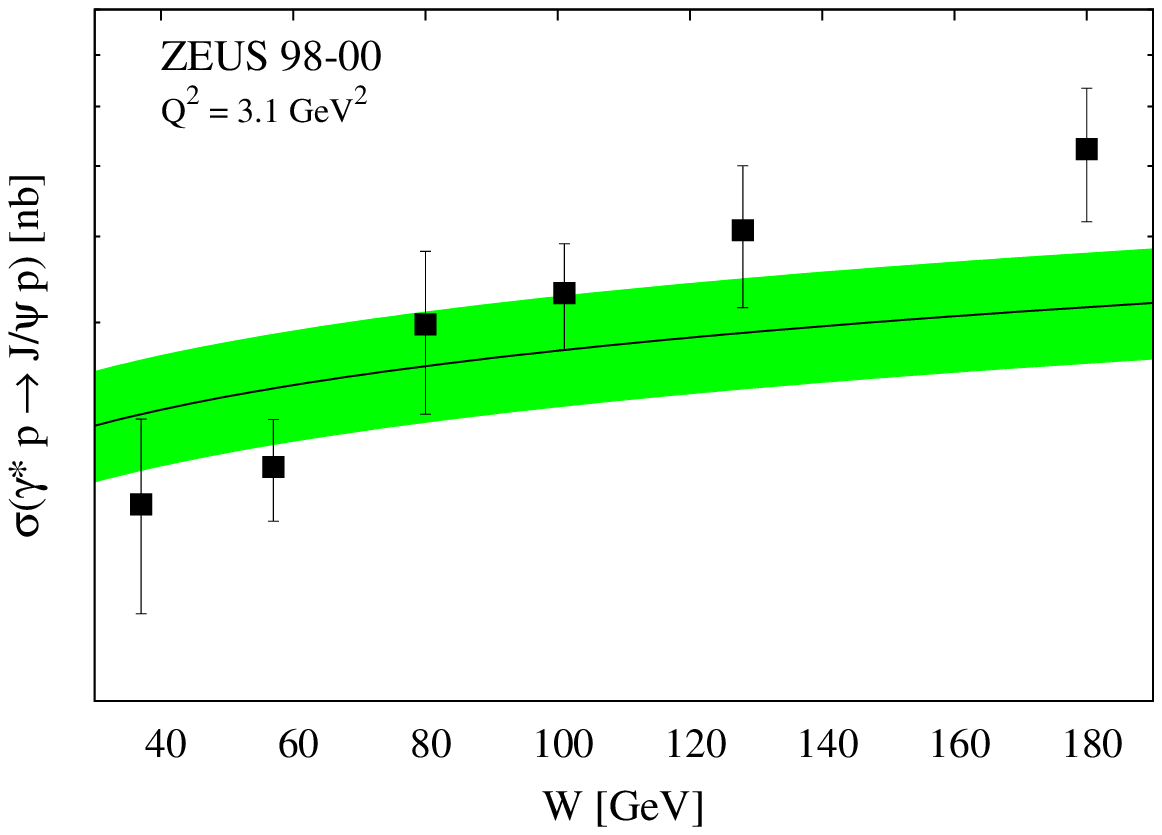}
\includegraphics[clip,scale=0.427]{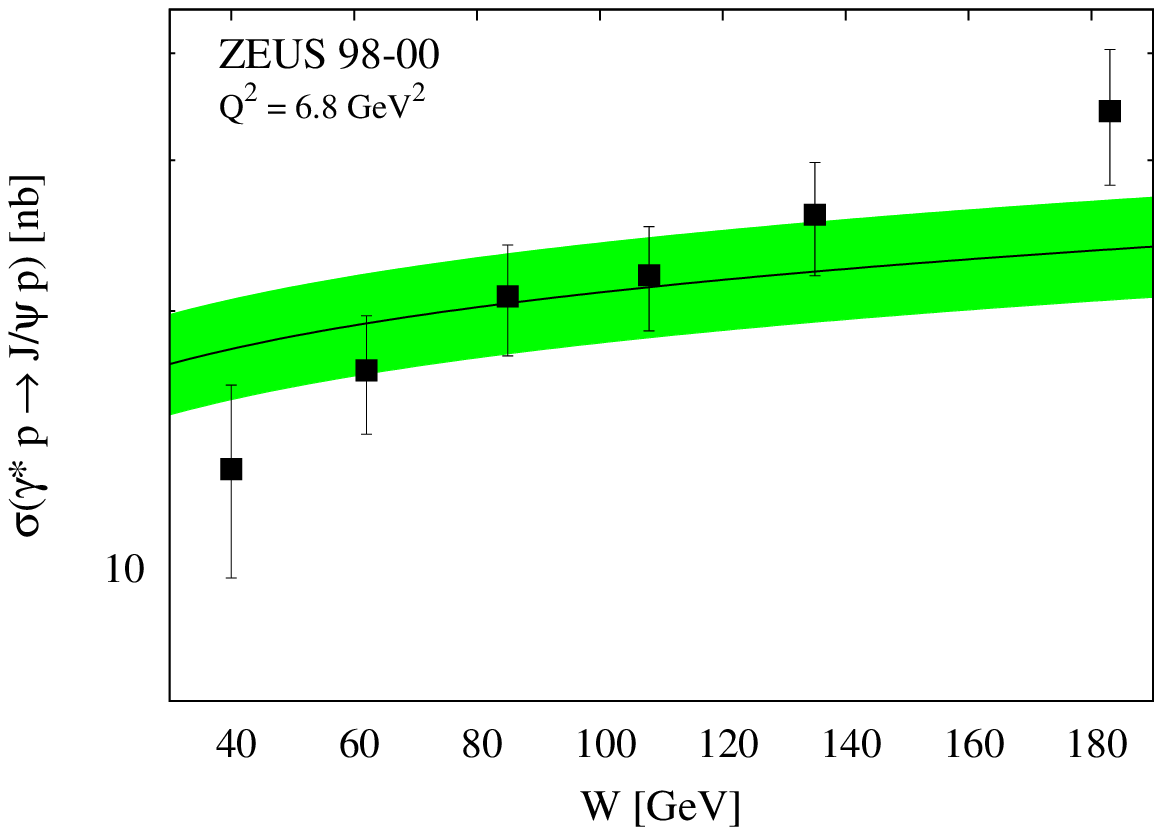}
\includegraphics[clip,scale=0.427]{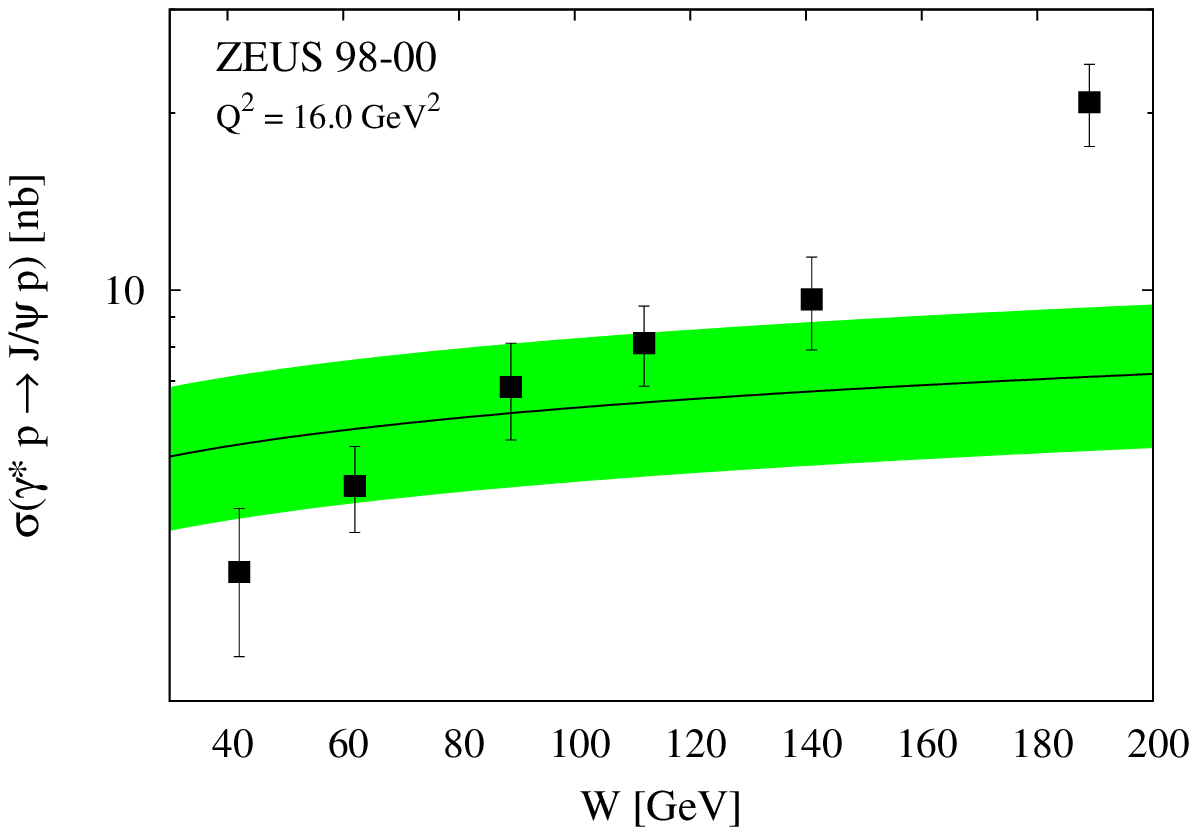}
\caption{\label{fig:jpsiw}The behaviour according to our model of $\gamma^*p\rightarrow J/\psi p$ total cross section as function of $W$ is compared with data from Ref.~\cite{j1,j2} measured by the H1 and ZEUS Collaborations for several values of $Q^2$. The green bands are calculated accordingly with the uncertainties on the free parameter $|A_0|$.}
\end{figure*}

% sigma(W) ZEUS 98-00
%\begin{figure*}
%\includegraphics[clip,scale=0.427]{figure_dvcs_vmp-2011/jpsi/int/jpsi_int_w/sigmael_ZEUS_98-00_w_Q2_0.4_jpsi.eps}
%\includegraphics[clip,scale=0.427]{figure_dvcs_vmp-2011/jpsi/int/jpsi_int_w/sigmael_ZEUS_98-00_w_Q2_3.1_jpsi.eps}
%\includegraphics[clip,scale=0.427]{figure_dvcs_vmp-2011/jpsi/int/jpsi_int_w/sigmael_ZEUS_98-00_w_Q2_6.8_jpsi.eps}
%\includegraphics[clip,scale=0.427]{figure_dvcs_vmp-2011/jpsi/int/jpsi_int_w/sigmael_ZEUS_98-00_w_Q2_16.0_jpsi.eps}
%\caption{\label{fig:jpsiw_zeus}The behaviour according to our model of $\gamma^*p\rightarrow J/\psi p$ total cross section as function of $W$ is compared with data from Ref.~\cite{j1}  measured by the ZEUS Collaboration for several values of $Q^2$. The green bands are calculated accordingly with the uncertainties on the free parameter $|A_0|$.}
%\end{figure*}

% dsigma/dt H1 99-00
\begin{figure*}
\includegraphics[clip,scale=0.427]{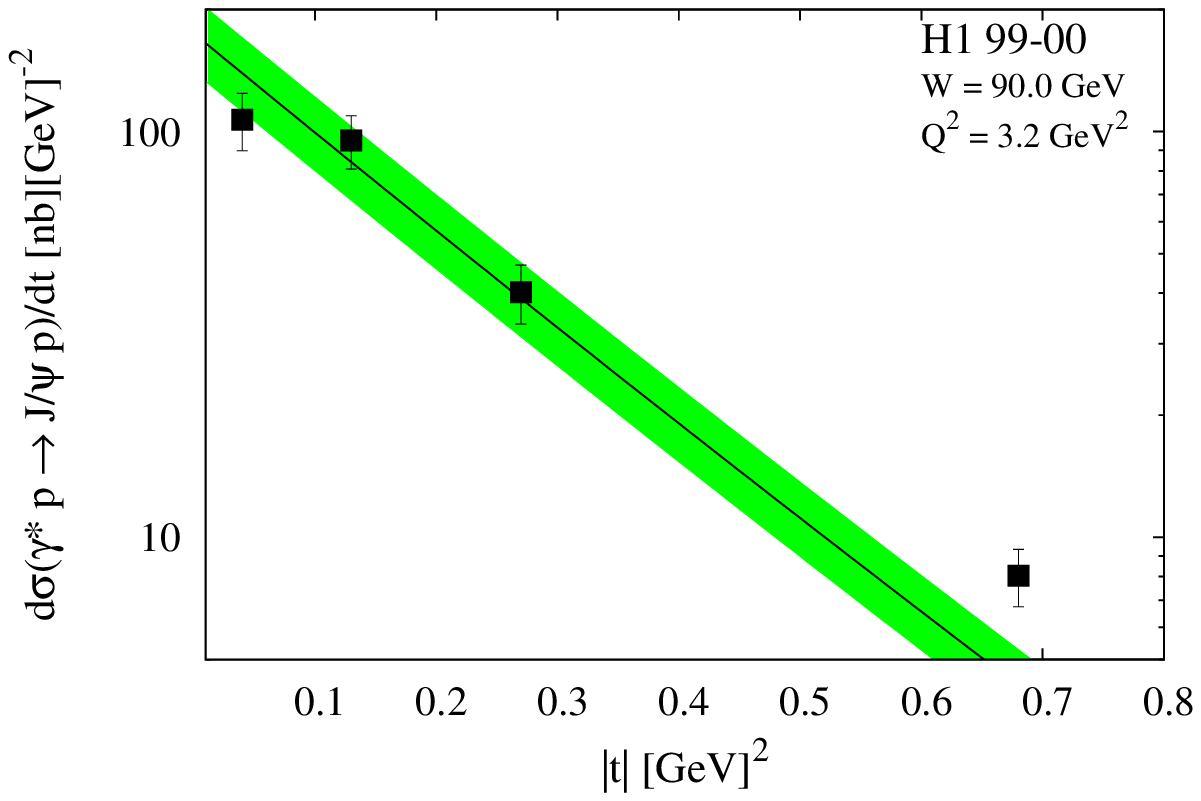}
\includegraphics[clip,scale=0.427]{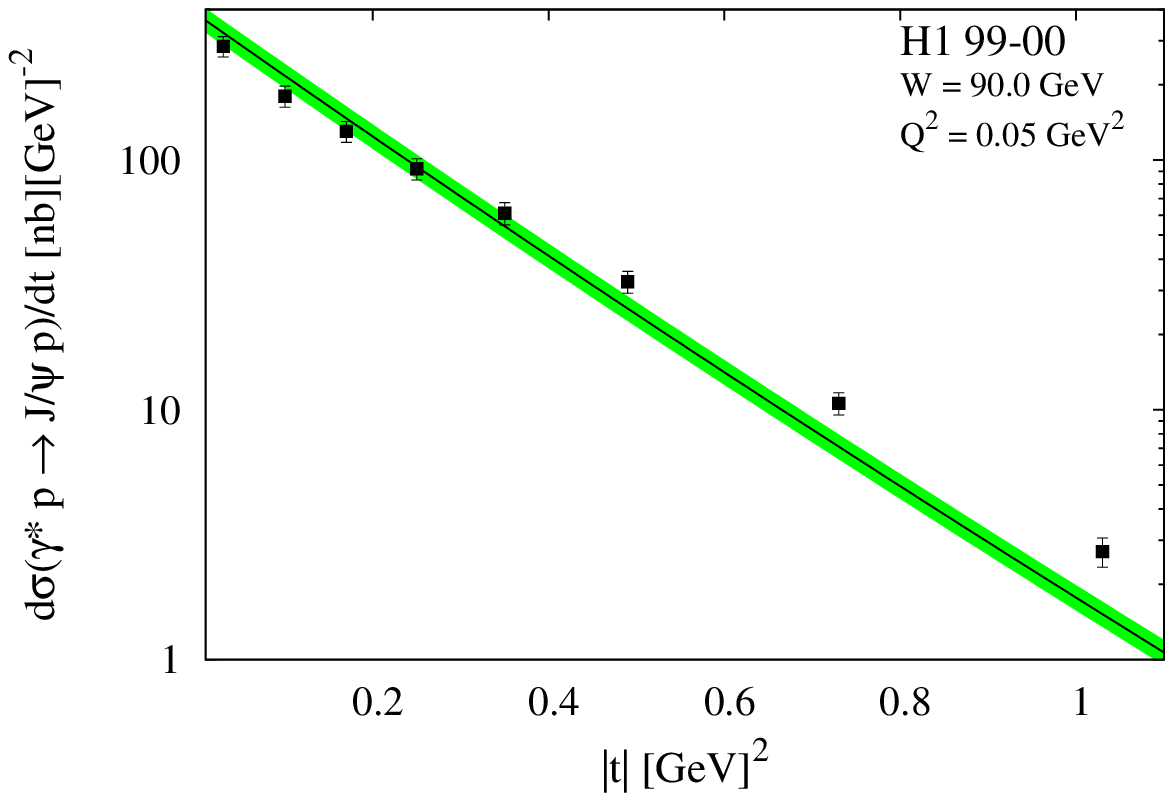}
\includegraphics[clip,scale=0.427]{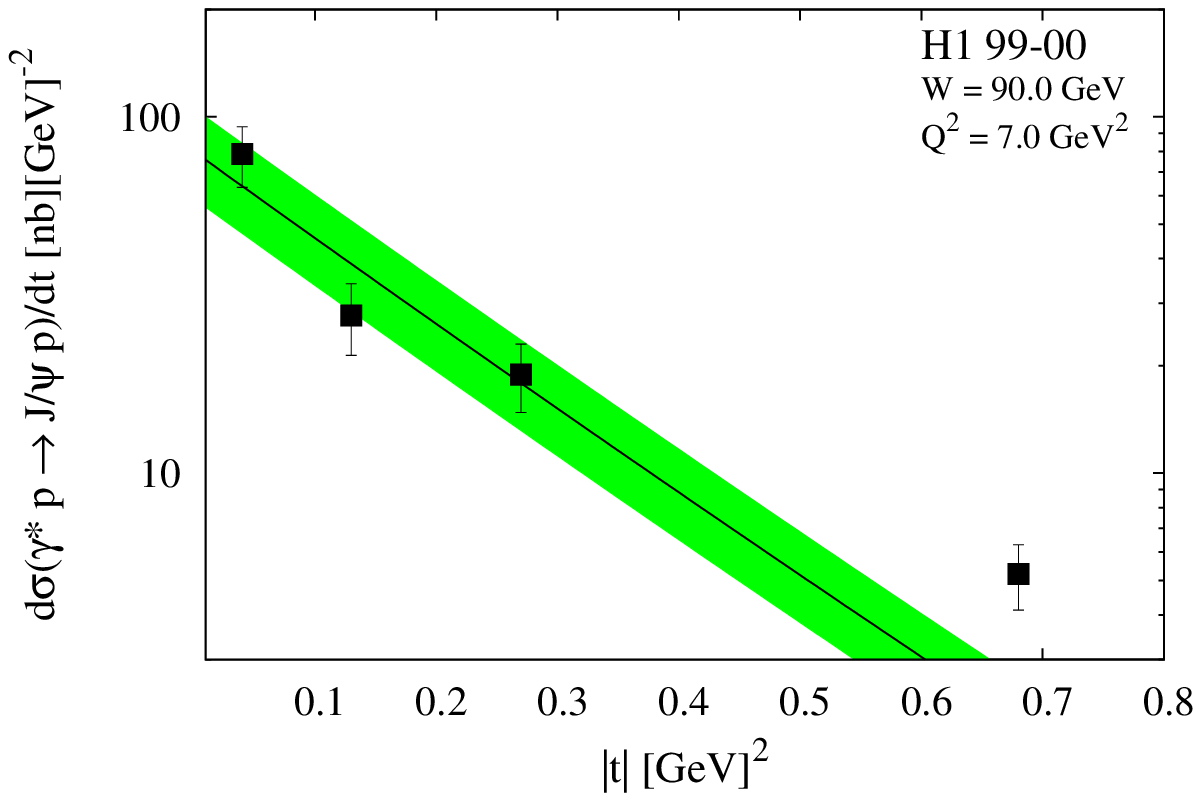}
\includegraphics[clip,scale=0.427]{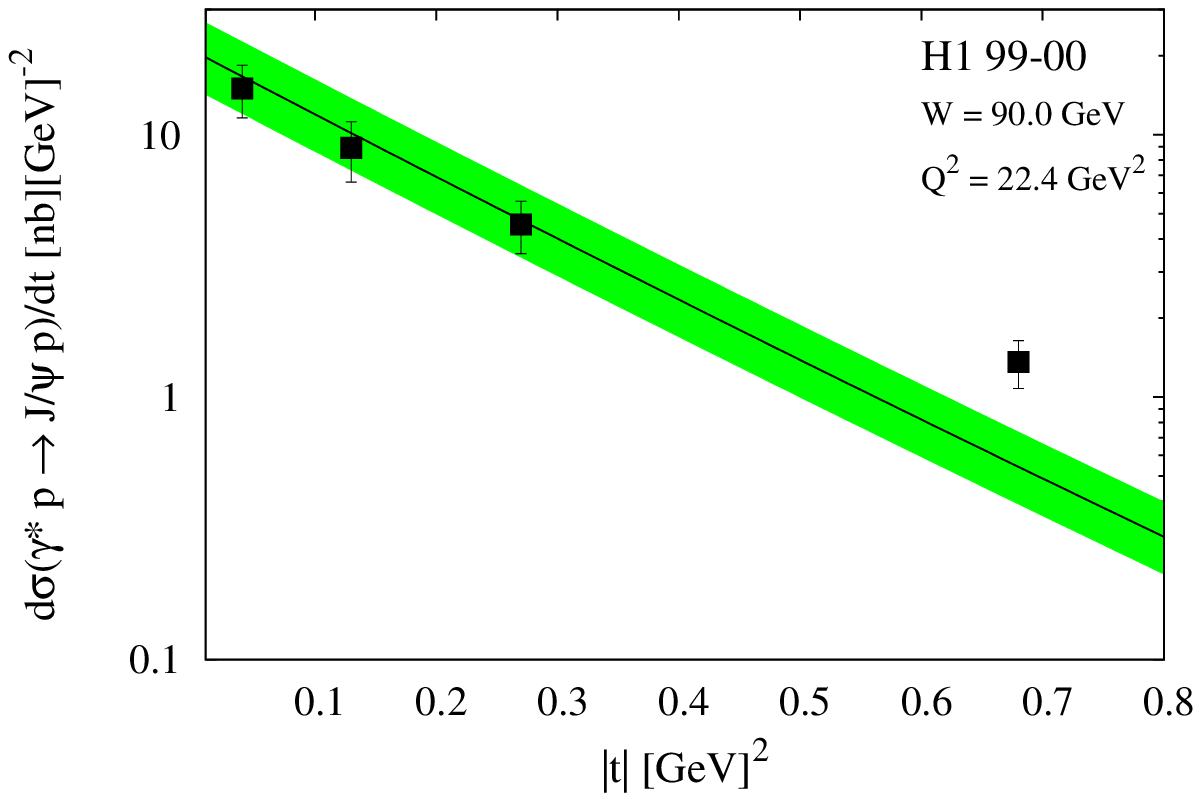}
\includegraphics[clip,scale=0.427]{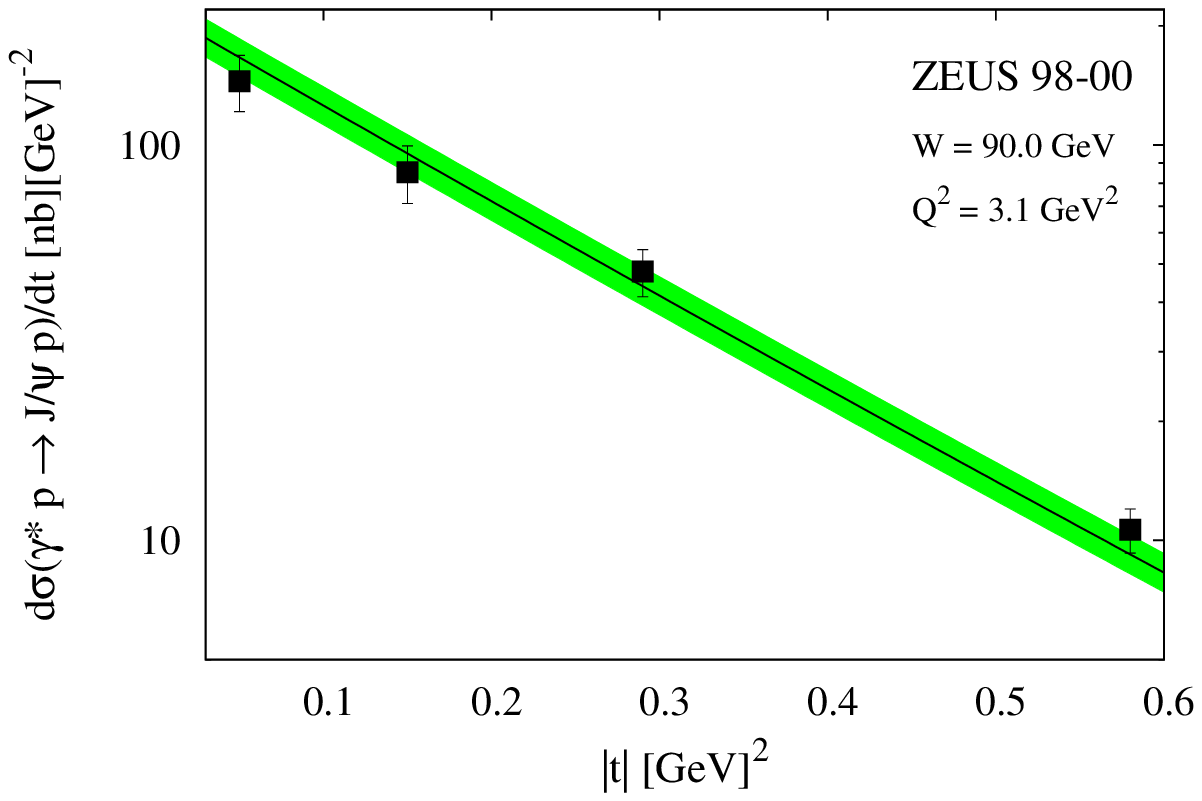}
\includegraphics[clip,scale=0.427]{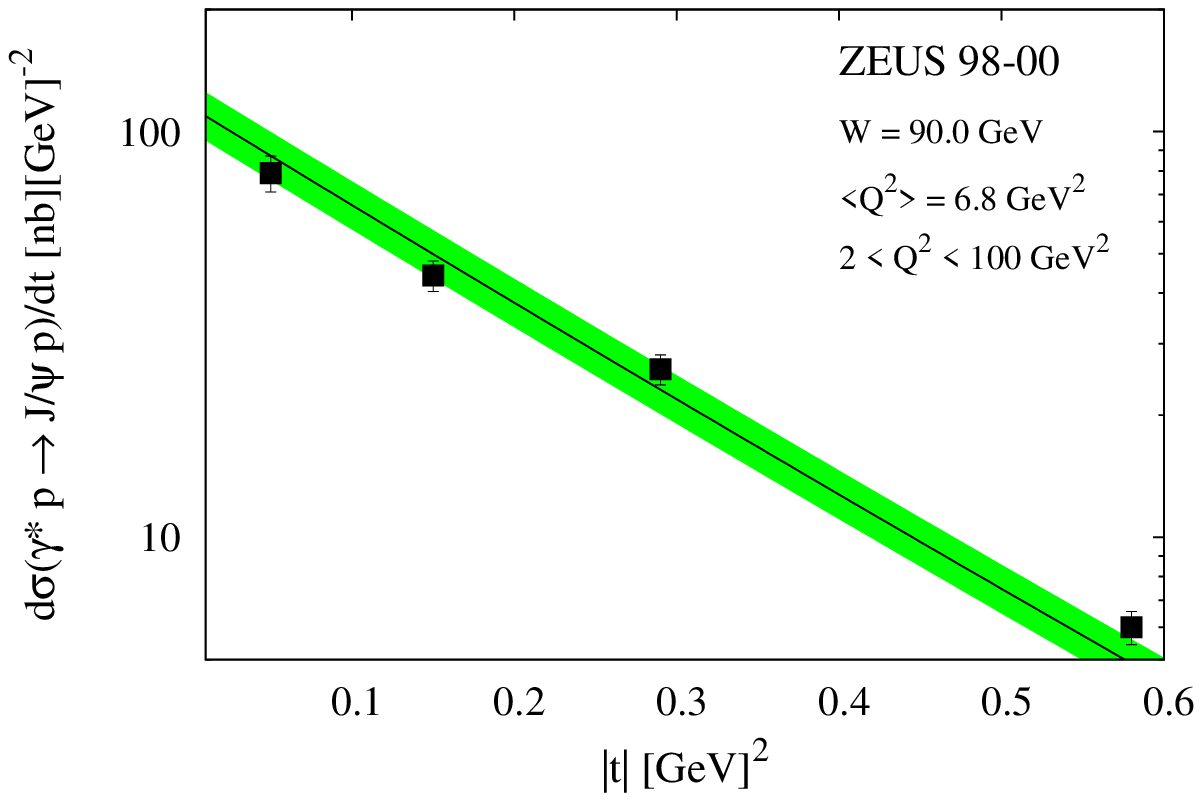}
\includegraphics[clip,scale=0.427]{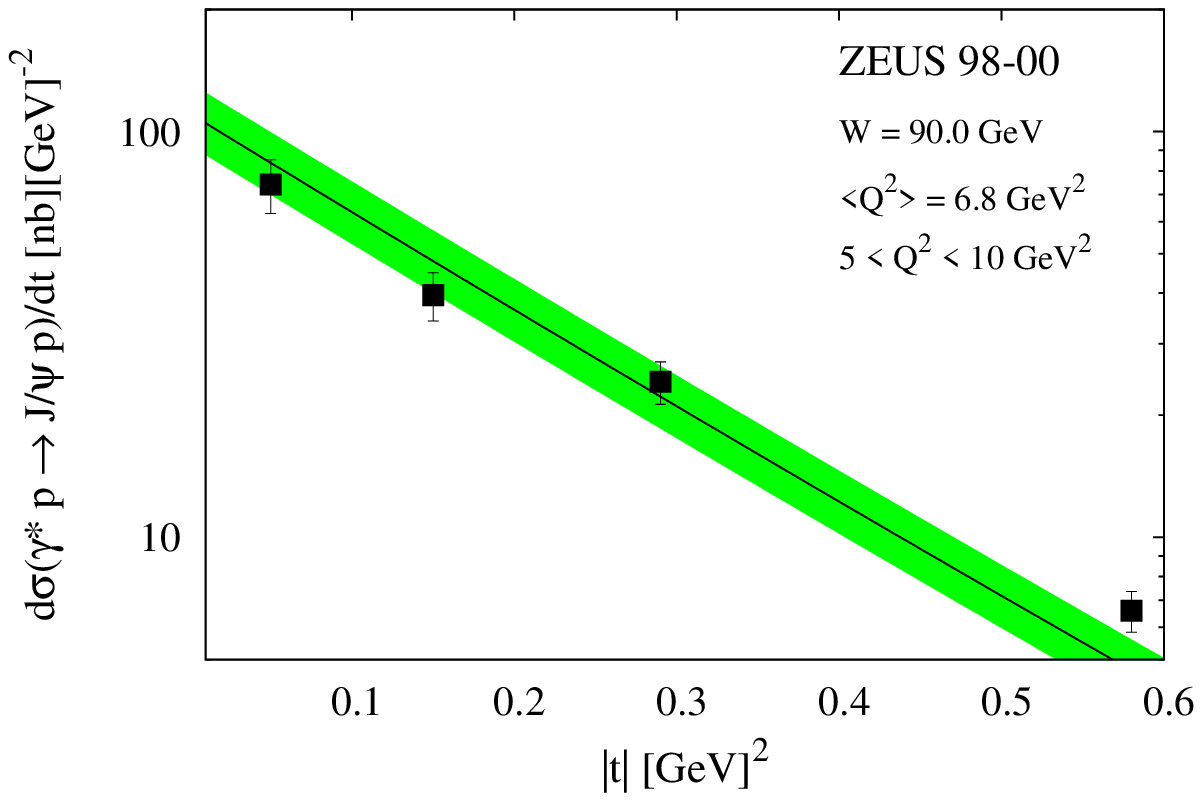}
\includegraphics[clip,scale=0.427]{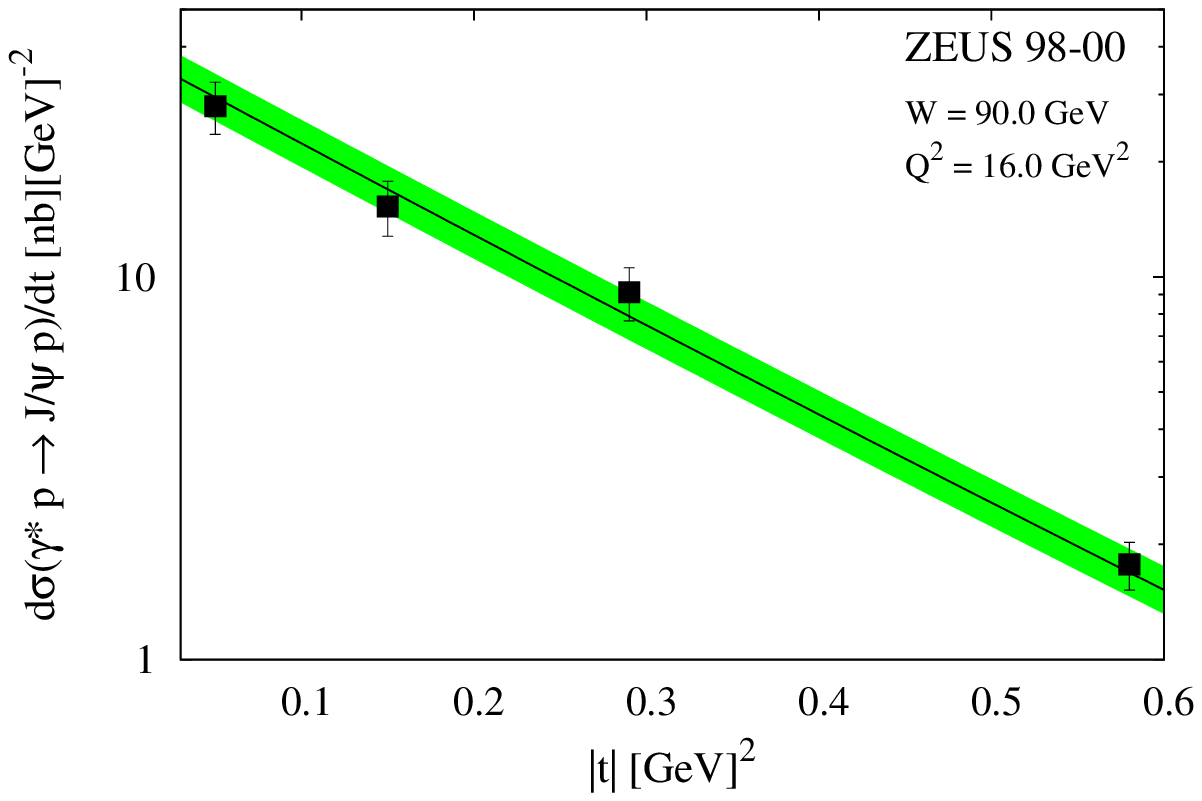}
\caption{\label{fig:jpsidsdt}The behaviour according to our model of $\gamma^*p\rightarrow J/\psi p$ differential cross section as function of $t$ is compared with data from Ref.~\cite{j1,j2} measured by the H1 and ZEUS Collaborations for several values of $Q^2$ and fixed $W$. The green bands are calculated accordingly with the uncertainties on the free parameter $|A_0|$.}
\end{figure*}

% dsigma/dt ZEUS 98-00
%\begin{figure*}
%\includegraphics[clip,scale=0.427]{figure_dvcs_vmp-2011/jpsi/diff_b2_int/jpsi_diff_zeus_98-00/dsigmadt_ZEUS_98-00_W-90_Q2-3.1_jpsi.eps}
%\includegraphics[clip,scale=0.427]{figure_dvcs_vmp-2011/jpsi/diff_b2_int/jpsi_diff_zeus_98-00/dsigmadt_ZEUS_98-00_W-90_Q2-6.8_2-100_jpsi.eps}
%\includegraphics[clip,scale=0.427]{figure_dvcs_vmp-2011/jpsi/diff_b2_int/jpsi_diff_zeus_98-00/dsigmadt_ZEUS_98-00_W-90_Q2-6.8_5-10_jpsi.eps}
%\includegraphics[clip,scale=0.427]{figure_dvcs_vmp-2011/jpsi/diff_b2_int/jpsi_diff_zeus_98-00/dsigmadt_ZEUS_98-00_W-90_Q2-16.0_jpsi.eps}
%\caption{\label{fig:jpsidsdt_zeus}The behaviour according to our model of $\gamma^*p\rightarrow J/\psi p$ differential cross section as function of $t$ is compared with data from Ref.~\cite{j1} measured by the ZEUS Collaboration for several values of $Q^2$ and fixed $W$. The green bands are calculated accordingly with the uncertainties on the free parameter $|A_0|$.}
%\end{figure*}

\end{document}